\documentclass{aa} 

\usepackage[breaklinks, colorlinks, citecolor=blue, linkcolor=blue]{hyperref}
\usepackage{amsmath}
\usepackage{mathtools}
\usepackage{graphicx}
\usepackage{placeins}
\usepackage{txfonts}
\usepackage{float}
\usepackage{lscape}
\usepackage{subcaption}

\makeatletter
\DeclareRobustCommand{\ion}[2]{%
  \text{#1\,\check@mathfonts\fontsize\sf@size\z@\selectfont #2}%
}
\makeatother

\begin{document}

   \title{Link between the properties of circumgalactic outflows and the recent star formation history of their host galaxies}

   \author{
   Ismael Pessa\inst{\ref{AIP}} \and 
   Lutz Wisotzki\inst{\ref{AIP}} \and
   Constanza Mu\~noz L\'opez\inst{\ref{AIP}} \and 
    Tanya Urrutia \inst{\ref{AIP}} \and 
    Ramona Augustin \inst{\ref{AIP}} \and
    Thierry Contini\inst{\ref{Tolouse}} \and
    Yucheng Guo\inst{\ref{Arizona}} \and
    Daria Kozlova \inst{\ref{AIP}} \and
    Haruka Kusakabe \inst{\ref{UTokyo}} \and 
    Sowgat Muzahid \inst{\ref{IUCAA}} \and
    John Pharo \inst{\ref{AIP}}
   }
   \institute{   
   Leibniz Institute for Astrophysics Potsdam (AIP), An der Sternwarte 16, 14482 Potsdam, Germany\label{AIP}\\
        \email{ipessa@aip.de}
   \and
   Université de Toulouse, CNES, CNRS, IRAP, Toulouse, France\label{Tolouse}
   \and
   School of Earth \& Space Exploration, Arizona State University, 781 Terrace Mall, Tempe, AZ 85287, USA\label{Arizona}
   \and Department of General Systems Studies, Graduate School of Arts and Sciences, The University of Tokyo, 3-8-1 Komaba, Meguro-ku, Tokyo, 153-8902, Japan\label{UTokyo}
   \and IUCAA, Postbag 4, Ganeshkind, Pune 411007, India\label{IUCAA}
    }
   \date{Received XX YY ZZZZ; Accepted XX YY ZZZZ}
  \abstract
  {Galactic outflows play a key role in galaxy evolution, yet their connection to the star formation history (SFH) of their host galaxies remains poorly constrained. Building on previous work modeling the extended \ion{Mg}{II} emission of galaxies observed with VLT/MUSE to infer outflow properties, we investigate how these properties depend on the recent SFH. We derive non-parametric SFHs by fitting spectroscopic and photometric data with linear combinations of single stellar population models, and compute star formation rates (SFRs) over different timescales sensitive to recent stochastic bursts. We find moderate correlations between these SFRs (and sSFRs) and the kinematic properties of the outflows, namely their maximum velocity and radial acceleration rate. Such correlations are not recovered using SFR tracers based on [\ion{O}{II}] luminosity or parametric SFHs, highlighting the importance of capturing bursty star formation in timescales consistent with the traveling time of ejected gas into the circumgalactic medium. Using the extent and kinematics of the \ion{Mg}{II} halos, we further estimate the outflow launching time for each galaxy and compute a corresponding SFR averaged over the relevant timescale. We identify a correlation between this SFR and the maximum outflow velocity, although it is not significantly stronger than the correlations obtained using fixed averaging timescales. This suggests that the properties of \ion{Mg}{II} halos are shaped by a series of star formation events rather than by a single dominant burst. Lastly, we find faster outflows in galaxies whose recent star formation is dominated by one or a few concentrated episodes, whereas galaxies with star formation spread more evenly over time exhibit slower outflows. These results indicate that the circumgalactic medium acts as a dynamical archive of recent star formation, linking bursty galaxy growth to outflow kinematics and providing a direct observational constraint on feedback models.}
   \keywords{galaxies: general --
                galaxies: evolution --
                galaxies: halos --
                galaxies: structure
               }

\titlerunning{}
\authorrunning{I. Pessa}
\maketitle
\section{Introduction}

Galactic-scale outflows are widely considered a key driver of the kinematics of the circumgalactic medium \citep[CGM, for a review, see][]{Tumlinson2017} in the vicinity of galaxies \citep[$\lesssim60$ kpc,][]{Weng2023} and play a central role in galaxy evolution \citep[][]{Veilleux2020}. By expelling metal-enriched gas from the interstellar medium (ISM) into the CGM, these outflows shape the distribution of heavy elements both within galaxies and in their surrounding environments, while also removing gas that could otherwise form stars and thereby regulating the baryon cycle through ejective feedback. The energy and momentum they deposit into the surrounding gas can further suppress subsequent star formation through preventive feedback \citep[see, e.g.,][]{Wang2009}. Under some conditions, feedback may also promote star formation through gas compression \citep[see, e.g.,][]{Luisi2021}. Outflows further contribute to the redistribution of baryons within galaxies, thereby influencing their structural evolution. Their importance is highlighted by cosmological simulations, which rely on efficient feedback-driven winds to reproduce key observables such as the galaxy stellar mass function \citep[e.g.,][]{Somerville2015,Schaye2015,Pillepich2018}.

In galaxies with stellar masses below $\log M_{*}/M_{\odot}\lesssim10.5$, such large-scale outflows are predominantly powered by stellar feedback \citep{Buchner2024}. Star formation occurs in clustered environments within giant molecular clouds (GMCs), where gravitational collapse and fragmentation lead to the formation of stellar populations rather than individual stars. The resulting collective feedback from stellar winds and numerous supernova explosions injects substantial energy and momentum into the ISM on relatively short timescales, ultimately driving galactic-scale outflows.

Absorption-line spectroscopy has been the primary observational approach for investigating the CGM, including circumgalactic outflows. Using bright background sources such as quasars or galaxies, this technique probes intervening gas along the line of sight and constrains its column densities, metallicities, and kinematics. Extensive samples of QSO--galaxy pairs have established several fundamental characteristics of the CGM. For example, \citet{Bouche2012} found that absorbing systems preferentially align with either the projected minor axes of galaxies, interpreted as signatures of outflows, or with their major axes, tracing inflowing material. This result provided early evidence that gas flows around galaxies are anisotropic, and similar trends have since been reported in other studies \citep{Zabl:2019ija, Schroetter:2019es}. More recently, \citet{Dutta2026} showed that while cool gas exhibits this anisotropic distribution, the warm--hot, highly ionized \ion{O}{VI}-bearing gas ($T\sim10^{5}$--$10^{6}$ K) appears to be distributed more uniformly, with no significant azimuthal dependence. However, other studies have found no evidence for an azimuthal dependence of \ion{Mg}{II} absorption \citep[e.g.,][]{Dutta2020}, suggesting that the spatial distribution of circumgalactic metals is complex and may depend on additional factors, such as galactocentric distance, or the presence of multiple galaxies associated with the same absorption system.

Correlations between cool ($T\sim10^{4}$ K) outflow characteristics and host-galaxy properties have also been explored through absorption-line studies. For instance, \cite{Schroetter:2019es} used data from the MEGAFLOW survey \citep{Bouche2025} to measure outflow velocities, mass outflow rates, and mass-loading factors. They found mass outflow rates comparable to the host-galaxy star formation rate (SFR) and concluded that cool gas in high-mass systems is likely to fall back because outflow velocities generally remain below the escape velocity of the host halo. \citet{Schroetter2024} used the same dataset and reported that outflow velocities scale with both the SFR and the SFR surface density, with the tightest relations found in galaxies exhibiting the most powerful winds. These trends are consistent with expectations for outflows driven by stellar feedback \citep[e.g.,][]{Muratov2015, Nelson2019}. Constraints on cool outflows have also been obtained from down-the-barrel absorption-line studies, which probe gas directly along the line of sight to the host galaxy. These studies have revealed correlations between outflow properties and host-galaxy characteristics such as stellar mass, star formation rate, and stellar surface density \citep[see, e.g.,][]{Martin:2012dx,Chisholm2015,Rong2026}.

While absorption-line studies are sensitive to extremely low gas column densities \citep[e.g.,][]{Werk2013}, making them ideal for probing diffuse gas in the CGM, they sample only discrete sightlines through a halo. As a result, they do not provide a complete picture of the CGM and outflow structure around individual galaxies. Several studies have therefore investigated cool galactic outflows through their direct emission rather than through absorption against background sources.

In particular, \ion{Mg}{II} is an excellent tracer of cool outflowing gas. As a resonant transition, \ion{Mg}{II} photons can scatter off ions in the CGM, producing spatially extended emission. In addition, magnesium is a relatively abundant metal, and the second ionization potential of Mg (15.0 eV) is slightly higher than that of hydrogen, allowing \ion{Mg}{II} to remain abundant in predominantly neutral gas. These properties make \ion{Mg}{II} uniquely suited for directly mapping cool circumgalactic gas in emission.

Because the surface brightness of extended \ion{Mg}{II} emission is typically very low, some recent studies have relied on stacking, combining large samples of galaxies to enhance the low-surface-brightness signal of extended \ion{Mg}{II} halos \citep{Guo2023b,Dutta2023}. \citet{Guo2023b} stacked the \ion{Mg}{II} emission from 172 star-forming galaxies in the MUSE Hubble Ultra Deep Field surveys after aligning them with respect to their semi-major and semi-minor axes. They detected a significant excess of \ion{Mg}{II} emission along the semi-minor axis, consistent with bipolar outflows being a widespread feature among star-forming galaxies, extending up to 10 kpc from the central galaxies.

Although stacking provides a powerful statistical view of circumgalactic \ion{Mg}{II} emission, it inevitably averages over the diverse morphologies and kinematic properties of individual systems. Deep integral-field observations have therefore been used to investigate \ion{Mg}{II} halos around individual galaxies. Using observations obtained with the Multi Unit Spectroscopic Explorer \citep[MUSE;][]{Bacon:2010jn}, \citet{Zabl2021} and \citet{Leclercq2022} detected extended \ion{Mg}{II} emission reaching projected distances of several tens of kiloparsecs from their host galaxies. Similarly, \citet{Burchett2021} used the Keck Cosmic Web Imager (KCWI) to map the extended \ion{Mg}{II} emission around a galaxy previously studied with slit spectroscopy by \citet{Rubin:2011bm}, and compared the observations with the radiative transfer models of \citet{Prochaska:2011eqa}.

We recently reported the discovery of a prominent \ion{Mg}{II}-emitting halo around a galaxy at $z=0.737$ \citep{Pessa2024}. The emission extends to projected distances of $\sim30$--$40$ kpc, far beyond the stellar component ($\sim1.5$ kpc), and exhibits strong central absorption together with redshifted emission, producing a P-Cygni profile characteristic of resonant scattering in an optically thick outflow. By modeling the spatially resolved \ion{Mg}{II} emission with a semi-analytic outflow model, we inferred a nearly face-on biconical outflow with a wide opening angle ($\sim60^\circ$) and a radially increasing velocity reaching $\sim120$ km s$^{-1}$ at 10 kpc.

These studies demonstrate the power of integral-field spectroscopy to resolve the morphology and kinematics of cool circumgalactic outflows in individual galaxies. However, they are based on only a handful of objects and therefore do not provide a statistical framework for investigating how the properties of \ion{Mg}{II} halos vary across the galaxy population or how they relate to the characteristics of their host galaxies.

In a recent study \citep{Pessa2026}, we extended the analysis of \citet{Pessa2024} to a sample of nearly 50 galaxies drawn from the MUSE Cosmic Assembly Survey Targeting Extragalactic Legacy Fields (MUSCATEL; PIs: L. Wisotzki, R. Bacon, T. Contini). We modeled their extended \ion{Mg}{II} emission to infer the properties of their galactic outflows. This was the first study to investigate galactic outflows traced by extended \ion{Mg}{II} emission in a statistically significant sample, rather than focusing on a single exceptional case. We found tentative trends linking host-galaxy and outflow properties, with more massive galaxies appearing to host faster outflows and possibly larger optical depths. The outflow geometry also seems to depend on host-galaxy mass, with lower-mass galaxies primarily hosting isotropic outflows, whereas biconical geometries are more common among more massive systems, consistent with the results from \citet{Guo2023b}.

Beyond their dependence on directly measured galaxy parameters, such as stellar mass or star formation rate, the properties of the CGM may also be linked to the recent star formation history (SFH) of galaxies. \cite{Harvey2025}, for example, showed that post-starburst galaxies exhibit stronger \ion{Mg}{II} absorption in nearby quasar sightlines than actively star-forming systems, suggesting the presence of larger cool-gas reservoirs in their halos. Such behavior is anticipated from theoretical models, as winds launched by star formation require time to propagate to CGM scales, introducing a delay between the onset of feedback and its observable impact in the halo. This suggests that the extended emission (or absorption observed in background sightlines) around a galaxy today may be more closely linked to earlier episodes of star formation than to its present-day SFR. At higher redshift, JWST spectroscopy has revealed that massive galaxies can host fast, cool outflows traced by \ion{Mg}{II}, \ion{Fe}{II}, and \ion{Na}{I} absorption even after star formation has already declined, suggesting that winds launched during earlier star-forming phases persist into the quenching stage \citep{Valentino2025}.

On the theoretical side, \cite{Kang2025} showed that different prescriptions for star formation on submolecular cloud scales can significantly alter the timing, efficiency, and burstiness of star formation, which in turn strongly affects the generation of galactic winds. In particular, burstier SFHs driven by clustered star formation and efficient feedback lead to stronger and more variable outflows, whereas smoother SFHs produce weaker winds and reduced gas expulsion. These results highlight the close connection between the cool CGM and the recent SFH of galaxies, as the strength and temporal variability of outflows depend sensitively on how and when stars form. Consequently, the CGM may retain imprints of past star formation episodes, providing a valuable observational window into the feedback history of galaxies.

However, a causal link between the recent SFH of galaxies and the properties of galactic outflows remains largely unexplored, particularly from the perspective of emission-line studies. In this paper, we investigate whether outflow properties inferred from observed \ion{Mg}{II} halos are modulated by past episodes of star formation that would not be captured by present-day SFR measurements. The paper is structured as follows. In Sec.~\ref{sec:obs}, we describe the data used in this work, and in Sec.~\ref{sec:sample} we present the galaxy sample. In Sec.~\ref{sec:methods}, we describe the methodology, including how we infer the SFHs and galactic outflow properties of our sample galaxies. We present the results in Sec.~\ref{sec:results}, discuss their implications in Sec.~\ref{sec:discussion}, and summarize our findings in Sec.~\ref{sec:summary}. We adopt a flat concordance cosmology with $H_0 = 70$ km s$^{-1}$ Mpc$^{-1}$ and $\Omega_m = 0.3$. All quoted transverse distances are physical rather than comoving.

\section{Observational data} 
\label{sec:obs}

\subsection{The MUSCATEL survey}
\label{sec:muscatel}

We use integral-field spectroscopic observations obtained with MUSE on the ESO Very Large Telescope \citep{Bacon:2010jn} as part of the MUSCATEL survey. In wide-field mode, MUSE provides a contiguous $1'\times1'$ field of view sampled at $0\farcs2$ and covers the wavelength range 4700--9350 $\AA$ with a spectral sampling of 1.25 $\AA$. The spectral resolution varies from $\sim$190 to $\sim$80 km s$^{-1}$ (FWHM) from the blue to the red end of the wavelength range.

A detailed presentation of the survey is given in \citet{Pessa2026}; here, we summarize the elements most relevant to this work. MUSCATEL targets parallel fields of the Hubble Frontier Fields \citep[HFF;][]{Pagul2021}, focusing on four clusters accessible from the VLT: Abell 2744, MACS J0416, Abell S1063, and Abell 370. The observations follow a tiered "wedding-cake" strategy that balances area and depth. A shallow $3'\times3'$ mosaic (100 min per pointing) traces the footprint of the HST/ACS imaging, while an intermediate $2'\times2'$ region reaches a depth of 5 h, matching the HST/WFC3 coverage. At the center, a single $1'\times1'$ pointing achieves a depth of 25 h (10 h for Abell 370), enabling the detection of faint emission-line sources. Nearly all observations were obtained using the VLT adaptive optics facility \citep[AO;][]{Leibundgut2017}, with only a small number of exposures acquired without AO assistance.

The data were reduced using the standard MUSE pipeline \citep{Weilbacher2014, Weilbacher2020}, complemented by additional post-processing steps (e.g., superflat correction; \citealt{Bacon2023}), following an approach similar to that adopted by \citet{Urrutia2019}. An important post-processing step was the replacement of the combined variance cube with a self-calibrated "effective variance" cube corrected for resampling\footnote{The voxel-by-voxel variances obtained by formal error propagation in the MUSE pipeline systematically underestimate the true uncertainties in the cube because of the resampling required to construct it, which transfers part of the variance into covariance terms.}. See \citet{Urrutia2019} and \citet{Weilbacher2020} for details of this correction. Emission-line sources were identified using LSDCat \citep{Herenz:2017er} on the reduced datacubes. We adopt a detection threshold of S/N = 5 (S/N = 6 for the shallow fields) to minimize spurious detections.

This process yields a catalog of 3245 emission-line-selected galaxies with robust redshift measurements obtained through supervised inspection. Further details on the catalog construction, redshift determination, and overall survey characterization will be presented in a dedicated MUSCATEL survey paper.

\subsection{Ancillary Photometric data}
\label{sec:photometry}

In addition to the MUSE spectroscopy, the MUSCATEL fields benefit from extensive archival imaging spanning optical to mid-infrared wavelengths. Deep observations from the Hubble Space Telescope (HST) provide coverage in the F435W, F606W, F775W, F814W, and F850LP bands with ACS, and in the F105W, F125W, F140W, and F160W bands with WFC3 \citep{Shipley2018, Pagul2021}. Complementary VLT/HAWK-I $K_s$ and \textit{Spitzer}/IRAC (3.6 and 4.5 $\mu$m) imaging are also available. Recent James Webb Space Telescope observations are available as well \citep[e.g.,][]{Bezanson2024, Rihtar2025}, but are not included in this work because their processing is still ongoing.

We associate each emission-line-selected galaxy with its HST counterpart through positional cross-matching followed by visual inspection to ensure reliable identifications. Given the relatively bright continua of sources in our sample (median F814W $\sim$ 23.4 mag), counterpart assignment is generally unambiguous. We use the multi-band photometry to perform both photometric spectral energy distribution (SED) fitting with FAST (see Sec.~\ref{sec:SFR_photo_only}) and joint spectro-photometric SED fitting (see Sec.~\ref{sec:SFH_det}).

\section{Galaxy sample}
\label{sec:sample}

\begin{figure}
\centering
        \includegraphics[width=1\columnwidth]{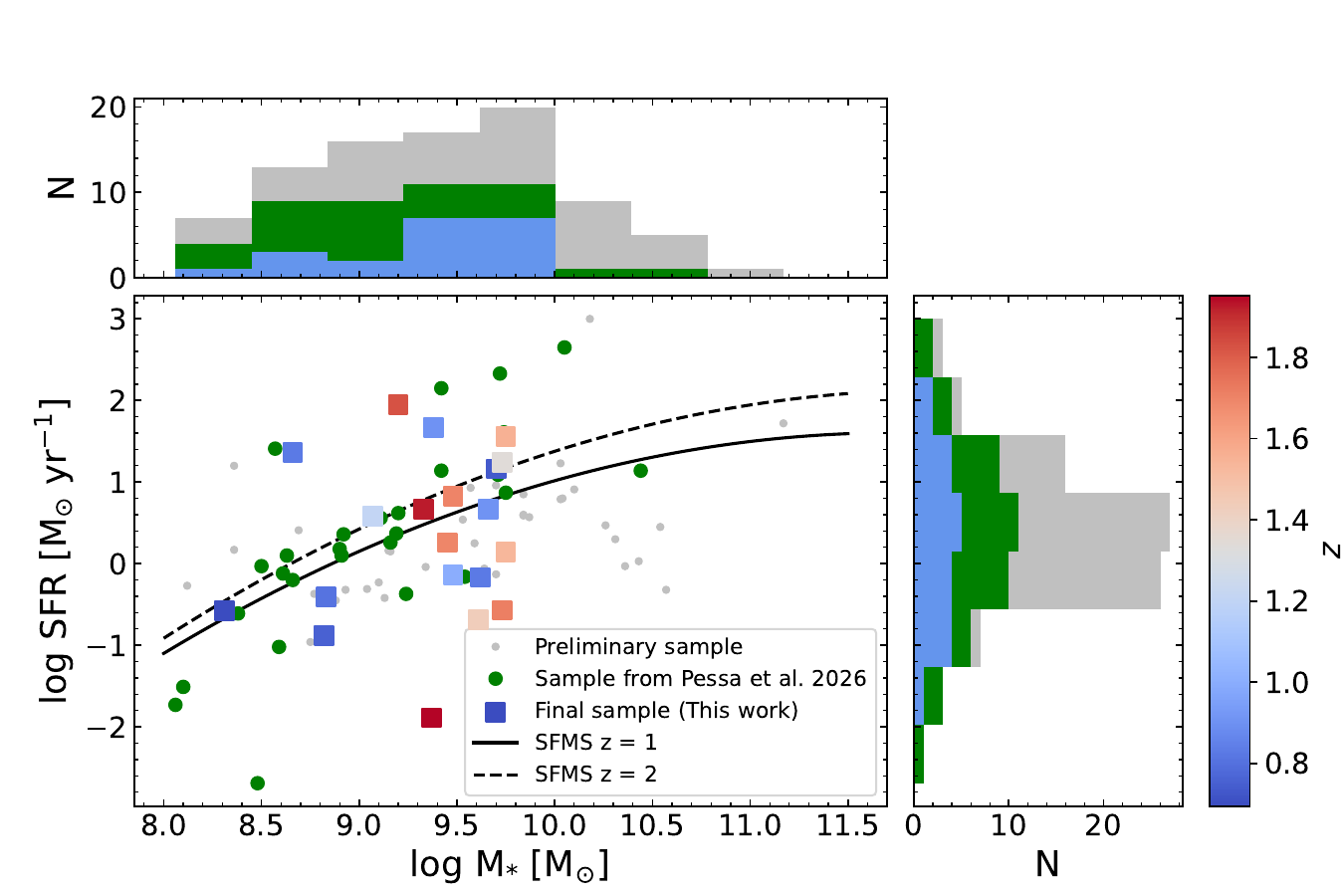}
\caption{Star formation rate of our sample galaxies as a function of their stellar mass. The SFR and stellar masses come from the SED fitting described in Sec.~\ref{sec:SFR_photo_only}. The sample used in this work is shown with squares color-coded by galaxy redshift (central panel) and by the blue histograms. The gray and green data points and histograms show the SFR and stellar mass distributions of the preliminary and final sample used in our previous work \citet{Pessa2026}, from which the sample used here was selected. The solid and dashed lines show the star formation main sequence at redshifts one and two, for reference, as calculated by \citet{Popesso2023}.}
\label{fig:sample_SFMS}
\end{figure}

This work is based on the sample presented by \citet{Pessa2026}, which consists of star-forming galaxies exhibiting circumgalactic outflows traced by P-Cygni \ion{Mg}{II} profiles and significant extended \ion{Mg}{II} emission beyond the extent of their stellar component. For details of the sample selection and characterization, we refer the reader to \citet{Pessa2026}. The preliminary sample consisted of 89 galaxies drawn from the MUSCATEL survey that exhibit \ion{Mg}{II} either in absorption, emission, or as a P-Cygni profile. We retained only those galaxies showing a P-Cygni profile in \ion{Mg}{II} (50 out of the 89 galaxies), as this feature is commonly interpreted as a signature of resonant scattering in an optically thick outflow, with absorption arising in the approaching (blueshifted) side of the outflow and redshifted emission produced by backscattering in the receding material \citep[e.g.,][]{Castor1970, Castor1979, Scuderi1992}. This clear outflow signature makes these systems particularly well-suited for studying the physical properties of galactic winds. Requiring galaxies to exhibit a P-Cygni profile effectively excluded several of the most massive galaxies from the sample (see Fig.~\ref{fig:sample_SFMS}), as \ion{Mg}{II} is expected to appear predominantly in absorption in such systems \citep[see][]{Feltre:2018in}.

We then selected galaxies with \ion{Mg}{II} emission above a given significance threshold \citep[see][for details of the exact procedure]{Pessa2026}. This selection leaves us with a sample of 47 galaxies drawn from the three depth tiers of the MUSCATEL survey: 22 from the deep fields, 17 from the medium-depth fields, and 8 from the shallow fields. The galaxies lie in the redshift range where \ion{Mg}{II} falls within the MUSE wavelength coverage, namely $0.7 \lesssim z \lesssim 2.3$, and all have a redshift confidence of 3, corresponding to secure redshift determinations based on multiple spectral features. The redshift range in which the \ion{Mg}{II} doublet falls within the MUSE AO wavelength gap is naturally excluded from the sample (approximately $1.07 \lesssim z \lesssim 1.17$).

Although the sample of \citet{Pessa2026} contains 47 galaxies, here we focus on a subset suitable for deriving recent SFHs through joint spectral and broadband photometric fitting. We require a minimum continuum S/N of five, as preliminary tests indicate that above this threshold the median uncertainties of the SFH-derived SFRs (see Sec.~\ref{sec:SFR_measurements}) decrease to below approximately $20\%$. The continuum S/N is computed as the median value across the full MUSE wavelength range. Spectra were extracted within apertures corresponding to twice the continuum half-light radius (HLR) measured with \textsc{GALFIT} \citep{Galfit_a, Galfit_b}.

This selection yields a sample of 23 galaxies. We further require a reduced chi-square of $\chi^{2}_{\rm red} < 3$ for the spectral fitting (see Sec.~\ref{sec:SFH_det}) to ensure an acceptable fit quality. We note that the median value of the $\chi^{2}_{\rm red}$ distribution is $\sim1$, without any need to rescale the uncertainties beyond the use of the effective variance described in Sec.~\ref{sec:muscatel}. This final selection leaves us with 20 galaxies for which the recent SFH can be robustly inferred. We additionally visually inspected all best-fitting models. For these galaxies, the properties of the galactic winds have already been derived through spatially resolved modeling of their extended \ion{Mg}{II} emission \citep[see][]{Pessa2024, Pessa2026}\footnote{One of these galaxies (A370-DF-099) was excluded from \citet{Pessa2026} because it fell slightly below the S/N threshold adopted in that work (S/N $\sim2.7$, compared to a threshold of 3) despite satisfying the P-Cygni criterion. However, given its proximity to the threshold, we opted not to exclude it a priori from the present sample. Since it subsequently satisfied all the selection criteria adopted in this work, it is included in our final sample.}. Table~\ref{tab:sample} summarizes the main characteristics of the galaxies used in this work.

Figure~\ref{fig:sample_SFMS} compares the star formation rates and stellar masses of the galaxies used in this work with those of both the preliminary \ion{Mg}{II}-selected sample and the final outflow sample presented by \citet{Pessa2026}, from which all but one of the galaxies in our final sample were drawn. Stellar masses and star formation rates were derived through SED fitting with FAST (see Sec.~\ref{sec:SFR_photo_only} for details on how these quantities were calculated). The figure shows that the galaxies selected for this work (blue histogram) preferentially occupy the high-stellar-mass end of the distribution defined by the \citet{Pessa2026} sample (green histogram). This is a direct consequence of the continuum S/N requirement described above. Although this selection biases the sample against lower-mass galaxies, it is necessary to derive statistically robust star formation histories. For comparison, the median stellar masses for the three samples are 9.42 (gray), 9.24 (green), and 9.47 (blue).

\begin{table*}
\fontsize{7pt}{7pt}\selectfont
\caption{Summary of the main properties of the galaxies from our final sample.}
\centering
\begin{tabular}{llccccccccccr}
\hline
ID & Name & log(M$_{*}$) & $z$  & $v_\mathrm{max}$   &  $\gamma$  &  log SFR$_{100}$  &  log SFR$_{200}$ & log SFR$_{400}$ & log SFR$_{\mathrm{SED}}$ &  log SFR$_{\mathrm{[OII]}}$ &  log SFR$_{\mathrm{HLR}}$ & MM$_{200}$ \\ 
 & & [$M_{\odot}$] & & [km s$^{-1}$]& & [$M_{\odot}$ yr$^{-1}$] & [$M_{\odot}$ yr$^{-1}$] & [$M_{\odot}$ yr$^{-1}$]& [$M_{\odot}$ yr$^{-1}$]& [$M_{\odot}$ yr$^{-1}$]& [$M_{\odot}$ yr$^{-1}$]&  \\ 
(1) & (2) & (3) & (4) & (5) & (6) & (7) & (8) & (9) & (10) & (11) & (12) & (13) \\ 
\hline
1 & M0416-DF-005 & 8.3 & 0.69428 & 419$\pm$145 & 0.9$\pm$0.4 & -1.6$\pm$0.1& -1.9$\pm$0.1& -2.2$\pm$0.1 &  -0.58 &  -0.35 &  -2.8$\pm$1.2 & 0.6$\pm$0.1\\ 
2 & AS1063-MF3-091 & 8.7 & 0.83008 & 319$\pm$91 & 1.4$\pm$0.4 & -0.7$\pm$0.1& -1.0$\pm$0.1& -1.3$\pm$0.1 &  1.36 &  0.89 &  -0.7$\pm$0.1 & 0.5$\pm$0.1\\ 
3 & M0416-DF-024 & 8.8 & 0.75516 & 519$\pm$108 & 0.6$\pm$0.3 & -1.5$\pm$0.1& -1.8$\pm$0.1& -2.1$\pm$0.1 &  -0.88 &  -0.02 &  - & 0.6$\pm$0.2\\ 
4 & A370-DF-099 & 8.8 & 0.81014 & 284$\pm$102 & 1.0$\pm$0.5 & -1.2$\pm$0.3& -1.5$\pm$0.3& -1.8$\pm$0.3 &  -0.41 &  -0.05 &  -1.1$\pm$0.7 & 0.6$\pm$0.2\\ 
5 & M0416-DF-295 & 9.1 & 1.20755 & 399$\pm$53 & 0.7$\pm$0.3 & -0.6$\pm$0.3& -0.4$\pm$0.1& -0.7$\pm$0.2 &  0.58 &  0.20 &  -0.6$\pm$0.4 & 0.4$\pm$0.2\\ 
6 & A2744-DF-002 & 9.2 & 1.82717 & 718$\pm$23 & 1.8$\pm$0.1 & 0.6$\pm$0.1& 0.3$\pm$0.1& 0.0$\pm$0.1 &  1.95 &  - &  0.5$\pm$0.0 & 0.1$\pm$0.0\\ 
7 & A2744-DF-202 & 9.3 & 1.92721 & 555$\pm$81 & 1.5$\pm$0.3 & 1.0$\pm$0.1& 0.8$\pm$0.0& 0.5$\pm$0.0 &  0.67 &  - &  1.2$\pm$0.1 & 0.4$\pm$0.1\\ 
8 & AS1063-DF-240 & 9.4 & 1.95230 & 479$\pm$99 & 1.1$\pm$0.3 & 0.0$\pm$0.3& -0.2$\pm$0.3& 0.6$\pm$0.1 &  -1.89 &  - &  0.1$\pm$0.3 & 0.7$\pm$0.2\\ 
9 & M0416-SF7-012 & 9.4 & 0.90457 & 425$\pm$15 & 0.8$\pm$0.2 & 0.6$\pm$0.1& 0.3$\pm$0.1& 0.0$\pm$0.1 &  1.67 &  0.99 &  1.0$\pm$0.1 & 0.2$\pm$0.0\\ 
10 & A2744-DF-227 & 9.4 & 1.69302 & 401$\pm$41 & 1.2$\pm$0.3 & -0.2$\pm$0.4& 0.1$\pm$0.3& 0.5$\pm$0.1 &  0.26 &  - &  -0.8$\pm$0.8 & 0.3$\pm$0.1\\ 
11 & AS1063-MF4-024 & 9.5 & 1.69904 & 519$\pm$78 & 1.5$\pm$0.3 & 0.7$\pm$0.1& 0.4$\pm$0.1& 0.9$\pm$0.0 &  0.83 &  - &  1.1$\pm$0.1 & 0.1$\pm$0.1\\ 
12 & AS1063-SF1-015 & 9.5 & 1.00275 & 574$\pm$101 & 0.6$\pm$0.3 & 0.8$\pm$0.0& 0.5$\pm$0.0& 0.2$\pm$0.0 &  -0.14 &  0.96 &  1.0$\pm$0.0 & 0.0$\pm$0.0\\ 
13 & M0416-MF3-043 & 9.6 & 1.43364 & 582$\pm$70 & 1.2$\pm$0.4 & 0.2$\pm$0.2& 0.4$\pm$0.0& 0.1$\pm$0.0 &  -0.69 &  0.75 &  0.6$\pm$0.2 & 0.4$\pm$0.1\\ 
14 & AS1063-MF3-004 & 9.6 & 0.83087 & 430$\pm$79 & 1.2$\pm$0.2 & 0.3$\pm$0.1& 0.0$\pm$0.0& -0.3$\pm$0.0 &  -0.17 &  0.58 &  -0.8$\pm$0.4 & 0.6$\pm$0.2\\ 
15 & M0416-DF-032 & 9.7 & 0.90712 & 585$\pm$78 & 1.7$\pm$0.3 & -0.9$\pm$0.4& -0.3$\pm$0.1& -0.5$\pm$0.0 &  0.67 &  0.72 &  -0.4$\pm$0.0 & 0.2$\pm$0.1\\ 
16 & AS1063-MF3-009 & 9.7 & 0.73470 & 490$\pm$121 & 0.5$\pm$0.3 & -0.7$\pm$0.3& -0.2$\pm$0.0& -0.5$\pm$0.0 &  1.16 &  0.58 &  -0.6$\pm$0.4 & 0.2$\pm$0.1\\ 
17 & AS1063-DF-029 & 9.7 & 1.71289 & 685$\pm$13 & 2.4$\pm$0.1 & 0.8$\pm$0.1& 0.5$\pm$0.1& 0.8$\pm$0.0 &  -0.57 &  - &  0.7$\pm$0.1 & 0.5$\pm$0.1\\ 
18 & M0416-DF-147 & 9.7 & 1.33893 & 555$\pm$113 & 1.3$\pm$0.4 & 0.2$\pm$0.4& 0.8$\pm$0.5& 0.5$\pm$0.4 &  1.24 &  0.49 &  -0.1$\pm$0.9 & 0.1$\pm$0.1\\ 
19 & AS1063-MF2-011 & 9.8 & 1.53938 & 546$\pm$90 & 0.5$\pm$0.1 & 0.5$\pm$0.0& 0.2$\pm$0.0& -0.1$\pm$0.0 &  0.14 &  - &  0.8$\pm$0.0 & 0.6$\pm$0.0\\ 
20 & AS1063-MF3-055 & 9.8 & 1.55457 & 578$\pm$57 & 1.8$\pm$0.4 & 1.1$\pm$0.0& 0.8$\pm$0.0& 0.5$\pm$0.0 &  1.56 &  - &  1.5$\pm$0.0 & 0.2$\pm$0.0\\ 
\hline
\end{tabular}
\tablefoot{The name of each galaxy shown in column (2) indicates the MUSCATEL field and depth level of the data. Galaxies are sorted by stellar mass in ascending order, shown in column (3). The stellar masses are calculated via the SED fitting described in Sec.~\ref{sec:SFR_photo_only}.  Column (4) shows the redshift of each galaxy, with an expected uncertainty of $\sim50$ km s$^{-1}$. Columns (5) and (6) show the maximum outflow velocity and outflow acceleration rate, respectively, determined by modeling the extended \ion{Mg}{II} emission following the methodology of \citet{Pessa2026}. Columns (7), (8), and (9) show the SFR computed from the non-parametric SFHs, in timescales of $100$, $200$, and $400$ Myr, respectively, as described in Sec.~\ref{sec:SFR_measurements}. Column (10) shows the SFR computed from the parametric SFH modeling of the photometric data of our sample galaxies, as described in Sec.~\ref{sec:SFR_photo_only}. Column (11) shows the SFR based on the [\ion{O}{II}] luminosity for those galaxies where the [\ion{O}{II}] line is inside the MUSE wavelength range, calculated following \citet{Kewley2004}, with typical uncertainties of $0.1$ dex. Column (12) shows the SFR computed from the non-parametric SFH, in timescales that depend on the \ion{Mg}{II} halo size, velocity, and acceleration rate, aiming to coincide with the ejection time of the galactic outflows, as explained in Sec.~\ref{sec:SFR_halo_size}. Column (13) corresponds to the multimodality parameter, which parametrizes the shape of the recent $200$ Myr of the SFH of our sample galaxies, as described in Sec.~\ref{sec:multimodality}. Uncertainties in columns (3) and (10) are of the order of $0.2$ and $0.3$ dex, respectively. The uncertainties in columns (5) and (6) come from the posterior distributions for the parameters computed via an MCMC fitting (see Sec.~\ref{sec:outflow_model}). Uncertainties in columns (7), (8), (9), (12), and (13) are computed by performing 100 MC iterations, perturbing the observed spectrum in each iteration (see Sec.~\ref{sec:SFH_det}).}
\label{tab:sample}
\end{table*}

\section{Methods}
\label{sec:methods}
\subsection{Full spectral and broadband fitting}
\label{sec:SFH_det}

The main goal of this paper is to connect the properties of galactic winds, inferred from modeling their extended \ion{Mg}{II} emission, with recent episodes of star formation in their host galaxies. To achieve this, we use the Penalized PiXel-Fitting method \citep[pPXF;][]{Cappellari:2004gm, Cappellari2017} to simultaneously fit the MUSE spectra and the available photometric data of our galaxies, deriving their SFHs in a non-parametric manner. pPXF performs a $\chi^2$-minimization to determine the linear combination of single stellar population (SSP) templates that best reproduces the observations, with the contribution of each SSP being interpreted in terms of the SFH of the galaxy. The resulting SFHs allow us to measure star formation rates averaged over different timescales and thereby quantify the strength of recent star formation episodes.

For each galaxy, we fitted the spectrum extracted from the MUSE datacube using an aperture with a diameter equal to twice the HLR. The fitting procedure closely follows that described in \citet{MunozLopez2025}. We model the observed spectra using Flexible Stellar Population Synthesis templates \citep[FSPS;][]{Conroy2009, Conroy2010}, adopting MIST stellar evolutionary tracks \citep{Choi2016} and a Salpeter initial mass function \citep{Salpeter1955}. In the optical regime ($\lambda \sim 3600$--$7400\,\AA$), the FSPS models are based on the MILES stellar library \citep{Vazdekis2015}. At wavelengths outside this range, they rely on the BaSel stellar library \citep{Lastennet2002}, extending the model coverage to $91\,\AA < \lambda < 160\,\mu$m. Prior to the fitting, all templates were convolved to match the wavelength-dependent spectral resolution of MUSE.

The fitting was performed using nine metallicity bins, [$M/H$] = [-1.75, -1.5, -1.25, -1.0, -0.75, -0.5, -0.25, 0.0, 0.25], and a logarithmically spaced age grid with a spacing of 0.1 dex. The minimum age was set to 1 Myr, while the maximum age corresponds to the age of the Universe at the redshift of each galaxy. We adopt the FSPS models instead of alternatives such as eMILES \citep{Vazdekis:2016ho}, which also provide broad wavelength coverage, because of their more extensive age coverage. This is particularly useful for the analysis presented in Sec.~\ref{sec:SFR_halo_size}, where it is relevant to probe stellar populations younger than the estimated launching times of the outflows. We note, however, that the results presented throughout this paper are largely unchanged when eMILES templates are used instead.

Because our primary interest is in recent star formation episodes, we do not apply regularization during the fitting, as regularization is designed to produce smoother SFHs. For each galaxy, we performed 100 Monte Carlo (MC) realizations, by perturbing both the observed spectrum and photometric measurements according to their corresponding uncertainties, assuming Gaussian error distributions. The final SFH is taken as the average over the 100 realizations. As discussed in Sec.~\ref{sec:obs}, the uncertainties of the MUSE datacubes were corrected for covariance effects through the use of effective variance cubes, providing a more realistic estimate of the flux uncertainties within a given aperture. Uncertainties on all derived quantities (e.g., SFRs measured over different timescales) were computed from the standard deviation of the values obtained across the 100 MC realizations. Via extensive testing, we found that this number of realizations provides stable estimates of the recovered stellar population properties and their uncertainties \citep[see also, e.g.,][]{Pessa2023}.

Since our goal is to recover stellar population properties, we masked several nebular emission lines, including [\ion{O}{II}] $\lambda\lambda3726,3728$, H$\delta$ $\lambda4101$, H$\gamma$ $\lambda4340$, H$\beta$ $\lambda4861$, and [\ion{O}{III}] $\lambda\lambda4958,5006$. Because the observations were obtained in AO mode, we also masked the wavelength range affected by the laser dichroic ($5800$--$5980\,\AA$). The spectral region around the \ion{Mg}{II} doublet was additionally excluded, since the line profile is dominated by outflow kinematics rather than stellar population signatures. In addition to these masks, we applied an iterative $5\sigma$ clipping procedure to remove residual sky features and emission lines not included in the masking scheme. Finally, we used fourth-order multiplicative polynomials, and no additive polynomials, to adjust the continuum shape of the templates and account for dust attenuation. Thus, the number of free parameters depends on the redshift of each galaxy, with lower-redshift galaxies having a larger number of age bins due to the older age of the Universe at their observed redshift. The total number of free parameters therefore ranges from 330 (36 age bins $\times$ 9 metallicity bins, four coefficients for the fourth-order multiplicative polynomial, and the first two Gaussian-Hermite moments) to 357 (39 age bins $\times$ 9 metallicity bins, with all other parameters kept the same). 

\begin{figure}
\centering
        \includegraphics[width=1\columnwidth]{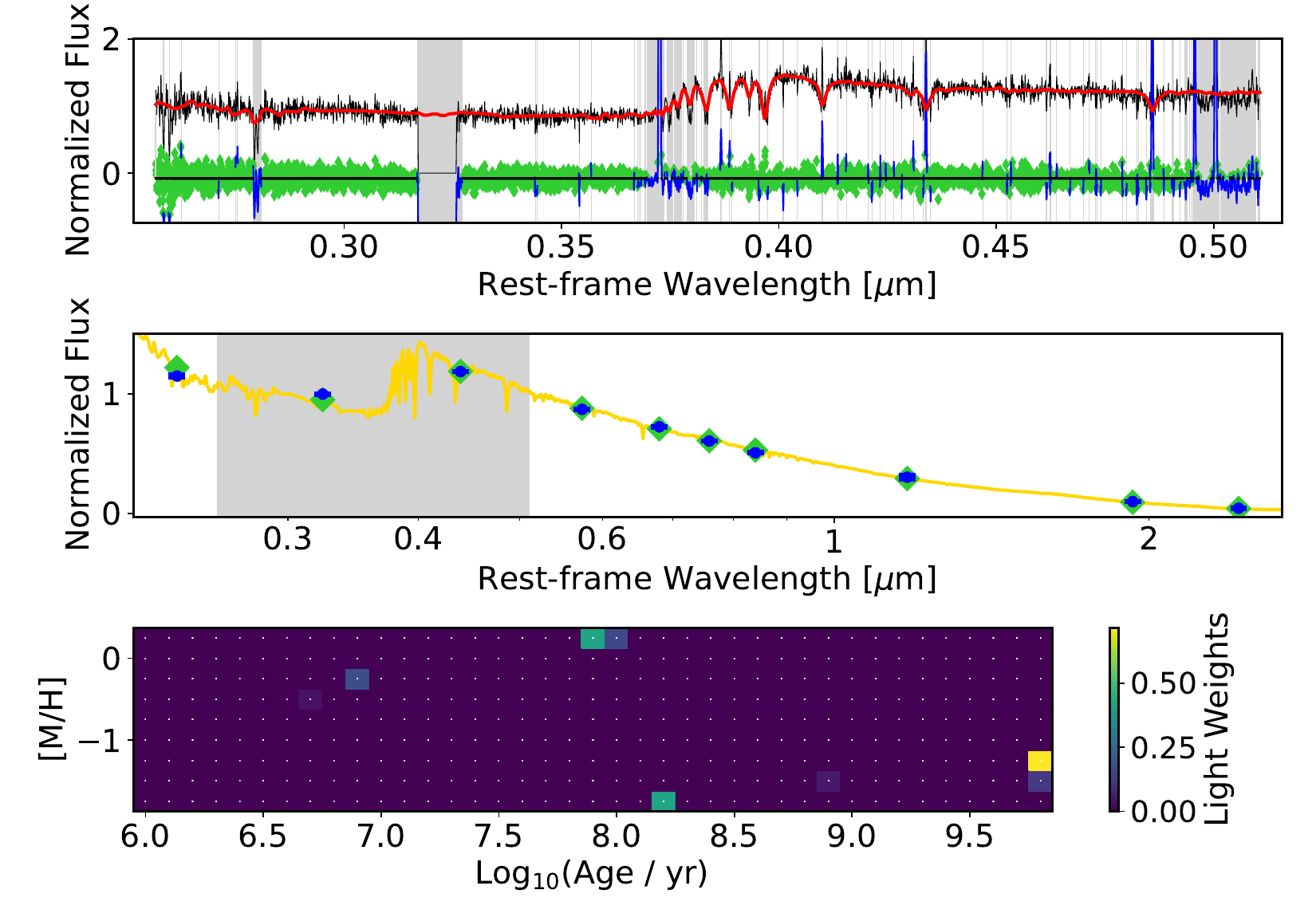}
\caption{Illustrative example of the spectral and photometric fitting carried out for our sample galaxies. The top panel shows the observed spectrum (black) and best-fitting model (red) for one of our sample galaxies (AS1063-MF3-004). The residuals of the fit are shown in green. The blue lines indicate the masked wavelength channels, either by our explicit masking of emission lines, the AO gap, and the \ion{Mg}{II} line, or by the iterative $\sigma$-clipping. The central panel shows the best-fitting model (yellow) and predicted photometry (green diamonds) versus the actual photometric data available for this galaxy (blue symbols), the shaded area shows the wavelength range of the spectral fitting. The bottom panel shows the inferred SFH via our spectral and photometric modeling. The grid represents the age and metallicity values of our models, and the color shows the light fraction assigned to each combination of age and metallicity.}
\label{fig:fit_example}
\end{figure}

Figure~\ref{fig:fit_example} presents an example fit for one of the galaxies in our sample. The top panel shows the fit to the MUSE spectrum, the middle panel shows the simultaneous broad-band fitting, and the bottom panel presents the resulting age--metallicity distribution of the star formation episodes recovered in the SFH (in Appendix~\ref{sec:all_fits_appendix} we show the data available, best-fitting models, and recovered SFHs for all of our sample galaxies).

For the purposes of this work, we are primarily interested in star formation episodes occurring within the last few hundred megayears. Since all galaxies in our sample are actively star-forming systems selected through their emission lines in the MUSE datacubes, their spectra are generally dominated by luminous young stellar populations. As a result, older stellar components can be more difficult to constrain robustly through spectral fitting because younger populations may outshine a substantial underlying old stellar mass component. This limitation does not significantly affect our analysis, however, since our goal is to characterize the stellar populations that dominate the emitted optical light and shape the recent spectral energy distribution of the galaxies.

Furthermore, our data provide broad and complementary wavelength coverage that is particularly well suited to constraining the recent SFHs targeted in this work. Depending on redshift, the MUSE spectra probe the rest-frame UV and blue-optical emission of the galaxies, and for the lower-redshift part of the sample they also cover the Balmer/4000 $\AA$ break region (11 out of 20 galaxies). At higher redshifts, where the 4000 $\AA$ break shifts beyond the MUSE wavelength range, this region is instead sampled by the near-infrared HST photometry. The available HST optical bands complement the MUSE spectra by sampling the rest-frame UV, while the HST near-infrared bands, together with the $K_s$ and IRAC photometry, extend the SED coverage to substantially redder rest-frame wavelengths. There are between 8 and 12 independent broadband photometric measurements spanning a wide wavelength baseline, in addition to the information contained in the MUSE spectrum. This combination is particularly relevant for the purpose of this work, which focuses on star formation over the last few hundred Myr: the rest-frame UV is sensitive to the luminous young stellar populations that dominate on these timescales, while the coverage across and redward of the Balmer/4000 $\AA$ break provides complementary constraints on more evolved stellar populations and helps to constrain the overall shape of the stellar SED. An exception in our sample is M0416-DF-147, for which the `\textit{use\_phot}' flag in the \citet{Shipley2018} catalog is set to zero, indicating potentially unreliable photometry due to a nearby bright star. We therefore perform a spectroscopy-only fit for this galaxy. We retain M0416-DF-147 in our sample because its MUSE spectrum covers the Balmer break and extends to the 4000~$\AA$ region, thereby retaining useful age-sensitive spectral information. Nevertheless, the absence of the photometric constraints reduces the wavelength baseline available for the fit, and the correspondingly larger uncertainties in its SFH-based SFR measurements are consistent with the larger random uncertainties expected for this galaxy. We stress, however, that our conclusions are not affected by the inclusion of this galaxy.

\subsection{Galactic wind properties}
\label{sec:outflow_model}
We derived the galactic wind properties of our galaxies by fitting a semi-analytic outflow model to the MUSE data. The model was first presented in \citet{Pessa2024} and later applied to a sample of galaxies drawn from the MUSCATEL survey \citep{Pessa2026}, including all galaxies analyzed in this work (with the exception noted in Sec.~\ref{sec:sample}).

Because both the model and the fitting results for the galaxies considered here have already been presented in previous work, we summarize only its main features and refer the reader to \citet{Pessa2024, Pessa2026} for a detailed description of the methodology, an assessment of its ability to reproduce the MUSE observations, and the distributions of the best-fitting parameters.

In brief, the outflow is modeled as an ensemble of spherical shells. The velocity of each shell increases with radius. Under the Sobolev approximation, continuum photons emitted by the central source interact with the outflowing material only at the radius where the absorbing \ion{Mg}{II} ions are brought into resonance by their Doppler shift. Owing to the varying velocity field and the assumption of mass conservation, both the gas density and the optical depth vary as a function of radius.

The model is characterized mainly by five parameters: the launching velocity of the wind ($v_0$), the launching radius ($R_0$), the central optical depth ($\tau_{0}$), the radial acceleration rate ($\gamma$), and the terminal velocity ($v_\mathrm{max}$). The radial velocity profile is parameterized as

\begin{equation}
\label{eq:vel_field}
\begin{aligned}
    &v = v_{0} \, \biggl(\frac{r}{R_{0}} \biggr)^{\gamma} &\mathrm{\;for\;} r < R_{\rm max} \\
    &v =v_{\rm max}   &\mathrm{\;for\;} r \geq R_{\rm max}\\
\end{aligned}
\end{equation}

The biconical geometry is described through three additional parameters: the opening angle (O.A.) of the bicone, a rotation angle (R.A.) that measures the rotation of the outflow perpendicular to the plane of the sky, toward the observer, and a position angle (P.A.) that corresponds to the angle of the outflow with respect to the horizontal axis, in the plane of the sky. Lastly, an additional contribution of nebular emission to the \ion{Mg}{II} is described by the $f_{C}$ parameter.

The best-fitting parameters and their associated uncertainties were derived through a Markov chain Monte Carlo (MCMC) analysis implemented with the \textsc{EMCEE} Python package \citep{ForemanMackey:2013io}. We sampled the posterior distributions using 22 MCMC chains, each evolved for 6000 iterations, initialized at randomly perturbed positions around initial parameter estimates obtained from a preliminary exploration of the parameter space. For each chain, the first 500 iterations were discarded as burn-in. In \citet{Pessa2024}, we present an example corner plot illustrating the posterior distributions of our best-fitting model parameters.

We adopted uniform priors for all model parameters, chosen to be sufficiently broad to encompass the posterior distributions without imposing strong constraints. The only exception is the launching radius, $R_0$, for which we adopted a normal prior. This choice is motivated by the fact that the approximate size of the galaxy is known from the HST imaging (see Sec.~\ref{sec:obs}) and it is expected to be comparable to $R_0$. The correspondence is not exact, however, because $R_0$ is a simplified parameterization of the model that assumes a spherical continuum source.

\subsection{Previously available star formation rate measurements}
\label{sec:previous_SFR}

In Sec.~\ref{sec:SFH_det}, we presented the methodology used to derive non-parametric SFHs for our galaxies and to measure SFRs averaged over different timescales that capture recent star formation episodes. Here, we briefly summarize the SFR estimates reported for these galaxies in \citet{Pessa2026}. One set of measurements is based on photometric SED fitting using a parametric SFH (Sec.~\ref{sec:SFR_photo_only}), while the other is derived from the observed [\ion{O}{II}] luminosity (Sec.~\ref{sec:SFR_OII}). In Sec.~\ref{sec:wind_vs_SFH}, we compare the results obtained from the non-parametric SFHs described in Sec.~\ref{sec:SFR_measurements} with those derived from these previously available SFR indicators.

\subsubsection{Photometry-based star formation rate}
\label{sec:SFR_photo_only}

The multi-band photometry described in Sec.~\ref{sec:photometry} is used to constrain the stellar populations through SED fitting. We employ the FAST code \citep{Kriek:2009cs}, adopting stellar population models from \citet{Bruzual2003}, a \citet{Chabrier2003} initial mass function, and exponentially declining star formation histories. The fits explore four stellar metallicities ($Z = 0.004$--$0.05$) and include dust attenuation following \citet{Calzetti:2000iy}.

From this analysis, we derive stellar masses, star formation rates, ages, and global dust attenuation for all galaxies in the sample. The star formation rates are calculated on timescales of 100 Myr. Uncertainties are estimated from 300 Monte Carlo realizations, yielding typical errors of $\sim$0.2 dex in stellar mass and $\sim$0.3 dex in SFR. Throughout the remainder of this paper, we refer to this quantity as the SED-based SFR (or SFR$_{\mathrm{SED}}$).

\subsubsection{[O\,II]-based star formation rate}
\label{sec:SFR_OII}

For galaxies at redshifts below 1.48, the [\ion{O}{II}] $\lambda\lambda$3727,3729 doublet falls within the MUSE wavelength range. For these systems, we can therefore derive an independent estimate of the SFR from the [\ion{O}{II}] $\lambda\lambda$3727,3729 luminosity, which traces star formation on timescales of a few megayears. The [\ion{O}{II}] line is detected for all the galaxies below this redshift.

To compute the SFR from the [\ion{O}{II}] $\lambda\lambda$3727,3729 luminosity, we adopt the calibration of \citet{Kewley2004}:
\begin{equation}
    \mathrm{SFR}_\mathrm{[OII]} \, (M_{\odot} \mathrm{yr}^{-1}) = 6.58 \times 10^{-42} \, \mathrm{L}_\mathrm{[OII]} \,(\mathrm{erg}\,\mathrm{s}^{-1})
\end{equation}

with an expected uncertainty of $\sim0.1$ dex. The spectra of the sample galaxies have been corrected by dust extinction before computing their SFR$_\mathrm{[OII]}$, using the extinction derived from the SED fitting, as described in Sec.~\ref{sec:SFR_photo_only} \citep[generally, the emission lines will be subject to higher extinction values than the stellar continuum, so this mainly provides a lower limit on the emission line extinction, see, e.g.,][]{Hemmati2015, Emsellem2022}. Also, because of the lack of additional emission lines to perform further diagnosis of the ionization mechanism \citep[e.g., Baldwin–Phillips–Terlevich diagram][]{BPT}, we assume that all the [\ion{O}{II}] emission of these galaxies traces SFR, which might not be entirely true, as some [\ion{O}{II}] emission can be produced by other mechanisms such as shock ionization \citep[see, e.g.,][]{Reynaldi2013}. In \citet{Pessa2026}, we show the direct comparison between the [\ion{O}{II}]-based and photometry-based SFR, for the galaxies where both measurements are available. In the rest of the paper, we will refer to this measurement as [\ion{O}{II}]-based SFR (or SFR$_{\mathrm{[OII]}}$).

\section{Results}
\label{sec:results}

\subsection{Star formation rates on different timescales}
\label{sec:SFR_measurements}

Our goal is to connect recent episodes of star formation occurring on timescales of hundreds of megayears with the observed properties of galactic winds. Using the SFHs derived as described in Sec.~\ref{sec:SFH_det}, we computed time-averaged SFRs over timescales of 100 Myr, 200 Myr, and 400 Myr. These timescales are motivated by the results of \citet{Pessa2024}, where we found that the time required for an outflow to accelerate from its launching velocity to its terminal velocity is typically on the order of a few hundred megayears.

We computed the time-averaged SFRs by summing the normalized mass weights of the SFH (which add up to unity for the full SFH) for all age bins younger than a given timescale. The resulting fraction was then multiplied by the total stellar mass of each galaxy listed in Table~\ref{tab:sample} and divided by the duration of the corresponding timescale. Although the IMF adopted for the full spectral fitting differs from that used in the photometry-based SED fitting, we account for this by normalizing the recovered SFHs to the stellar masses derived from the photometric SED fitting reported in \citet{Pessa2026}. This places the different SFR tracers on a consistent stellar-mass normalization, ensuring that comparisons between them primarily reflect differences in the inferred shape of the SFH rather than differences in the overall stellar-mass normalization.

These time-integrated SFRs are, on one hand, sensitive to discrete bursts of star formation, since they are calculated by integrating the mass formed in a non-parametric SFH. At the same time, because they span relatively large time windows, they provide robust measurements, in the sense that most of them do not show large uncertainties across the MC iterations. This indicates that the combination of spectroscopy and broad-band photometry provides enough information to constrain the time-averaged SFH of galaxies on timescales of hundreds of megayears, whereas averaging over shorter timescales leads to substantially less reliable estimates. Furthermore, these SFR tracers do not change greatly when calculated using the eMILES templates instead of the FSPS models, indicating that even if the detailed shape and individual peaks of the SFH vary across MC realizations and stellar population models, the corresponding time-integrated SFR measurements remain comparatively stable (see Appendix~\ref{sec:comparison_templates}). Therefore, we stress that although the SFHs are sampled using a fine age grid of 0.1 dex, this should not be interpreted as the effective temporal resolution of the recovered SFHs, and individual age bins should not be regarded as independently resolved star-formation episodes. The SFR tracers employed in our analysis are instead obtained by integrating the recovered star formation over substantially broader timescales, encompassing many individual age bins.

\begin{figure}
\centering
    \begin{minipage}{0.5\textwidth}
        \centering
        \includegraphics[width=\columnwidth]{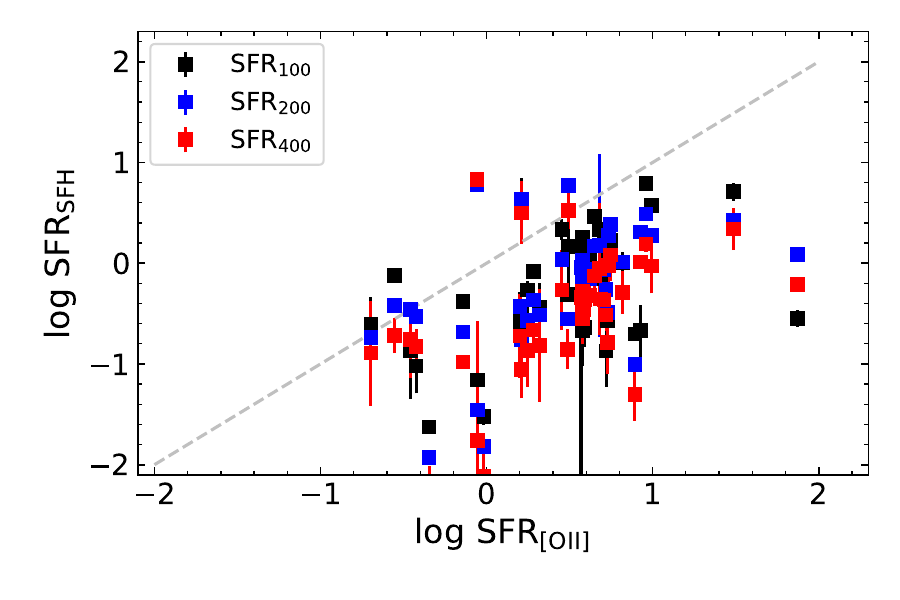}
        \includegraphics[width=\columnwidth]{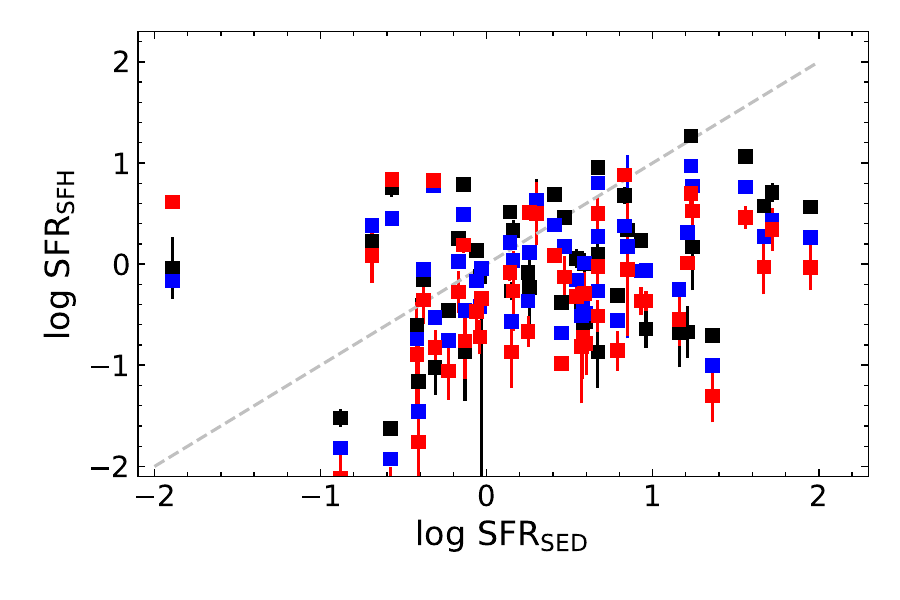}
    \end{minipage}

\caption{Comparison between the non-parametric SFH-based SFR tracers and those based on the [\ion{O}{II}] luminosity for galaxies in our preliminary sample, above the S/N threshold (top panel, see Sec.~\ref{sec:SFR_OII}) and those computed via a photometric SED fitting using a parametric SFH (bottom panel, see Sec.~\ref{sec:SFR_photo_only}). The three different colors correspond to the three different timescales explored, described in Sec.~\ref{sec:SFR_measurements}; 100 Myr in black, 200 Myr in blue, and 400 Myr in red. The dashed line shows the identity. We find consistency between the different measurements, as there is a clear positive slope in the comparison with [\ion{O}{II}]. The slope is much shallower (although still present) when comparing with the SED fitting-based measurements.}
\label{fig:SFR_comparison_with_OII_SED}
\end{figure}

Figure~\ref{fig:SFR_comparison_with_OII_SED} compares the SFRs derived from the non-parametric SFHs on different averaging timescales with the SED-based (FAST) and [\ion{O}{II}]-based SFR measurements. Since these quantities are independent of the outflow modeling, the comparison includes all galaxies from the preliminary sample that satisfy the S/N threshold discussed in Sec.~\ref{sec:sample}, including galaxies exhibiting \ion{Mg}{II} in either absorption or emission that are not part of the final outflow sample. Overall, the different tracers show correlated SFR estimates, despite systematic offsets between them. In the comparison with the [\ion{O}{II}]-based SFRs, a clear positive correlation is evident, reflecting the consistency between the two measurements (p-values in the range of $0.04$--0.0004 for the different timescales). A positive trend is also present when comparing with the SED-based SFRs, although it is noticeably shallower (p-values of 0.02-0.14). Larger differences in this case are not unexpected. On the one hand, the SED-based SFRs have relatively large random uncertainties ($\sim0.3$ dex). On the other hand, they are derived assuming a smooth, exponentially declining SFH, which differs substantially from the non-parametric SFHs used here to recover stochastic episodes of star formation.

The SFRs inferred from the non-parametric SFHs are systematically lower than those obtained with the alternative tracers. For the [\ion{O}{II}]-based measurements, this behavior is expected because [\ion{O}{II}] traces star formation on short timescales of only a few megayears, effectively probing periods of active star formation. Once star formation declines, the [\ion{O}{II}] luminosity rapidly decreases. In contrast, the SFH-based measurements represent averages over much longer timescales and therefore include both active and quiescent periods, naturally yielding lower average SFRs. Furthermore, as discussed in Sec.~\ref{sec:SFR_OII}, not all of the observed [\ion{O}{II}] emission is necessarily associated with ongoing star formation. Consequently, the [\ion{O}{II}]-based SFR estimates could be regarded as upper limits.

For the SED-based SFRs, the offset arises from the fundamentally different SFH parameterization. By construction, both approaches are normalized to the same total stellar mass. Consequently, a systematically lower SFR at recent times implies that a larger fraction of the stellar mass was formed at earlier epochs in the parametric SED-fitting models. This is a direct consequence of assuming an exponentially declining SFH. The SFR values measured for each galaxy on the different averaging timescales are listed in Table~\ref{tab:sample}.

\subsection{Galactic wind kinematics versus star formation rate derived from the SFHs}
\label{sec:wind_vs_SFH}

We now explore the connection between the observed properties of the galactic winds reported by \citet{Pessa2026} and recent episodes of star formation inferred from the SFHs. To do so, we investigate correlations between the outflow parameters and the SFR measurements described in the previous section. We focus on the parameters most directly related to the wind kinematics: the maximum outflow velocity ($v_{\rm max}$) and the radial acceleration rate ($\gamma$). Our outflow model also constrains the launching velocity ($v_0$; see Sec.~\ref{sec:outflow_model}). However, we do not find any significant or tentative correlation between $v_0$ and the quantities explored in this work, and therefore do not explicitly include it further in the following analysis.

\begin{figure*}
\centering

\begin{minipage}[t]{0.495\textwidth}
    \centering
    \includegraphics[width=\linewidth]{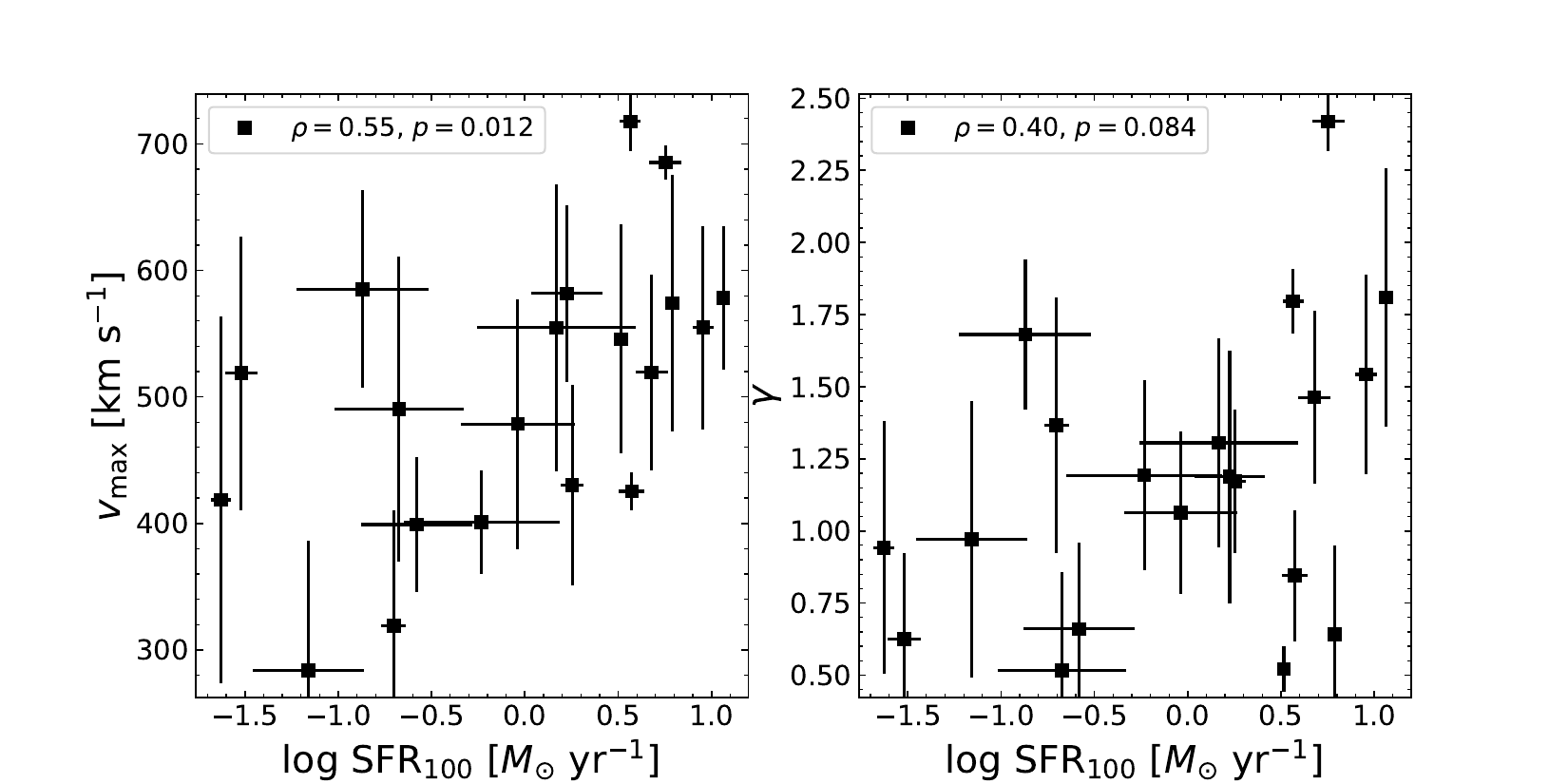}

    \vspace{0.25em}

    \includegraphics[width=\linewidth]{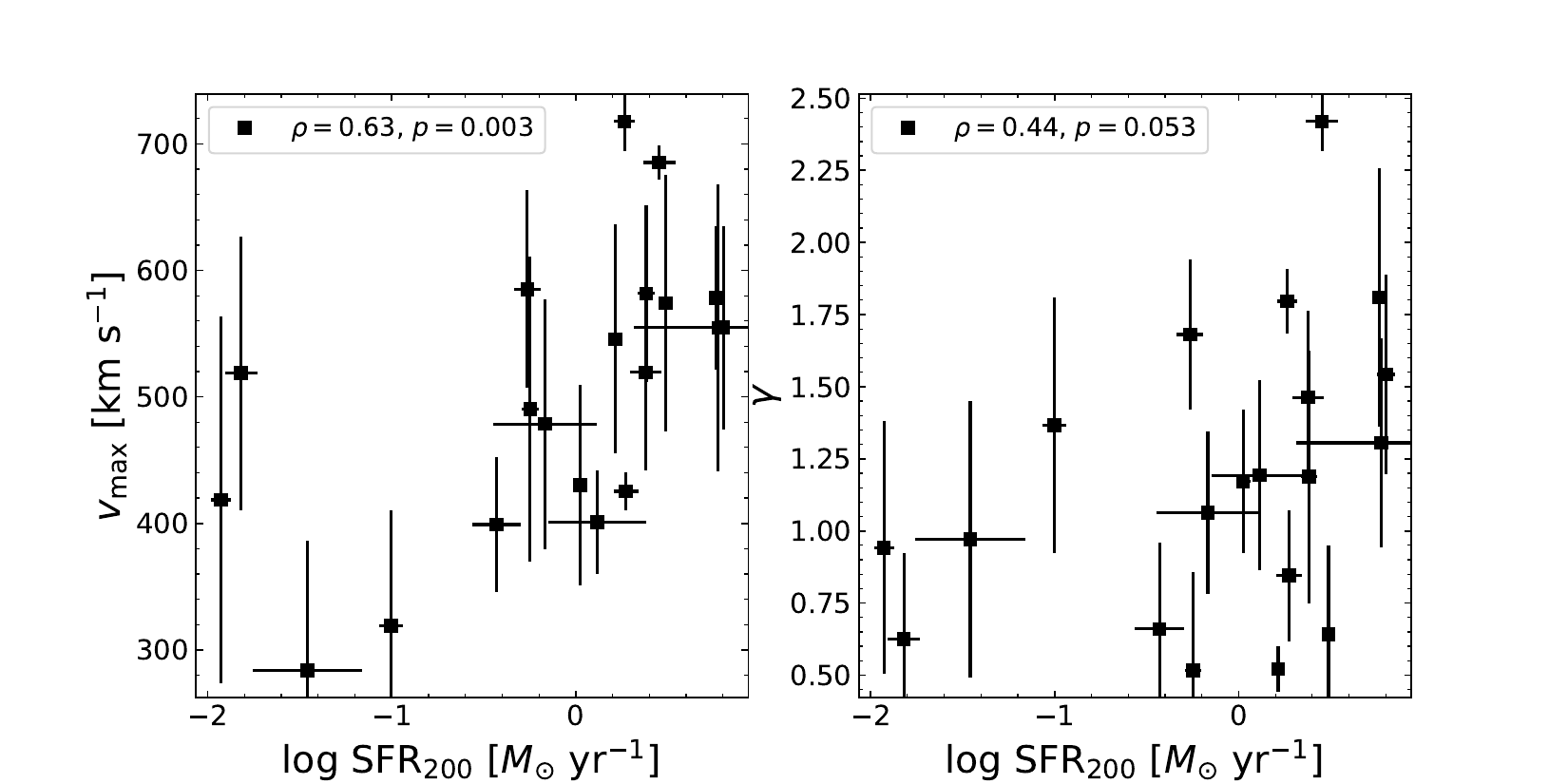}

    \vspace{0.25em}

    \includegraphics[width=\linewidth]{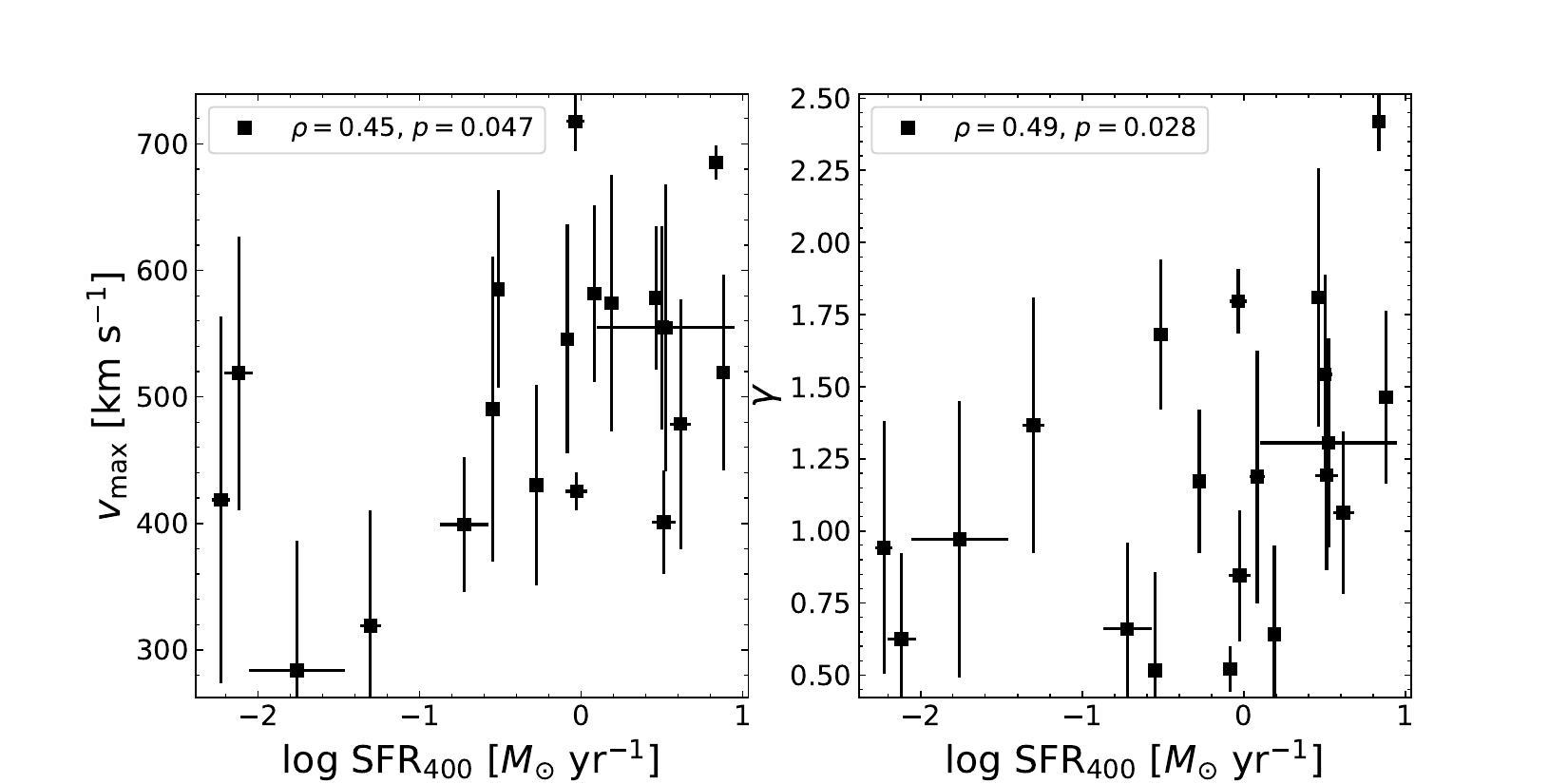}
\end{minipage}%
\hspace{0.01\textwidth}%
\begin{minipage}[t]{0.495\textwidth}
    \centering
    \includegraphics[width=\linewidth]{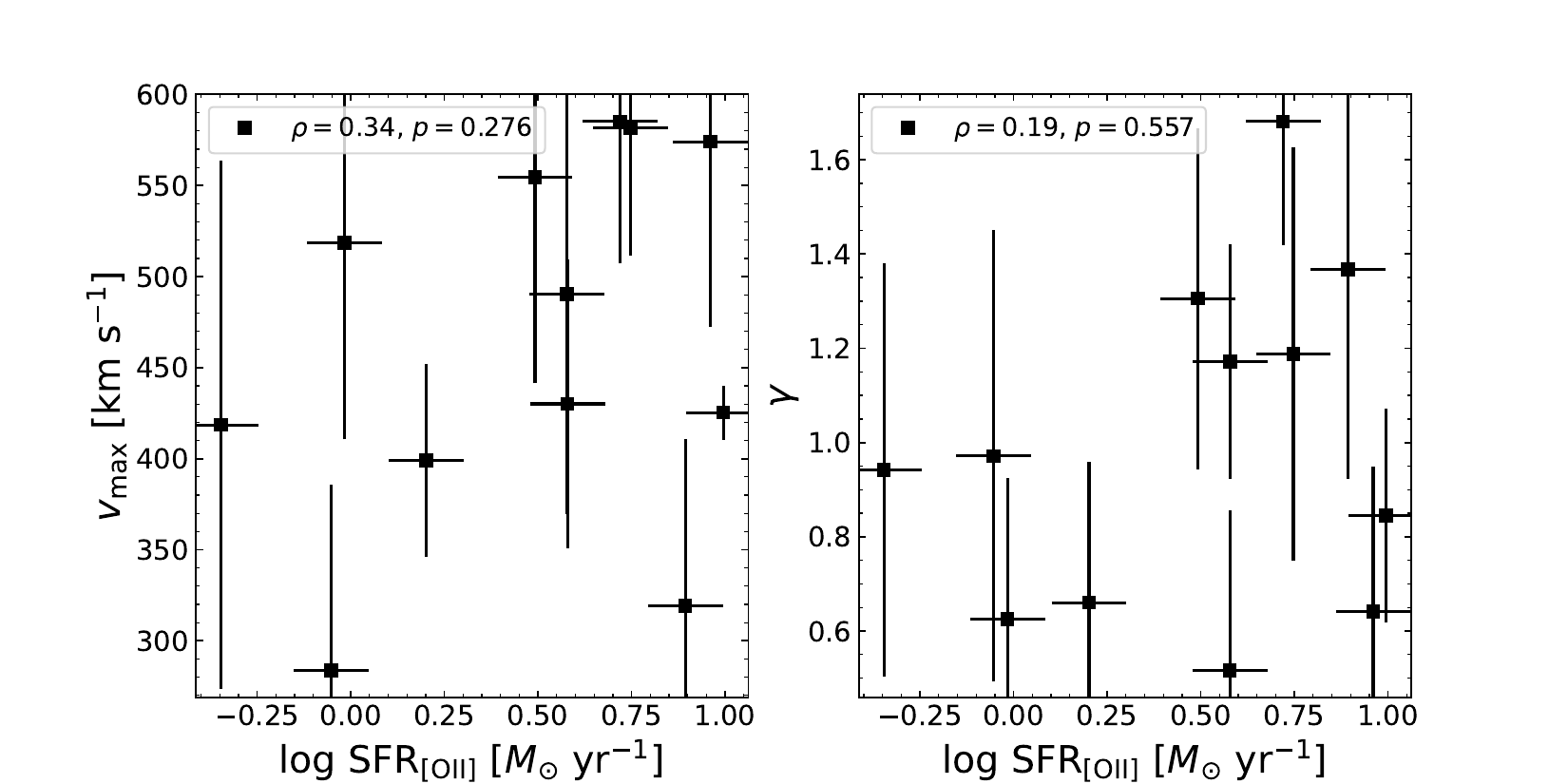}

    \vspace{0.25em}

    \includegraphics[width=\linewidth]{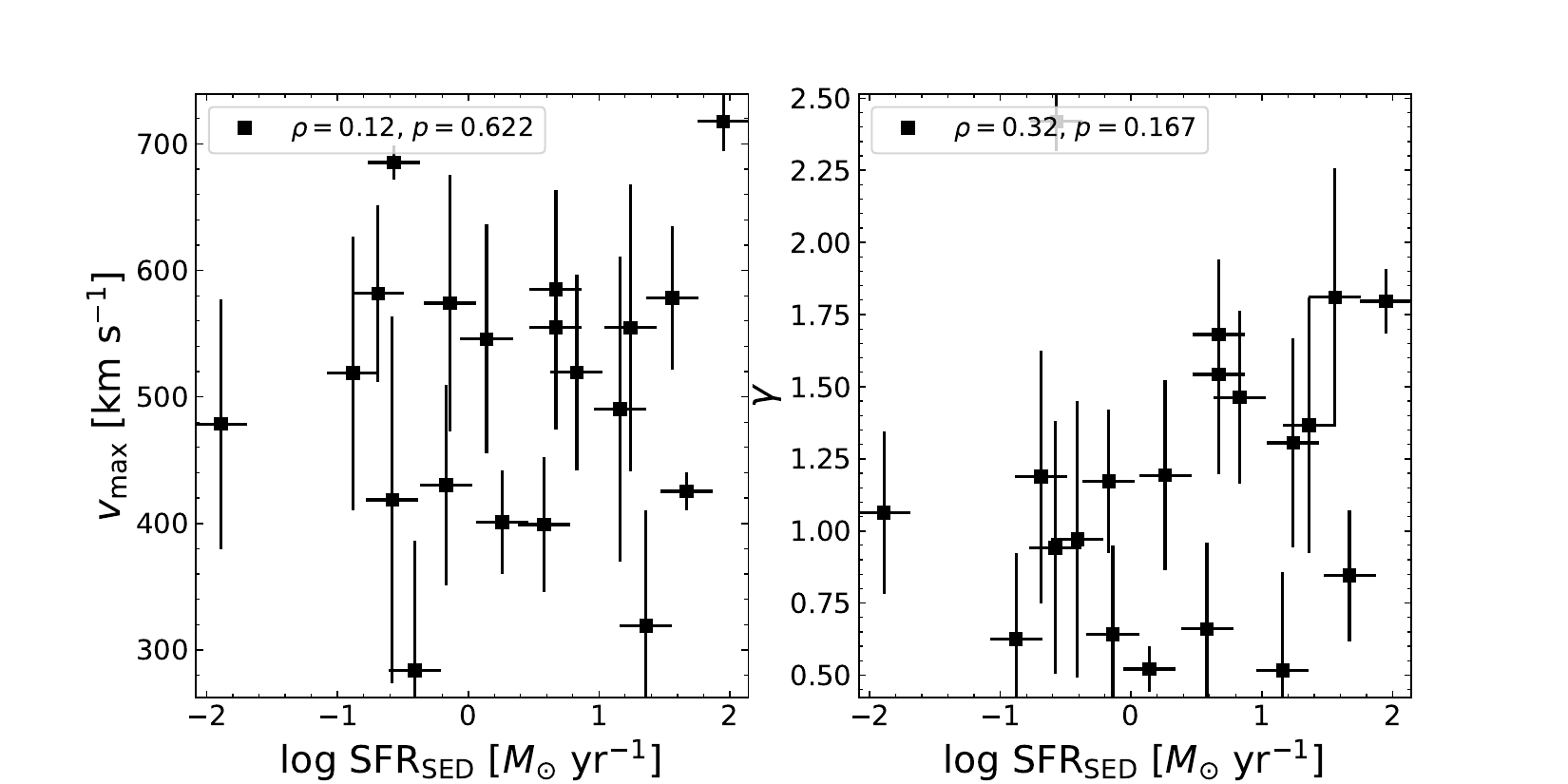}
\end{minipage}

\caption{Maximum outflow velocity ($v_{\rm max}$) and acceleration rate ($\gamma$) as a function of the SFR derived using different methods sensitive to various timescales. \textit{Left column:} SFRs averaged over 100, 200, and 400 Myr from non-parametric SFHs (Sec.~\ref{sec:SFR_measurements}). \textit{Right column:} SFRs derived from the [\ion{O}{II}] emission and SED fitting assuming an exponentially declining SFH (Sec.~\ref{sec:previous_SFR}). Spearman rank coefficients and $p$-values are shown in each panel. The SFRs computed from the non-parametric SFHs correlate with the parameters that describe the kinematics of the outflow, whereas the SFRs computed from the [\ion{O}{II}] emission or from the SED fitting assuming a parametric SFH do not show any clear correlation.}
\label{fig:SFR_SFH_vs_wind}
\end{figure*}

The left column of Fig.~\ref{fig:SFR_SFH_vs_wind} shows the relation between the three SFH-based SFR measurements and the outflow kinematic parameters, together with the corresponding Spearman correlation coefficients ($\rho$) and p-values. Tentative correlations are found between all three SFR tracers and both $v_{\rm max}$ and $\gamma$. For $v_{\rm max}$, the correlations have $\rho$ values between 0.45 and 0.63, with p-values below 0.05. The correlations with $\gamma$ are somewhat weaker, although still statistically suggestive, with $\rho \sim 0.4$--0.5 and p-values in the range $\sim0.03$--0.08. The overall trends remain unchanged when alternative correlation metrics, such as the Pearson correlation coefficient or Kendall's $\tau$, are used. In Sec.~\ref{sec:robustness} we discuss in more detail the statistical significance of these correlations.

For comparison, we also show the same relations using the previously available SFR tracers based on the [\ion{O}{II}] emission line and on photometric SED fitting with FAST in the right column of Fig.~\ref{fig:SFR_SFH_vs_wind}. In contrast to the SFH-based measurements, we find no evidence for significant correlations between these SFR tracers and the outflow properties considered here. This result demonstrates that SFRs derived from non-parametric SFHs provide a more physically meaningful connection between the recent star formation activity of a galaxy and its outflow properties than conventional SFR estimators, which do not probe the timescales most relevant for the observed outflows. We discuss the physical interpretation of this result in more detail in Sec.~\ref{sec:SFH_modulating}.
In Appendix~\ref{app:wind_vs_sSFR}, we present the analogous relations using the specific star formation rate (sSFR) instead of the SFR. The sSFR is obtained by dividing each SFR estimate by the total stellar mass of the corresponding galaxy. The resulting trends are broadly similar to those shown in the left column of Fig.~\ref{fig:SFR_SFH_vs_wind}. However, the correlations with $v_{\rm max}$ are slightly weaker, whereas those with $\gamma$ are somewhat stronger, both in terms of their Spearman correlation coefficients and p-values. We also examined the offset from the star-forming main sequence, $\Delta$SFMS, defined as the logarithmic difference between the measured SFRs and the expected level of star formation, according to SFMS parameterization derived by \citet{Popesso2023}. We found qualitatively similar, although slightly weaker, correlations than for the sSFR. For SFR$_{200}$, for example, we obtain $\rho \approx 0.4$ with p-values of $\sim0.06$--0.08 (not shown).

\subsection{SFR on timescales based on the extent of the \ion{Mg}{II} halo}

\label{sec:SFR_halo_size}

In the previous subsection, we computed SFR tracers from the SFHs using fixed averaging timescales of hundreds of megayears, broadly consistent with the expected travel times of the outflowing gas. However, in \citet{Pessa2026} we not only constrained the kinematic and geometric properties of the galactic winds, but also characterized their spatial extent through the \ion{Mg}{II} half-light radius (HLR). By combining this size metric with the inferred outflow kinematics (i.e., the inferred velocity profile, given by Eq.~\ref{eq:vel_field}), we can estimate an approximate launching time for each galactic wind by calculating the time required for the outflowing material to travel from the launching radius to the measured HLR.

To account for projection effects in these travel-time estimates, we correct the observed HLRs using the inferred orientation of the outflow relative to the line of sight. Specifically, we divide the measured HLR by the sine of the outflow inclination angle. As a result, nearly edge-on outflows require only a minor correction, whereas nearly face-on outflows correspond to intrinsically larger travel distances and therefore longer inferred expected travel times.

\begin{figure}
\centering
        \includegraphics[width=0.80\columnwidth]{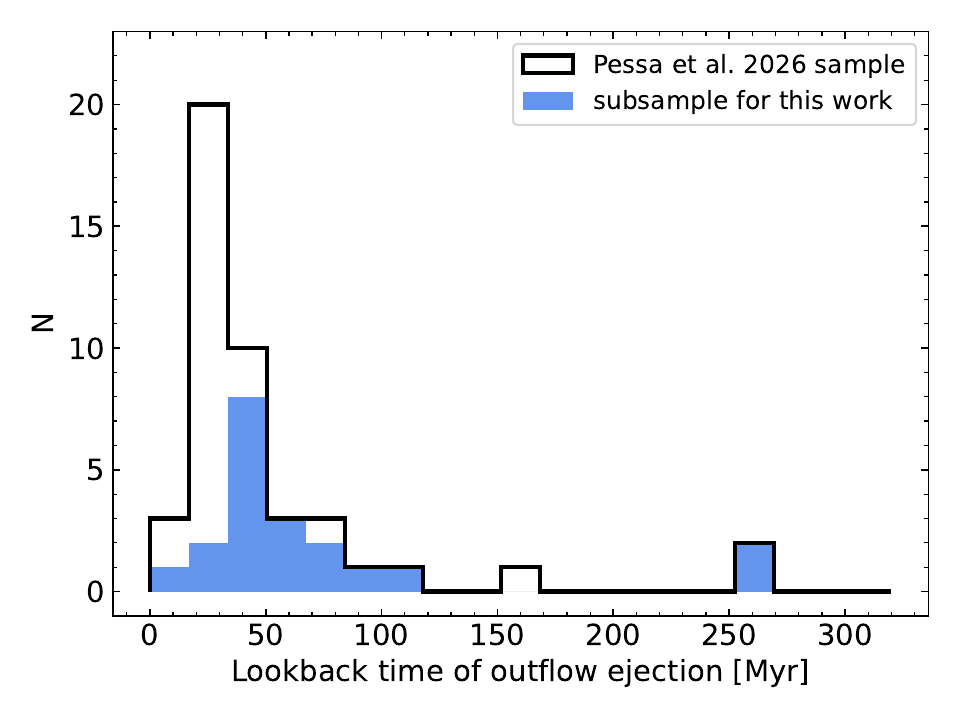}
\caption{Distribution of inferred outflow launching lookback times of the galactic winds in our sample. The launching times are calculated using the inferred kinematics of the cool galactic winds for our sample galaxies based on the extent of their extended \ion{Mg}{II} emission \citep[after correcting by their central \ion{Mg}{II} absorption, see][]{Pessa2026}. The black histogram shows the distribution of launching times for the full sample of \citet{Pessa2026}, for reference, and the blue histogram shows the distribution for the subset of galaxies that comprise the sample for this work.}
\label{fig:launch_time_dist}
\end{figure}

Figure~\ref{fig:launch_time_dist} shows the distribution of the inferred outflow launching times for the galaxies in our sample, as well as for the remaining galaxies in the sample of \citet{Pessa2026}. The distribution is broad, peaking at approximately 30 Myr and extending to expected travel times of up to 250 Myr. We note, however, that these expected travel times correspond only to the distance required for the outflow to reach the HLR, which by definition encloses half of the total \ion{Mg}{II} emission. Consequently, some of the emitting material must have been launched more recently than the inferred HLR expected travel time, while gas located beyond the HLR must have been launched earlier.

Rather than characterizing the outflow with a single discrete launching time, we therefore define a time window corresponding to the interval required for the gas to travel from 0.5 HLR to 1.5 HLR. We then compute the average SFR within this interval for each galaxy. This quantity is designed to probe star formation on timescales matched to the size and expansion velocity of each individual \ion{Mg}{II} halo and is therefore expected to better capture the star formation events responsible for launching the outflow. In Appendix~\ref{app:SFH_and_launching_times}, we compare the recent SFHs of our galaxies with their estimated outflow-launching times and use this information to derive SFR measurements tailored to each system.

\begin{figure}
\centering
        \includegraphics[width=0.99\columnwidth]{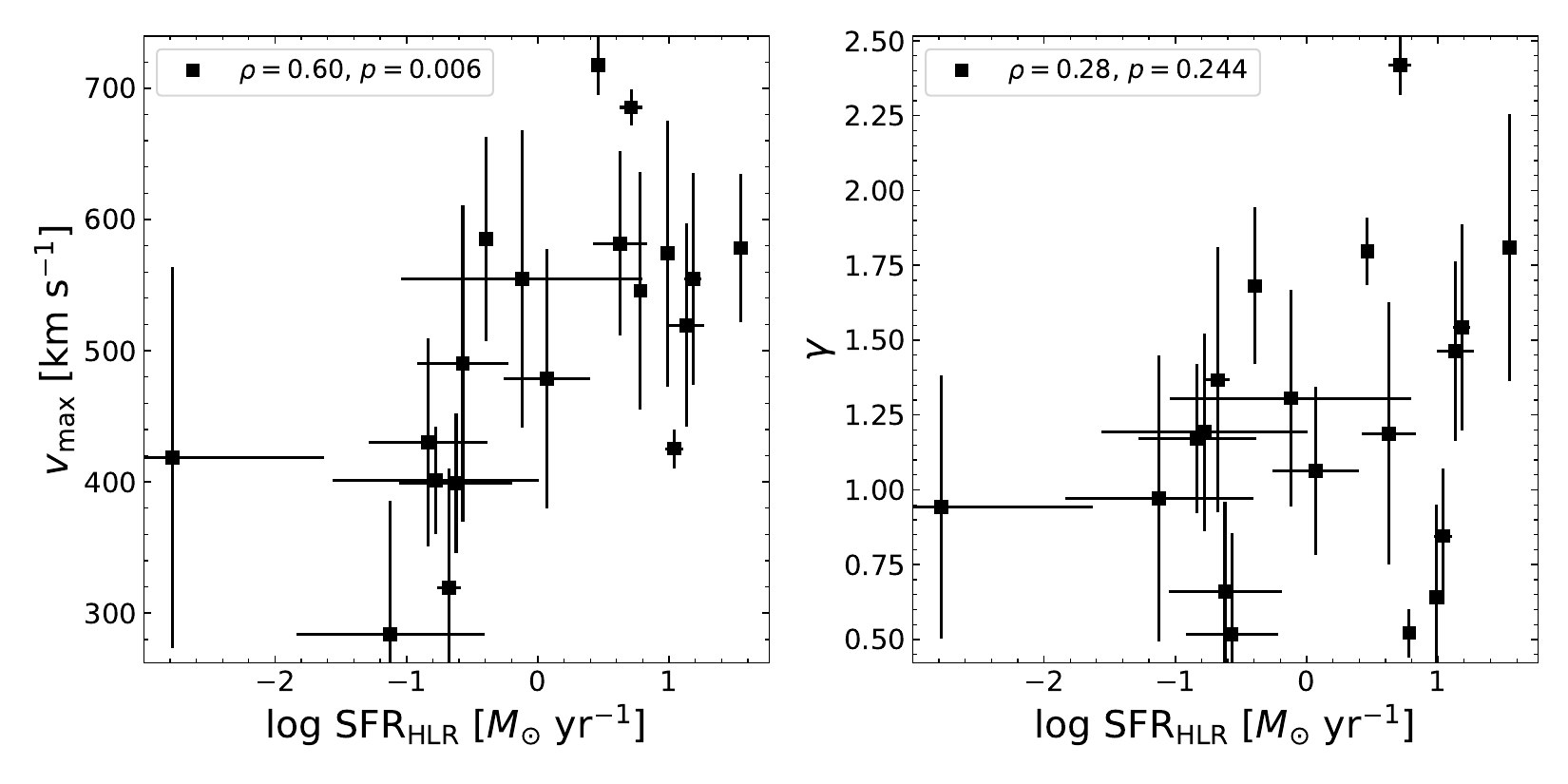}
\caption{Same as the panels in Fig.~\ref{fig:SFR_SFH_vs_wind}, but using a SFR computed over a timescale consistent with the expected travel times of the galactic outflow, instead of using fixed timescales. That is, the timescale used for each galaxy is specifically computed given the extent and kinematics of their \ion{Mg}{II} halo, see Sec.~\ref{sec:SFR_halo_size} for details.}
\label{fig:SFR_HLR_vs_wind}
\end{figure}

We quantify the strength of the star formation episodes potentially associated with the galactic winds by computing the time-averaged SFR within the corresponding HLR-based time window. Figure~\ref{fig:SFR_HLR_vs_wind} shows the relation between this halo-size-based SFR (SFR$_{\rm HLR}$) and the same outflow kinematic parameters explored previously. A clear positive trend is observed with $v_{\rm max}$, with $\rho \sim 0.6$. Although statistically significant, the correlation is not stronger than those obtained using the fixed-timescale SFR measurements, which show comparable correlation coefficients and $p$-values. In contrast, unlike the results shown in the left column of Fig.~\ref{fig:SFR_SFH_vs_wind}, we do not find a significant correlation between SFR$_{\rm HLR}$ and $\gamma$.

At face value, these results suggest that the outflows may be driven not by a single burst occurring within a narrow time interval, but rather by a sequence of star-forming episodes sustained over longer timescales. We cannot exclude additional mismatches between the inferred launching times and the true timescales of the outflows arising from simplifying assumptions in our model, such as the adopted velocity field. Nevertheless, for all but one galaxy (see Appendix~\ref{app:SFH_and_launching_times}), we identify star formation activity at lookback times consistent with the predicted wind-launching times. Furthermore, both the fixed-timescale and HLR-based SFR measurements correlate with the maximum outflow velocity. We therefore conclude that systematic uncertainties in the estimated launching times are unlikely to dominate our results.

\section{Discussion}
\label{sec:discussion}

\subsection{The SFH modulating the galactic wind kinematics}
\label{sec:SFH_modulating}
In the previous section, we showed that SFR tracers derived from non-parametric SFHs correlate much more strongly with wind kinematic properties than previously available SFR measurements based on [\ion{O}{II}] emission or photometric SED fitting assuming a parametric SFH. This suggests that the present-day SFR of a galaxy, as traced by [\ion{O}{II}] emission on timescales of $\lesssim10$ Myr, is not directly linked to the outflow properties observed today. Similarly, the intrinsically smooth SFHs assumed by FAST (i.e., exponentially declining $\tau$ models) are unable to capture the stochastic bursts of star formation that likely triggered the galactic outflows.

This interpretation is naturally supported by cosmological hydrodynamical simulations, which predict that star formation and galactic outflows are intrinsically time-dependent processes, with stellar feedback driving recurrent cycles of gas accretion, star formation, and gas ejection rather than maintaining a steady-state equilibrium \citep[e.g.,][]{Muratov2015,Faucher2018,Barbani2025}. In these simulations, stellar feedback injects momentum and energy into the surrounding medium tens of megayears following a star-formation episode, while the accelerated gas subsequently propagates into the CGM. Consequently, the strongest outflow episodes generally occur after the peak of the corresponding starburst, producing a natural temporal offset between the instantaneous SFR and the observed outflow properties.

This delayed response arises from two physical processes. First, the dominant mechanical feedback responsible for driving galactic winds is provided by core-collapse supernovae, which begin exploding only a few Myr after a star-formation episode and continue injecting energy and momentum into the ISM for several tens of megayears according to stellar evolution models \citep[e.g.,][]{Leitherer1999,Hopkins2018}. Second, the accelerated gas subsequently requires additional time to propagate from the ISM to circumgalactic scales. For our galaxies, we estimate this travel time to range from tens to hundreds of megayears (Fig.~\ref{fig:launch_time_dist}), although it depends strongly on the assumed velocity field, as discussed in the previous section. Consequently, the properties of the cool gas observed in the CGM are expected to be linked to star formation episodes that occurred well before the timescales probed by [\ion{O}{II}] emission.

Our observations are fully consistent with this theoretical picture. The time-averaged SFR tracers naturally account for both the delayed response of stellar feedback following a burst of star formation and the subsequent propagation of the outflowing gas into the circumgalactic medium. In contrast, instantaneous SFR tracers characterize only the current level of star formation, which need not be directly related to the gas presently observed in the CGM. Taken together, these results suggest that SFRs derived from non-parametric SFHs provide a more physically meaningful description of the stellar feedback history responsible for launching the observed galactic winds than instantaneous SFR indicators.

In this context, \citet{Harvey2025} find that post-starburst (PSB) galaxies, that is, galaxies that recently experienced a strong burst of star formation but exhibit little ongoing star formation, host more extended cool-gas halos. This is evidenced by significantly stronger \ion{Mg}{II} absorption out to at least $\sim300$ kpc compared to star-forming or quiescent galaxies of similar stellar mass. Their results are also consistent with evidence for strong outflows traced by blueshifted, down-the-barrel resonant absorption in PSB galaxies \citep{Taylor2024, Sun2024, Luo2026}. 

Using Sloan Digital Sky Survey data \citep{Aihara2011}, \citet{Sun2024} find that the velocity of galactic winds in PSB galaxies decreases with time since the last starburst event. Consistent with this picture, \citet{Luo2026} find that cool-gas outflow velocities traced by resonant \ion{Mg}{II} and \ion{Fe}{II} absorption in intermediate-redshift PSBs decrease with increasing time since quenching, and interpret these outflows as possible relics of previous starburst or AGN activity rather than being driven by the current star formation. Similarly, \citet{Taylor2024}, using optical spectra from the UKIRT Infrared Deep Sky Survey \citep[UKIDSS;][]{Lawrence2007}, investigated the evolution of outflow properties along the quenching sequence. By analyzing \ion{Mg}{II} absorption as a function of time since the last major starburst, they found that outflows remain detectable in stacked spectra of galaxies quenched up to $\sim1$ Gyr earlier. This provides further evidence for the delayed, or ``echoing'', response of outflows to past star formation episodes. Additional mechanisms may still contribute to driving outflows in galaxies quenched for such long periods. Nevertheless, our results are consistent with these previous findings and support a scenario in which the recent SFH of a galaxy plays a primary role in setting the outflow properties observed today, whereas the instantaneous SFR is of secondary importance.

\subsection{Impact of the shape of the SFH on the gas kinematics}
\label{sec:multimodality}

In addition to the strength of recent star formation episodes, quantified through the time-averaged SFH-based SFRs, the shape of the SFH itself may also influence the kinematics of galactic winds. In particular, we investigate whether the multimodality of the SFH, that is, whether the star formation within a given time interval is distributed among multiple episodes or concentrated in a single dominant burst, is related to the observed outflow properties.

To quantify the multimodality (MM) of the SFH within a timescale $\Delta t$, we identify the age bin containing the largest fraction of stellar mass formed during the time interval from the present day to a lookback time $\Delta t$. We then divide the mass formed in this dominant age bin by the total stellar mass formed within $\Delta t$ and subtract the resulting ratio from unity:

\begin{equation}
\label{eq:multimo}
    \mathrm{MM}_{\Delta t} = 1 -  \frac{max_{i,\Delta t}}{sum_{i,\Delta t}}
\end{equation}

where $max_{i,\Delta t}$ and $sum_{i,\Delta t}$ correspond to the maximum mass formed in a single time bin, and the total mass formed, within a timescale $\Delta t$. Thus, a MM $= 0$ means that all the mass was formed in a single star-forming episode, whereas a MM $\sim1$ means the stellar mass was formed across many comparable star-forming episodes.

We note that, by construction, the measured value of MM depends on the adopted age binning of the recovered SFH. A coarser temporal sampling can merge star formation episodes that would occupy separate bins at finer binning, increasing the relative contribution of the dominant bin and therefore tending to lower the measured MM. The MM values should therefore be interpreted in the context of our adopted SSP age grid.

\begin{figure}
\centering
        \includegraphics[width=0.99\columnwidth]{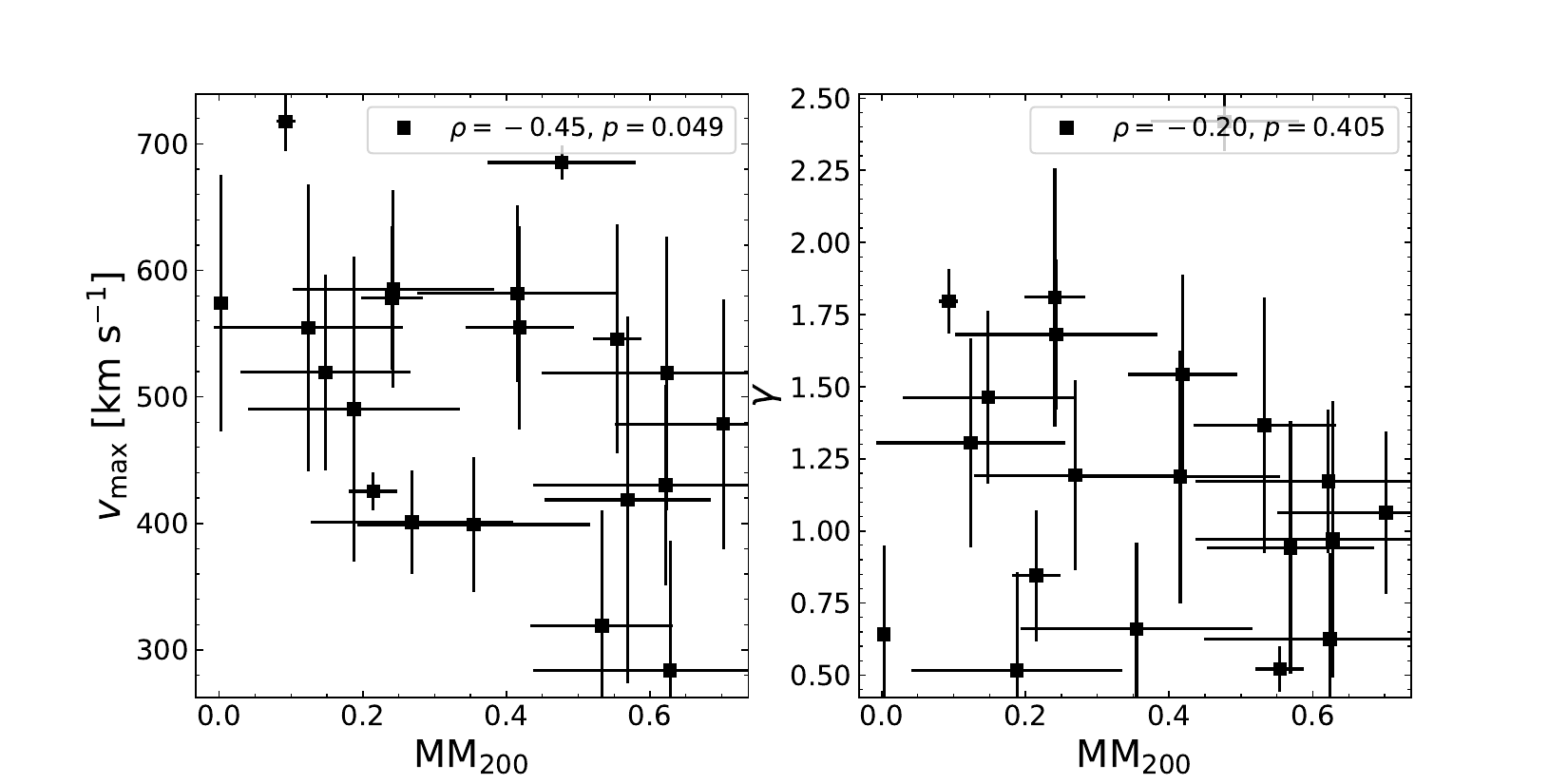}
\caption{Maximum velocity and acceleration rate of the cool gas outflows as a function of the multimodality parameter, defined as indicated in Eq.~\ref{eq:multimo}. The multimodality parameter characterizes the shape of the SFH of each galaxy, such that a multimodality close to zero indicates a star formation history dominated by a single burst of star formation. On the other hand, a multimodality close to one corresponds to the case where the SFH is composed of several comparable star formation episodes.}
\label{fig:Burstiness_vs_wind}
\end{figure}

Figure~\ref{fig:Burstiness_vs_wind} shows the same outflow kinematic parameters explored previously as a function of the multimodality measured over a timescale of 200 Myr. A clear negative correlation is observed between MM and $v_{\rm max}$ ($\rho \sim -0.45$), indicating that galaxies with multiple comparable star formation episodes tend to host slower outflows than galaxies in which most of the recent star formation is concentrated in a single dominant event. When considered together with the results of the previous subsection, this suggests that although galactic winds do not necessarily respond to a single star formation episode within a narrow time interval, the fastest winds are preferentially found in galaxies whose recent SFHs are dominated by one particularly strong burst (i.e., lower MM). In contrast, galaxies that experienced several star formation episodes of comparable strength tend to host slower outflows. We note that, although Fig.~\ref{fig:Burstiness_vs_wind} shows the multimodality measured over a 200 Myr timescale, we obtain qualitatively similar results when considering longer timescales ($\sim400$ Myr). Our findings are consistent with the cosmological zoom-in hydrodynamical simulations of \citet{Kang2025}, which show that outflows become stronger when star formation is more temporally clustered, owing to the greater coherence of stellar feedback, whereas smoother SFHs produce weaker winds.

Lastly, the burstiness of the SFH is often considered an important quantity in studies that seek to connect the SFHs of galaxies with their present-day properties \citep[see, e.g.,][]{Serrato-Alos2026}. We therefore also quantified the burstiness of the SFH following the methodology of \citet{Goh2006} and \citet{Maxime2026}, which compares the mean and standard deviation of the SFR measured across the different SFH time bins within a given timescale. We then searched for correlations between this alternative burstiness parameterization and the galactic wind properties, but found no statistically significant trends.

\subsection{Robustness and interpretation of the observed correlations}
\label{sec:robustness}

The different SFR tracers explored in this work probe overlapping timescales and are therefore not statistically independent. Consequently, the various correlations presented here should not be interpreted as independent statistical tests, and the reported significance levels should be regarded as indicative rather than definitive. Nevertheless, the fact that the strongest correlations are consistently recovered for SFH-based tracers probing timescales of hundreds of megayears, while the instantaneous and parametric SFR tracers do not exhibit comparable trends, suggests that the observed behavior is physically meaningful.

A potential concern is that the observed correlation between $v_{\rm max}$ and the SFH-based SFRs arises indirectly through the previously reported dependence of outflow velocity on stellar mass. In \citet{Pessa2026}, using the full sample of 47 galaxies, we reported a tentative correlation between stellar mass and $v_{\rm max}$, with a Pearson correlation coefficient of $\rho \sim 0.43$. Repeating the same analysis for the subset of galaxies used in this work yields a somewhat stronger correlation, with $\rho \sim 0.55$. However, this value is comparable to the correlations reported here between $v_{\rm max}$ and the SFH-based SFR tracers and is stronger than only one of them, namely the correlation measured for SFR$_{400}$. This suggests that the observed relation between $v_{\rm max}$ and the SFH-based SFRs is unlikely to be driven solely by the underlying stellar-mass dependence.

While this comparison suggests that stellar mass alone is unlikely to explain the observed relation between $v_{\rm max}$ and the SFH-based SFRs, it does not explicitly account for the covariance between SFR and stellar mass. To investigate this possibility more directly, we computed partial Spearman correlation coefficients while controlling for $\log M_*$. After accounting for stellar mass, the correlation of $v_{\rm max}$ with SFR$_{200}$ remains marginally significant, with $p \sim 0.057$, whereas the correlations with SFR$_{100}$ and SFR$_{400}$ become weaker, with $p \sim 0.113$ and $p \sim 0.341$, respectively. These results suggest that stellar mass contributes to the observed trends, particularly on the longest timescale. At the same time, the persistence of a marginal correlation for SFR$_{200}$ indicates that stellar mass alone may not fully explain the relation between $v_{\rm max}$ and the recent SFH of galaxies.

Additionally, in our previous work, we showed that although some of the outflow-model parameters exhibit mutual covariances, the $\gamma$ and $v_{\rm max}$ parameters do not display a strong or systematic dependence on one another. Therefore, the correlations reported here between the SFH-based SFRs and these two outflow parameters cannot be attributed simply to an intrinsic degeneracy within the outflow model.

Taken together, these tests suggest that the observed correlations between SFH-based SFR tracers and outflow kinematics are unlikely to arise solely from statistical fluctuations, stellar-mass scaling relations, or intrinsic parameter degeneracies in the outflow model. Furthermore, the fact that the strongest correlations are consistently recovered for SFH-based tracers probing timescales of hundreds of megayears, whereas instantaneous and parametric SFR tracers do not show comparable trends, supports a physical connection between recent star formation activity and the circumgalactic outflow kinematics observed today.

\section{Summary and conclusions}
\label{sec:summary}

In a previous work, we studied the properties of galactic outflows for a sample of nearly 47 galaxies drawn from the MUSCATEL survey, via their extended \ion{Mg}{II} emission \citep{Pessa2026}. In this paper, we selected a subset of galaxies from that sample, suitable to infer their recent SFH in a non-parametric manner via modeling their available spectroscopic and photometric data. From their SFHs, we investigated the presence and strength of recent star formation events and explored connections between them and the kinematics of their galactic outflow properties.

Overall, we find a statistically significant connection between the recent SFH of galaxies and the kinematics of their galactic outflows. This connection is inferred via clear positive correlations between the SFR and sSFR inferred from the non-parametric SFH in different timescales, which capture recent stochastic bursts of star formation, and the maximum velocity and acceleration rate of the galactic outflows (however, we did not find any significant correlation with the launching velocity of the outflows). 

On the other hand, these correlations are not observed when using the SFR derived via the [\ion{O}{II}] luminosity of our sample galaxies, or via modeling their SED distribution with a parametric exponentially declining SFH. In the case of the [\ion{O}{II}]-based SFRs, the lack of a correlation is expected because [\ion{O}{II}] traces SFR in the last few megayears, and given the expected travel times of gas ejected from the ISM into the CGM, the SFR in the last few megayears cannot be causally connected to the gas seen in the CGM of galaxies today. In the case of the SED-based SFR tracer, a lack of correlation is explained by the fact that a parametric SFH assumes, by design, a smooth SFH that would not capture the recent stochastic burst of star formation. Thus, with our methodology, we are able to connect these discrete past star formation episodes with the gas in the CGM of galaxies, and find that faster (and more rapidly accelerating) outflows are produced by stronger star formation episodes in the recent SFH of galaxies.

In addition to computing the SFR from the SFH on fixed timescales, we have carried out specific measurements for each galaxy, depending on the computed size and expansion velocity of their \ion{Mg}{II} halo. The goal is to estimate the launching time of the galactic outflow and compute a SFR on a time window that roughly matches this launching time, aiming to capture the star-forming episodes that could be more likely causally connected to the extended emission seen in the data. We find indeed a clear correlation between this SFR tracer and the maximum outflow velocity, but this correlation is not stronger than those computed in fixed timescales, and we interpret this as that the observed extended emission is not necessarily the result of one or a few star-forming episodes in the recent SFH of galaxies, but the result of a series of star-forming events, producing outflows and shaping the CGM properties gradually.

Finally, we explored whether the overall shape of the recent SFH, in addition to the time-integrated total SFR, could have an impact on the properties of galactic outflows, and we find that for SFHs in which a larger fraction of the star formation is concentrated in a single discrete episode, rather than distributed in several similar episodes, the maximum outflow velocity is higher. This means that, while the outflow properties are not necessarily driven by a single star formation episode, they are indeed faster when a single strong star formation episode (or a series of episodes separated by timescales smaller than our time resolution) is the main responsible for launching the wind. 

Our results confirm that the kinematic outflow properties respond to the recent SFH of galaxies, a result that is theoretically expected, since there is a delay between the ejection of the gas from the ISM and the time it reaches CGM scales. Such a connection has been probed before, although in a more indirect way, by comparing the \ion{Mg}{II} absorption strength between sightlines close to star-forming, quiescent, and post-starburst galaxies. Here, to our knowledge, for the first time, we have directly inferred the SFH of galaxies and linked these star-forming episodes with the observed properties of their galactic winds, probed by their \ion{Mg}{II} extended emission. The CGM thus emerges as a dynamical archive of recent star formation activity, preserving the imprint of galaxies’ bursty past in the kinematics of their outflows, and offering a direct observational pathway to constrain models of feedback and galaxy evolution in a temporally resolved framework.

\begin{acknowledgements}

This project has received funding from the European Research Council (ERC) under the European Union's Horizon 2020 research and innovation programme (grant agreement 101020943, SPECMAP-CGM). 

LW acknowledges funding by the Deutsche Forschungsgemeinschaft, Grant Wi 1369/32-1.

JP and LW acknowledge funding by the Deutsche Forschungsgemeinschaft, Grant Wi 1369/31-1.

We thank contributors to {\sc SciPy} \citep{Virtanen2020}, {\sc Matplotlib} \citep{Hunter2007}, {\sc NumPy} \citep{Harris2020}, {\sc Astropy} \citep{Astropy, Astropy2018}, and the PYTHON programming language; the free and open-source community; and the NASA Astrophysics Data System for software and services. We also thank contributors to linetools \citep{Prochaska2016} and PyMUSE \citep{PyMUSE}.

\end{acknowledgements}

\bibliographystyle{aa}
\bibliography{mg2halo}

@ARTICLE{Luo2026,
       author = {{Luo}, Yuanze and {Spilker}, Justin S. and {Setton}, David J. and {Heckman}, Timothy and {Bezanson}, Rachel and {D'Onofrio}, Vincenzo R. and {Geiger}, Alex and {Goulding}, Andy D. and {Greene}, Jenny E. and {Rowlands}, Kate and {Suess}, Katherine A. and {Taylor}, Elizabeth and {Verrico}, Margaret E. and {Weldon}, Andrew and {Wild}, Vivienne and {Zhu}, Pengpei},
        title = "{SQuIGG$\vec{L}$E: Ubiquitous Cool Gas Outflows with Large Mass-Loading Factors in $z\sim0.7$ Post-starburst Galaxies}",
      journal = {arXiv e-prints},
         year = 2026,
        month = sep,
          eid = {arXiv:2609.31308},
        pages = {arXiv:2609.31308},
          doi = {10.48550/arXiv.2609.31308},
archivePrefix = {arXiv},
       eprint = {2609.31308},
 primaryClass = {astro-ph.GA},
       adsurl = {https://ui.adsabs.harvard.edu/abs/2026arXiv260931308L}
}

@ARTICLE{Scuderi1992,
       author = {{Scuderi}, S. and {Bonanno}, G. and {di Benedetto}, R. and {Spadaro}, D. and {Panagia}, N.},
        title = "{H alpha Observations of Early-Type Stars}",
      journal = {\apj},
         year = 1992,
        month = jun,
       volume = {392},
        pages = {201},
          doi = {10.1086/171418},
       adsurl = {https://ui.adsabs.harvard.edu/abs/1992ApJ...392..201S}
}

@ARTICLE{Pessa2026,
       author = {{Pessa}, Ismael and {Wisotzki}, Lutz and {Urrutia}, Tanya and {Bouch{\'e}}, Nicolas F. and {Leclercq}, Floriane and {Augustin}, Ramona and {Guo}, Yucheng and {Kozlova}, Daria and {Kusakabe}, Haruka and {Pharo}, John},
        title = "{First statistical constraints on galactic-scale outflow properties traced by their extended Mg II emission with MUSE}",
      journal = {\aap},
         year = 2026,
        month = apr,
       volume = {708},
          eid = {A214},
        pages = {A214},
          doi = {10.1051/0004-6361/202557875},
archivePrefix = {arXiv},
       eprint = {2602.11280},
 primaryClass = {astro-ph.GA},
       adsurl = {https://ui.adsabs.harvard.edu/abs/2026A&A...708A.214P}
}

@ARTICLE{Wang2009,
       author = {{Wang}, L. and {Mao}, J. and {Xiang}, S. and {Yuan}, Y.-F.},
        title = "{Negative feedback effects on star formation history and cosmic reionization}",
      journal = {\aap},
         year = 2009,
        month = feb,
       volume = {494},
       number = {3},
        pages = {817-827},
          doi = {10.1051/0004-6361/200809916},
archivePrefix = {arXiv},
       eprint = {0812.4085},
 primaryClass = {astro-ph},
       adsurl = {https://ui.adsabs.harvard.edu/abs/2009A&A...494..817W}
}

@ARTICLE{Somerville2015,
       author = {{Somerville}, Rachel S. and {Dav{\'e}}, Romeel},
        title = "{Physical Models of Galaxy Formation in a Cosmological Framework}",
      journal = {\araa},
         year = 2015,
        month = aug,
       volume = {53},
        pages = {51-113},
          doi = {10.1146/annurev-astro-082812-140951},
archivePrefix = {arXiv},
       eprint = {1412.2712},
 primaryClass = {astro-ph.GA},
       adsurl = {https://ui.adsabs.harvard.edu/abs/2015ARA&A..53...51S}
}

@ARTICLE{Rong2026,
       author = {{Rong}, Yu and {Liu}, Shihong},
        title = "{Stellar Surface Density Modulates MgII Cool-gas Outflow Absorption in DESI Star-forming Galaxies}",
      journal = {arXiv e-prints},
         year = 2026,
        month = jun,
          eid = {arXiv:2606.28874},
        pages = {arXiv:2606.28874},
          doi = {10.48550/arXiv.2606.28874},
archivePrefix = {arXiv},
       eprint = {2606.28874},
 primaryClass = {astro-ph.GA},
       adsurl = {https://ui.adsabs.harvard.edu/abs/2026arXiv260628874R}
}

@ARTICLE{Bressan2012,
       author = {{Bressan}, Alessandro and {Marigo}, Paola and {Girardi}, L{\'e}o. and {Salasnich}, Bernardo and {Dal Cero}, Claudia and {Rubele}, Stefano and {Nanni}, Ambra},
        title = "{PARSEC: stellar tracks and isochrones with the PAdova and TRieste Stellar Evolution Code}",
      journal = {\mnras},
         year = 2012,
        month = nov,
       volume = {427},
       number = {1},
        pages = {127-145},
          doi = {10.1111/j.1365-2966.2012.21948.x},
archivePrefix = {arXiv},
       eprint = {1208.4498},
 primaryClass = {astro-ph.SR},
       adsurl = {https://ui.adsabs.harvard.edu/abs/2012MNRAS.427..127B}
}

@ARTICLE{Ge2019,
       author = {{Ge}, Junqiang and {Mao}, Shude and {Lu}, Youjun and {Cappellari}, Michele and {Yan}, Renbin},
        title = "{Recovering stellar population parameters via different population models and stellar libraries}",
      journal = {\mnras},
         year = 2019,
        month = may,
       volume = {485},
       number = {2},
        pages = {1675-1693},
          doi = {10.1093/mnras/stz418},
archivePrefix = {arXiv},
       eprint = {1902.03759},
 primaryClass = {astro-ph.GA},
       adsurl = {https://ui.adsabs.harvard.edu/abs/2019MNRAS.485.1675G}
}

@ARTICLE{Leitherer1999,
       author = {{Leitherer}, Claus and {Schaerer}, Daniel and {Goldader}, Jeffrey D. and {Delgado}, Rosa M. Gonz{\'a}lez and {Robert}, Carmelle and {Kune}, Denis Foo and {de Mello}, Du{\'\i}lia F. and {Devost}, Daniel and {Heckman}, Timothy M.},
        title = "{Starburst99: Synthesis Models for Galaxies with Active Star Formation}",
      journal = {\apjs},
         year = 1999,
        month = jul,
       volume = {123},
       number = {1},
        pages = {3-40},
          doi = {10.1086/313233},
archivePrefix = {arXiv},
       eprint = {astro-ph/9902334},
 primaryClass = {astro-ph},
       adsurl = {https://ui.adsabs.harvard.edu/abs/1999ApJS..123....3L}
}

@ARTICLE{Faucher2018,
       author = {{Faucher-Gigu{\`e}re}, Claude-Andr{\'e}},
        title = "{A model for the origin of bursty star formation in galaxies}",
      journal = {\mnras},
         year = 2018,
        month = jan,
       volume = {473},
       number = {3},
        pages = {3717-3731},
          doi = {10.1093/mnras/stx2595},
archivePrefix = {arXiv},
       eprint = {1701.04824},
 primaryClass = {astro-ph.GA},
       adsurl = {https://ui.adsabs.harvard.edu/abs/2018MNRAS.473.3717F}
}

@ARTICLE{Muratov2015,
       author = {{Muratov}, Alexander L. and {Kere{\v{s}}}, Du{\v{s}}an and {Faucher-Gigu{\`e}re}, Claude-Andr{\'e} and {Hopkins}, Philip F. and {Quataert}, Eliot and {Murray}, Norman},
        title = "{Gusty, gaseous flows of FIRE: galactic winds in cosmological simulations with explicit stellar feedback}",
      journal = {\mnras},
         year = 2015,
        month = dec,
       volume = {454},
       number = {3},
        pages = {2691-2713},
          doi = {10.1093/mnras/stv2126},
archivePrefix = {arXiv},
       eprint = {1501.03155},
 primaryClass = {astro-ph.GA},
       adsurl = {https://ui.adsabs.harvard.edu/abs/2015MNRAS.454.2691M}
}

@ARTICLE{Chisholm2015,
       author = {{Chisholm}, John and {Tremonti}, Christy A. and {Leitherer}, Claus and {Chen}, Yanmei and {Wofford}, Aida and {Lundgren}, Britt},
        title = "{Scaling Relations Between Warm Galactic Outflows and Their Host Galaxies}",
      journal = {\apj},
         year = 2015,
        month = oct,
       volume = {811},
       number = {2},
          eid = {149},
        pages = {149},
          doi = {10.1088/0004-637X/811/2/149},
archivePrefix = {arXiv},
       eprint = {1412.2139},
 primaryClass = {astro-ph.GA},
       adsurl = {https://ui.adsabs.harvard.edu/abs/2015ApJ...811..149C}
}

@ARTICLE{Pillepich2018,
       author = {{Pillepich}, Annalisa and {Springel}, Volker and {Nelson}, Dylan and {Genel}, Shy and {Naiman}, Jill and {Pakmor}, R{\"u}diger and {Hernquist}, Lars and {Torrey}, Paul and {Vogelsberger}, Mark and {Weinberger}, Rainer and {Marinacci}, Federico},
        title = "{Simulating galaxy formation with the IllustrisTNG model}",
      journal = {\mnras},
         year = 2018,
        month = jan,
       volume = {473},
       number = {3},
        pages = {4077-4106},
          doi = {10.1093/mnras/stx2656},
archivePrefix = {arXiv},
       eprint = {1703.02970},
 primaryClass = {astro-ph.GA},
       adsurl = {https://ui.adsabs.harvard.edu/abs/2018MNRAS.473.4077P}
}

@ARTICLE{Schaye2015,
       author = {{Schaye}, Joop and {Crain}, Robert A. and {Bower}, Richard G. and {Furlong}, Michelle and {Schaller}, Matthieu and {Theuns}, Tom and {Dalla Vecchia}, Claudio and {Frenk}, Carlos S. and {McCarthy}, I.~G. and {Helly}, John C. and {Jenkins}, Adrian and {Rosas-Guevara}, Y.~M. and {White}, Simon D.~M. and {Baes}, Maarten and {Booth}, C.~M. and {Camps}, Peter and {Navarro}, Julio F. and {Qu}, Yan and {Rahmati}, Alireza and {Sawala}, Till and {Thomas}, Peter A. and {Trayford}, James},
        title = "{The EAGLE project: simulating the evolution and assembly of galaxies and their environments}",
      journal = {\mnras},
         year = 2015,
        month = jan,
       volume = {446},
       number = {1},
        pages = {521-554},
          doi = {10.1093/mnras/stu2058},
archivePrefix = {arXiv},
       eprint = {1407.7040},
 primaryClass = {astro-ph.GA},
       adsurl = {https://ui.adsabs.harvard.edu/abs/2015MNRAS.446..521S}
}

@ARTICLE{Luisi2021,
       author = {{Luisi}, Matteo and {Anderson}, Loren D. and {Schneider}, Nicola and {Simon}, Robert and {Kabanovic}, Slawa and {G{\"u}sten}, Rolf and {Zavagno}, Annie and {Broos}, Patrick S. and {Buchbender}, Christof and {Guevara}, Cristian and {Jacobs}, Karl and {Justen}, Matthias and {Klein}, Bernd and {Linville}, Dylan and {R{\"o}llig}, Markus and {Russeil}, Delphine and {Stutzki}, J{\"u}rgen and {Tiwari}, Maitraiyee and {Townsley}, Leisa K. and {Tielens}, Alexander G.~G.~M.},
        title = "{Stellar feedback and triggered star formation in the prototypical bubble RCW 120}",
      journal = {Science Advances},
         year = 2021,
        month = apr,
       volume = {7},
       number = {15},
        pages = {eabe9511},
          doi = {10.1126/sciadv.abe9511},
archivePrefix = {arXiv},
       eprint = {2104.04568},
 primaryClass = {astro-ph.GA},
       adsurl = {https://ui.adsabs.harvard.edu/abs/2021SciA....7.9511L}
}

@ARTICLE{Conroy2010,
       author = {{Conroy}, Charlie and {Gunn}, James E.},
        title = "{The Propagation of Uncertainties in Stellar Population Synthesis Modeling. III. Model Calibration, Comparison, and Evaluation}",
      journal = {\apj},
         year = 2010,
        month = apr,
       volume = {712},
       number = {2},
        pages = {833-857},
          doi = {10.1088/0004-637X/712/2/833},
archivePrefix = {arXiv},
       eprint = {0911.3151},
 primaryClass = {astro-ph.CO},
       adsurl = {https://ui.adsabs.harvard.edu/abs/2010ApJ...712..833C}
}

@ARTICLE{Lawrence2007,
       author = {{Lawrence}, A. and {Warren}, S.~J. and {Almaini}, O. and {Edge}, A.~C. and {Hambly}, N.~C. and {Jameson}, R.~F. and {Lucas}, P. and {Casali}, M. and {Adamson}, A. and {Dye}, S. and {Emerson}, J.~P. and {Foucaud}, S. and {Hewett}, P. and {Hirst}, P. and {Hodgkin}, S.~T. and {Irwin}, M.~J. and {Lodieu}, N. and {McMahon}, R.~G. and {Simpson}, C. and {Smail}, I. and {Mortlock}, D. and {Folger}, M.},
        title = "{The UKIRT Infrared Deep Sky Survey (UKIDSS)}",
      journal = {\mnras},
         year = 2007,
        month = aug,
       volume = {379},
       number = {4},
        pages = {1599-1617},
          doi = {10.1111/j.1365-2966.2007.12040.x},
archivePrefix = {arXiv},
       eprint = {astro-ph/0604426},
 primaryClass = {astro-ph},
       adsurl = {https://ui.adsabs.harvard.edu/abs/2007MNRAS.379.1599L}
}

@ARTICLE{Aihara2011,
       author = {{Aihara}, Hiroaki and {Allende Prieto}, Carlos and {An}, Deokkeun and {Anderson}, Scott F. and {Aubourg}, {\'E}ric and {Balbinot}, Eduardo and {Beers}, Timothy C. and {Berlind}, Andreas A. and {Bickerton}, Steven J. and {Bizyaev}, Dmitry and {Blanton}, Michael R. and {Bochanski}, John J. and {Bolton}, Adam S. and {Bovy}, Jo and {Brandt}, W.~N. and {Brinkmann}, J. and {Brown}, Peter J. and {Brownstein}, Joel R. and {Busca}, Nicolas G. and {Campbell}, Heather and {Carr}, Michael A. and {Chen}, Yanmei and {Chiappini}, Cristina and {Comparat}, Johan and {Connolly}, Natalia and {Cortes}, Marina and {Croft}, Rupert A.~C. and {Cuesta}, Antonio J. and {da Costa}, Luiz N. and {Davenport}, James R.~A. and {Dawson}, Kyle and {Dhital}, Saurav and {Ealet}, Anne and {Ebelke}, Garrett L. and {Edmondson}, Edward M. and {Eisenstein}, Daniel J. and {Escoffier}, Stephanie and {Esposito}, Massimiliano and {Evans}, Michael L. and {Fan}, Xiaohui and {Femen{\'\i}a Castell{\'a}}, Bruno and {Font-Ribera}, Andreu and {Frinchaboy}, Peter M. and {Ge}, Jian and {Gillespie}, Bruce A. and {Gilmore}, G. and {Gonz{\'a}lez Hern{\'a}ndez}, Jonay I. and {Gott}, J. Richard and {Gould}, Andrew and {Grebel}, Eva K. and {Gunn}, James E. and {Hamilton}, Jean-Christophe and {Harding}, Paul and {Harris}, David W. and {Hawley}, Suzanne L. and {Hearty}, Frederick R. and {Ho}, Shirley and {Hogg}, David W. and {Holtzman}, Jon A. and {Honscheid}, Klaus and {Inada}, Naohisa and {Ivans}, Inese I. and {Jiang}, Linhua and {Johnson}, Jennifer A. and {Jordan}, Cathy and {Jordan}, Wendell P. and {Kazin}, Eyal A. and {Kirkby}, David and {Klaene}, Mark A. and {Knapp}, G.~R. and {Kneib}, Jean-Paul and {Kochanek}, C.~S. and {Koesterke}, Lars and {Kollmeier}, Juna A. and {Kron}, Richard G. and {Lampeitl}, Hubert and {Lang}, Dustin and {Le Goff}, Jean-Marc and {Lee}, Young Sun and {Lin}, Yen-Ting and {Long}, Daniel C. and {Loomis}, Craig P. and {Lucatello}, Sara and {Lundgren}, Britt and {Lupton}, Robert H. and {Ma}, Zhibo and {MacDonald}, Nicholas and {Mahadevan}, Suvrath and {Maia}, Marcio A.~G. and {Makler}, Martin and {Malanushenko}, Elena and {Malanushenko}, Viktor and {Mandelbaum}, Rachel and {Maraston}, Claudia and {Margala}, Daniel and {Masters}, Karen L. and {McBride}, Cameron K. and {McGehee}, Peregrine M. and {McGreer}, Ian D. and {M{\'e}nard}, Brice and {Miralda-Escud{\'e}}, Jordi and {Morrison}, Heather L. and {Mullally}, F. and {Muna}, Demitri and {Munn}, Jeffrey A. and {Murayama}, Hitoshi and {Myers}, Adam D. and {Naugle}, Tracy and {Neto}, Angelo Fausti and {Nguyen}, Duy Cuong and {Nichol}, Robert C. and {O'Connell}, Robert W. and {Ogando}, Ricardo L.~C. and {Olmstead}, Matthew D. and {Oravetz}, Daniel J. and {Padmanabhan}, Nikhil and {Palanque-Delabrouille}, Nathalie and {Pan}, Kaike and {Pandey}, Parul and {P{\^a}ris}, Isabelle and {Percival}, Will J. and {Petitjean}, Patrick and {Pfaffenberger}, Robert and {Pforr}, Janine and {Phleps}, Stefanie and {Pichon}, Christophe and {Pieri}, Matthew M. and {Prada}, Francisco and {Price-Whelan}, Adrian M. and {Raddick}, M. Jordan and {Ramos}, Beatriz H.~F. and {Reyl{\'e}}, C{\'e}line and {Rich}, James and {Richards}, Gordon T. and {Rix}, Hans-Walter and {Robin}, Annie C. and {Rocha-Pinto}, Helio J. and {Rockosi}, Constance M. and {Roe}, Natalie A. and {Rollinde}, Emmanuel and {Ross}, Ashley J. and {Ross}, Nicholas P. and {Rossetto}, Bruno M. and {S{\'a}nchez}, Ariel G. and {Sayres}, Conor and {Schlegel}, David J. and {Schlesinger}, Katharine J. and {Schmidt}, Sarah J. and {Schneider}, Donald P. and {Sheldon}, Erin and {Shu}, Yiping and {Simmerer}, Jennifer and {Simmons}, Audrey E. and {Sivarani}, Thirupathi and {Snedden}, Stephanie A. and {Sobeck}, Jennifer S. and {Steinmetz}, Matthias and {Strauss}, Michael A. and {Szalay}, Alexander S. and {Tanaka}, Masayuki and {Thakar}, Aniruddha R. and {Thomas}, Daniel and {Tinker}, Jeremy L. and {Tofflemire}, Benjamin M. and {Tojeiro}, Rita and {Tremonti}, Christy A. and {Vandenberg}, Jan and {Vargas Maga{\~n}a}, M. and {Verde}, Licia and {Vogt}, Nicole P. and {Wake}, David A. and {Wang}, Ji and {Weaver}, Benjamin A. and {Weinberg}, David H. and {White}, Martin and {White}, Simon D.~M. and {Yanny}, Brian and {Yasuda}, Naoki and {Yeche}, Christophe and {Zehavi}, Idit},
        title = "{The Eighth Data Release of the Sloan Digital Sky Survey: First Data from SDSS-III}",
      journal = {\apjs},
         year = 2011,
        month = apr,
       volume = {193},
       number = {2},
          eid = {29},
        pages = {29},
          doi = {10.1088/0067-0049/193/2/29},
archivePrefix = {arXiv},
       eprint = {1101.1559},
 primaryClass = {astro-ph.IM},
       adsurl = {https://ui.adsabs.harvard.edu/abs/2011ApJS..193...29A}
}

@ARTICLE{Sun2024,
       author = {{Sun}, Yang and {Lee}, Gwang-Ho and {Zabludoff}, Ann I. and {French}, K. Decker and {Helton}, Jakob M. and {Kerrison}, Nicole A. and {Tremonti}, Christy A. and {Yang}, Yujin},
        title = "{Evolution of gas flows along the starburst to post-starburst to quiescent galaxy sequence}",
      journal = {\mnras},
         year = 2024,
        month = mar,
       volume = {528},
       number = {4},
        pages = {5783-5803},
          doi = {10.1093/mnras/stae366},
archivePrefix = {arXiv},
       eprint = {2310.00424},
 primaryClass = {astro-ph.GA},
       adsurl = {https://ui.adsabs.harvard.edu/abs/2024MNRAS.528.5783S}
}

@ARTICLE{Taylor2024,
       author = {{Taylor}, Elizabeth and {Maltby}, David and {Almaini}, Omar and {Merrifield}, Michael and {Wild}, Vivienne and {Rowlands}, Kate and {Harrold}, Jimi},
        title = "{High-velocity outflows persist up to 1 Gyr after a starburst in recently quenched galaxies at z > 1}",
      journal = {\mnras},
         year = 2024,
        month = dec,
       volume = {535},
       number = {2},
        pages = {1684-1692},
          doi = {10.1093/mnras/stae2463},
archivePrefix = {arXiv},
       eprint = {2411.00102},
 primaryClass = {astro-ph.GA},
       adsurl = {https://ui.adsabs.harvard.edu/abs/2024MNRAS.535.1684T}
}

@ARTICLE{Hemmati2015,
       author = {{Hemmati}, Shoubaneh and {Mobasher}, Bahram and {Darvish}, Behnam and {Nayyeri}, Hooshang and {Sobral}, David and {Miller}, Sarah},
        title = "{Nebular and Stellar Dust Extinction Across the Disk of Emission-line Galaxies on Kiloparsec Scales}",
      journal = {\apj},
         year = 2015,
        month = nov,
       volume = {814},
       number = {1},
          eid = {46},
        pages = {46},
          doi = {10.1088/0004-637X/814/1/46},
archivePrefix = {arXiv},
       eprint = {1510.02506},
 primaryClass = {astro-ph.GA},
       adsurl = {https://ui.adsabs.harvard.edu/abs/2015ApJ...814...46H}
}

@ARTICLE{Pessa2024,
       author = {{Pessa}, Ismael and {Wisotzki}, Lutz and {Urrutia}, Tanya and {Pharo}, John and {Augustin}, Ramona and {Bouch{\'e}}, Nicolas F. and {Feltre}, Anna and {Guo}, Yucheng and {Kozlova}, Daria and {Krajnovic}, Davor and {Kusakabe}, Haruka and {Leclercq}, Floriane and {Salas}, H{\'e}ctor and {Schaye}, Joop and {Verhamme}, Anne},
        title = "{A galactic outflow traced by its extended Mg II emission out to a {\ensuremath{\sim}}30 kpc radius in the Hubble Ultra Deep Field with MUSE}",
      journal = {\aap},
         year = 2024,
        month = nov,
       volume = {691},
          eid = {A5},
        pages = {A5},
          doi = {10.1051/0004-6361/202450547},
archivePrefix = {arXiv},
       eprint = {2408.16067},
 primaryClass = {astro-ph.GA},
       adsurl = {https://ui.adsabs.harvard.edu/abs/2024A&A...691A...5P}
}

@ARTICLE{MunozLopez2025,
       author = {{Mu{\~n}oz L{\'o}pez}, C. and {Krajnovi{\'c}}, D. and {Epinat}, B. and {Urrutia}, T. and {Pessa}, I. and {Contini}, T. and {Nanayakkara}, T. and {Pharo}, J. and {Gon{\c{c}}alves}, T.~S. and {Thi Thai}, T. and {Bouch{\'e}}, N.~F.},
        title = "{Multiple star-forming episodes of intermediate-redshift galaxies}",
      journal = {\aap},
         year = 2025,
        month = oct,
       volume = {702},
          eid = {A146},
        pages = {A146},
          doi = {10.1051/0004-6361/202555197},
archivePrefix = {arXiv},
       eprint = {2509.01710},
 primaryClass = {astro-ph.GA},
       adsurl = {https://ui.adsabs.harvard.edu/abs/2025A&A...702A.146M}
}

@ARTICLE{Bouche2012,
       author = {{Bouch{\'e}}, N. and {Hohensee}, W. and {Vargas}, R. and {Kacprzak}, G.~G. and {Martin}, C.~L. and {Cooke}, J. and {Churchill}, C.~W.},
        title = "{Physical properties of galactic winds using background quasars}",
      journal = {\mnras},
         year = 2012,
        month = oct,
       volume = {426},
       number = {2},
        pages = {801-815},
          doi = {10.1111/j.1365-2966.2012.21114.x},
archivePrefix = {arXiv},
       eprint = {1110.5877},
 primaryClass = {astro-ph.CO},
       adsurl = {https://ui.adsabs.harvard.edu/abs/2012MNRAS.426..801B}
}

@ARTICLE{Kang2025,
       author = {{Kang}, Cheonsu and {Kimm}, Taysun and {Han}, Daniel and {Katz}, Harley and {Devriendt}, Julien and {Slyz}, Adrianne and {Teyssier}, Romain},
        title = "{Impact of star formation models on the growth of simulated galaxies at high redshifts}",
      journal = {\aap},
         year = 2025,
        month = jan,
       volume = {693},
          eid = {A149},
        pages = {A149},
          doi = {10.1051/0004-6361/202451502},
archivePrefix = {arXiv},
       eprint = {2407.12090},
 primaryClass = {astro-ph.GA},
       adsurl = {https://ui.adsabs.harvard.edu/abs/2025A&A...693A.149K}
}

@ARTICLE{Valentino2025,
       author = {{Valentino}, F. and {Heintz}, K.~E. and {Brammer}, G. and {Ito}, K. and {Kokorev}, V. and {Whitaker}, K.~E. and {Gallazzi}, A. and {de Graaff}, A. and {Weibel}, A. and {Frye}, B.~L. and {Kamieneski}, P.~S. and {Jin}, S. and {Ceverino}, D. and {Faisst}, A. and {Farcy}, M. and {Fujimoto}, S. and {Gillman}, S. and {Gottumukkala}, R. and {Hamadouche}, M. and {Harrington}, K.~C. and {Hirschmann}, M. and {Jespersen}, C.~K. and {Kakimoto}, T. and {Kubo}, M. and {Lagos}, C. d. P. and {Lee}, M. and {Magdis}, G.~E. and {Man}, A.~W.~S. and {Onodera}, M. and {Rizzo}, F. and {Shimakawa}, R. and {Setton}, D.~J. and {Tanaka}, M. and {Toft}, S. and {Wu}, P.-F. and {Zhu}, P.},
        title = "{Gas outflows in two recently quenched galaxies at z = 4 and 7}",
      journal = {\aap},
         year = 2025,
        month = jul,
       volume = {699},
          eid = {A358},
        pages = {A358},
          doi = {10.1051/0004-6361/202553908},
archivePrefix = {arXiv},
       eprint = {2503.01990},
 primaryClass = {astro-ph.GA},
       adsurl = {https://ui.adsabs.harvard.edu/abs/2025A&A...699A.358V}
}

@ARTICLE{Buchner2024,
       author = {{Buchner}, Johannes},
        title = "{Impediments to the cosmic growth of galaxies: The outflow budget from Star Formation and Active Galactic Nuclei}",
      journal = {\aap},
         year = 2024,
        month = sep,
       volume = {689},
          eid = {L2},
        pages = {L2},
          doi = {10.1051/0004-6361/202451170},
archivePrefix = {arXiv},
       eprint = {2407.11127},
 primaryClass = {astro-ph.GA},
       adsurl = {https://ui.adsabs.harvard.edu/abs/2024A&A...689L...2B}
}

@ARTICLE{Serrato-Alos2026,
       author = {{Sarrato-Al{\'o}s}, J. and {Bullock}, J. and {Di Cintio}, A. and {Brook}, C. and {Valenciano}, F. and {Macci{\`o}}, A.~V.},
        title = "{The impact of star formation histories on the inner dark matter density slopes of galaxies}",
      journal = {\aap},
         year = 2026,
        month = jun,
       volume = {710},
          eid = {A95},
        pages = {A95},
          doi = {10.1051/0004-6361/202558718},
archivePrefix = {arXiv},
       eprint = {2601.17113},
 primaryClass = {astro-ph.GA},
       adsurl = {https://ui.adsabs.harvard.edu/abs/2026A&A...710A..95S}
}

@ARTICLE{Galfit_b,
       author = {{Peng}, Chien Y. and {Ho}, Luis C. and {Impey}, Chris D. and {Rix}, Hans-Walter},
        title = "{Detailed Decomposition of Galaxy Images. II. Beyond Axisymmetric Models}",
      journal = {\aj},
         year = 2010,
        month = jun,
       volume = {139},
       number = {6},
        pages = {2097-2129},
          doi = {10.1088/0004-6256/139/6/2097},
archivePrefix = {arXiv},
       eprint = {0912.0731},
 primaryClass = {astro-ph.CO},
       adsurl = {https://ui.adsabs.harvard.edu/abs/2010AJ....139.2097P}
}

@ARTICLE{Galfit_a,
       author = {{Peng}, Chien Y. and {Ho}, Luis C. and {Impey}, Chris D. and {Rix}, Hans-Walter},
        title = "{Detailed Structural Decomposition of Galaxy Images}",
      journal = {\aj},
         year = 2002,
        month = jul,
       volume = {124},
       number = {1},
        pages = {266-293},
          doi = {10.1086/340952},
archivePrefix = {arXiv},
       eprint = {astro-ph/0204182},
 primaryClass = {astro-ph},
       adsurl = {https://ui.adsabs.harvard.edu/abs/2002AJ....124..266P}
}

@ARTICLE{Reynaldi2013,
       author = {{Reynaldi}, V. and {Feinstein}, C.},
        title = "{Shock ionization in the extended emission-line region of 3C 305: the last piece of the (optical) puzzle}",
      journal = {\mnras},
         year = 2013,
        month = oct,
       volume = {435},
       number = {2},
        pages = {1350-1357},
          doi = {10.1093/mnras/stt1377},
archivePrefix = {arXiv},
       eprint = {1307.7489},
 primaryClass = {astro-ph.CO},
       adsurl = {https://ui.adsabs.harvard.edu/abs/2013MNRAS.435.1350R}
}

@ARTICLE{Nelson2019,
       author = {{Nelson}, Dylan and {Pillepich}, Annalisa and {Springel}, Volker and {Pakmor}, R{\"u}diger and {Weinberger}, Rainer and {Genel}, Shy and {Torrey}, Paul and {Vogelsberger}, Mark and {Marinacci}, Federico and {Hernquist}, Lars},
        title = "{First results from the TNG50 simulation: galactic outflows driven by supernovae and black hole feedback}",
      journal = {\mnras},
         year = 2019,
        month = dec,
       volume = {490},
       number = {3},
        pages = {3234-3261},
          doi = {10.1093/mnras/stz2306},
archivePrefix = {arXiv},
       eprint = {1902.05554},
 primaryClass = {astro-ph.GA},
       adsurl = {https://ui.adsabs.harvard.edu/abs/2019MNRAS.490.3234N}
}

@ARTICLE{Hopkins2018,
       author = {{Hopkins}, Philip F. and {Wetzel}, Andrew and {Kere{\v{s}}}, Du{\v{s}}an and {Faucher-Gigu{\`e}re}, Claude-Andr{\'e} and {Quataert}, Eliot and {Boylan-Kolchin}, Michael and {Murray}, Norman and {Hayward}, Christopher C. and {Garrison-Kimmel}, Shea and {Hummels}, Cameron and {Feldmann}, Robert and {Torrey}, Paul and {Ma}, Xiangcheng and {Angl{\'e}s-Alc{\'a}zar}, Daniel and {Su}, Kung-Yi and {Orr}, Matthew and {Schmitz}, Denise and {Escala}, Ivanna and {Sanderson}, Robyn and {Grudi{\'c}}, Michael Y. and {Hafen}, Zachary and {Kim}, Ji-Hoon and {Fitts}, Alex and {Bullock}, James S. and {Wheeler}, Coral and {Chan}, T.~K. and {Elbert}, Oliver D. and {Narayanan}, Desika},
        title = "{FIRE-2 simulations: physics versus numerics in galaxy formation}",
      journal = {\mnras},
         year = 2018,
        month = oct,
       volume = {480},
       number = {1},
        pages = {800-863},
          doi = {10.1093/mnras/sty1690},
archivePrefix = {arXiv},
       eprint = {1702.06148},
 primaryClass = {astro-ph.GA},
       adsurl = {https://ui.adsabs.harvard.edu/abs/2018MNRAS.480..800H}
}

@ARTICLE{BPT,
       author = {{Baldwin}, J.~A. and {Phillips}, M.~M. and {Terlevich}, R.},
        title = "{Classification parameters for the emission-line spectra of extragalactic objects.}",
      journal = {\pasp},
         year = 1981,
        month = feb,
       volume = {93},
        pages = {5-19},
          doi = {10.1086/130766},
       adsurl = {https://ui.adsabs.harvard.edu/abs/1981PASP...93....5B}
}

@ARTICLE{Emsellem2022,
       author = {{Emsellem}, Eric and {Schinnerer}, Eva and {Santoro}, Francesco and {Belfiore}, Francesco and {Pessa}, Ismael and {McElroy}, Rebecca and {Blanc}, Guillermo A. and {Congiu}, Enrico and {Groves}, Brent and {Ho}, I. -Ting and {Kreckel}, Kathryn and {Razza}, Alessandro and {Sanchez-Blazquez}, Patricia and {Egorov}, Oleg and {Faesi}, Chris and {Klessen}, Ralf S. and {Leroy}, Adam K. and {Meidt}, Sharon and {Querejeta}, Miguel and {Rosolowsky}, Erik and {Scheuermann}, Fabian and {Anand}, Gagandeep S. and {Barnes}, Ashley T. and {Be{\v{s}}li{\'c}}, Ivana and {Bigiel}, Frank and {Boquien}, M{\'e}d{\'e}ric and {Cao}, Yixian and {Chevance}, M{\'e}lanie and {Dale}, Daniel A. and {Eibensteiner}, Cosima and {Glover}, Simon C.~O. and {Grasha}, Kathryn and {Henshaw}, Jonathan D. and {Hughes}, Annie and {Koch}, Eric W. and {Kruijssen}, J.~M. Diederik and {Lee}, Janice and {Liu}, Daizhong and {Pan}, Hsi-An and {Pety}, J{\'e}r{\^o}me and {Saito}, Toshiki and {Sandstrom}, Karin M. and {Schruba}, Andreas and {Sun}, Jiayi and {Thilker}, David A. and {Usero}, Antonio and {Watkins}, Elizabeth J. and {Williams}, Thomas G.},
        title = "{The PHANGS-MUSE survey. Probing the chemo-dynamical evolution of disc galaxies}",
      journal = {\aap},
         year = 2022,
        month = mar,
       volume = {659},
          eid = {A191},
        pages = {A191},
          doi = {10.1051/0004-6361/202141727},
archivePrefix = {arXiv},
       eprint = {2110.03708},
 primaryClass = {astro-ph.GA},
       adsurl = {https://ui.adsabs.harvard.edu/abs/2022A&A...659A.191E}
}

@ARTICLE{Kewley2004,
       author = {{Kewley}, Lisa J. and {Geller}, Margaret J. and {Jansen}, Rolf A.},
        title = "{[O II] as a Star Formation Rate Indicator}",
      journal = {\aj},
         year = 2004,
        month = apr,
       volume = {127},
       number = {4},
        pages = {2002-2030},
          doi = {10.1086/382723},
archivePrefix = {arXiv},
       eprint = {astro-ph/0401172},
 primaryClass = {astro-ph},
       adsurl = {https://ui.adsabs.harvard.edu/abs/2004AJ....127.2002K}
}

@ARTICLE{Maxime2026,
       author = {{Rey}, Maxime and {Blaizot}, J{\'e}r{\'e}my and {Kimm}, Taysun and {Rosdahl}, Joakim and {Michel-Dansac}, L{\'e}o and {Mauerhofer}, Valentin},
        title = "{ARCHITECTS II: Impact of subgrid physics on the observable properties of the circumgalactic medium}",
      journal = {arXiv e-prints},
         year = 2026,
        month = feb,
          eid = {arXiv:2602.13394},
        pages = {arXiv:2602.13394},
          doi = {10.48550/arXiv.2602.13394},
archivePrefix = {arXiv},
       eprint = {2602.13394},
 primaryClass = {astro-ph.GA},
       adsurl = {https://ui.adsabs.harvard.edu/abs/2026arXiv260213394R}
}

@ARTICLE{Goh2006,
       author = {{Goh}, K.-I. and {Barab{\'a}si}, A.-L.},
        title = "{Burstiness and memory in complex systems}",
      journal = {EPL (Europhysics Letters)},
         year = 2008,
        month = feb,
       volume = {81},
       number = {4},
        pages = {48002},
          doi = {10.1209/0295-5075/81/48002},
       adsurl = {https://ui.adsabs.harvard.edu/abs/2008EL.....8148002G}
}

@ARTICLE{Castor1970,
       author = {{Castor}, J.~I.},
        title = "{Spectral line formation in Wolf-Rayet envelopes.}",
      journal = {\mnras},
         year = 1970,
        month = jan,
       volume = {149},
        pages = {111},
          doi = {10.1093/mnras/149.2.111},
       adsurl = {https://ui.adsabs.harvard.edu/abs/1970MNRAS.149..111C}
}

@ARTICLE{Burchett2021,
       author = {{Burchett}, Joseph N. and {Rubin}, Kate H.~R. and {Prochaska}, J. Xavier and {Coil}, Alison L. and {Vaught}, Ryan Rickards and {Hennawi}, Joseph F.},
        title = "{Circumgalactic Mg II Emission from an Isotropic Starburst Galaxy Outflow Mapped by KCWI}",
      journal = {\apj},
         year = 2021,
        month = mar,
       volume = {909},
       number = {2},
          eid = {151},
        pages = {151},
          doi = {10.3847/1538-4357/abd4e0},
archivePrefix = {arXiv},
       eprint = {2005.03017},
 primaryClass = {astro-ph.GA},
       adsurl = {https://ui.adsabs.harvard.edu/abs/2021ApJ...909..151B}
}

@ARTICLE{Tumlinson2017,
       author = {{Tumlinson}, Jason and {Peeples}, Molly S. and {Werk}, Jessica K.},
        title = "{The Circumgalactic Medium}",
      journal = {\araa},
         year = 2017,
        month = aug,
       volume = {55},
       number = {1},
        pages = {389-432},
          doi = {10.1146/annurev-astro-091916-055240},
archivePrefix = {arXiv},
       eprint = {1709.09180},
 primaryClass = {astro-ph.GA},
       adsurl = {https://ui.adsabs.harvard.edu/abs/2017ARA&A..55..389T}
}

@ARTICLE{Leibundgut2017,
       author = {{Leibundgut}, B. and {Bacon}, R. and {Jaff{\'e}}, Y.~L. and {Johnston}, E. and {Kuntschner}, H. and {Selman}, F. and {Valenti}, E. and {Vernet}, J. and {Vogt}, F.},
        title = "{MUSE WFM AO Science Verification}",
      journal = {The Messenger},
         year = 2017,
        month = dec,
       volume = {170},
        pages = {20-25},
          doi = {10.18727/0722-6691/5050},
       adsurl = {https://ui.adsabs.harvard.edu/abs/2017Msngr.170...20L}
}

@ARTICLE{Guo2023b,
       author = {{Guo}, Yucheng and {Bacon}, Roland and {Bouch{\'e}}, Nicolas F. and {Wisotzki}, Lutz and {Schaye}, Joop and {Blaizot}, J{\'e}r{\'e}my and {Verhamme}, Anne and {Cantalupo}, Sebastiano and {Boogaard}, Leindert A. and {Brinchmann}, Jarle and {Cherrey}, Maxime and {Kusakabe}, Haruka and {Langan}, Ivanna and {Leclercq}, Floriane and {Matthee}, Jorryt and {Michel-Dansac}, L{\'e}o and {Schroetter}, Ilane and {Wendt}, Martin},
        title = "{Bipolar outflows out to 10 kpc for massive galaxies at redshift z {\ensuremath{\approx}} 1}",
      journal = {\nat},
         year = 2023,
        month = dec,
       volume = {624},
       number = {7990},
        pages = {53-56},
          doi = {10.1038/s41586-023-06718-w},
archivePrefix = {arXiv},
       eprint = {2312.05167},
 primaryClass = {astro-ph.GA},
       adsurl = {https://ui.adsabs.harvard.edu/abs/2023Natur.624...53G}
}

@ARTICLE{Bacon2023,
       author = {{Bacon}, Roland and {Brinchmann}, Jarle and {Conseil}, Simon and {Maseda}, Michael and {Nanayakkara}, Themiya and {Wendt}, Martin and {Bacher}, Raphael and {Mary}, David and {Weilbacher}, Peter M. and {Krajnovi{\'c}}, Davor and {Boogaard}, Leindert and {Bouch{\'e}}, Nicolas F and {Contini}, Thierry and {Epinat}, Beno{\^\i}t and {Feltre}, Anna and {Guo}, Yucheng and {Herenz}, Christian and {Kollatschny}, Wolfram and {Kusakabe}, Haruka and {Leclercq}, Floriane and {Michel-Dansac}, L{\'e}o and {Pello}, Roser and {Richard}, Johan and {Roth}, Martin and {Salvignol}, Gregory and {Schaye}, Joop and {Steinmetz}, Matthias and {Tresse}, Laurence and {Urrutia}, Tanya and {Verhamme}, Anne and {Vitte}, Eloise and {Wisotzki}, Lutz and {Zoutendijk}, Sebastiaan L.},
        title = "{The MUSE Hubble Ultra Deep Field surveys: Data release II}",
      journal = {\aap},
         year = 2023,
        month = feb,
       volume = {670},
          eid = {A4},
        pages = {A4},
          doi = {10.1051/0004-6361/202244187},
archivePrefix = {arXiv},
       eprint = {2211.08493},
 primaryClass = {astro-ph.GA},
       adsurl = {https://ui.adsabs.harvard.edu/abs/2023A&A...670A...4B}
}

@ARTICLE{Leclercq2022,
       author = {{Leclercq}, Floriane and {Verhamme}, Anne and {Epinat}, Benoit and {Simmonds}, Charlotte and {Matthee}, Jorryt and {Bouch{\'e}}, Nicolas F. and {Garel}, Thibault and {Urrutia}, Tanya and {Wisotzki}, Lutz and {Zabl}, Johannes and {Bacon}, Roland and {Abril-Melgarejo}, Valentina and {Boogaard}, Leindert and {Brinchmann}, Jarle and {Cantalupo}, Sebastiano and {Contini}, Thierry and {Kerutt}, Josephine and {Kusakabe}, Haruka and {Maseda}, Michael and {Michel-Dansac}, L{\'e}o and {Muzahid}, Sowgat and {Nanayakkara}, Themiya and {Richard}, Johan and {Schaye}, Joop},
        title = "{The MUSE eXtremely deep field: first panoramic view of an Mg II emitting intragroup medium}",
      journal = {\aap},
         year = 2022,
        month = jul,
       volume = {663},
          eid = {A11},
        pages = {A11},
          doi = {10.1051/0004-6361/202142179},
archivePrefix = {arXiv},
       eprint = {2203.05614},
 primaryClass = {astro-ph.GA},
       adsurl = {https://ui.adsabs.harvard.edu/abs/2022A&A...663A..11L}
}

@ARTICLE{Zabl2021,
       author = {{Zabl}, Johannes and {Bouch{\'e}}, Nicolas F. and {Wisotzki}, Lutz and {Schaye}, Joop and {Leclercq}, Floriane and {Garel}, Thibault and {Wendt}, Martin and {Schroetter}, Ilane and {Muzahid}, Sowgat and {Cantalupo}, Sebastiano and {Contini}, Thierry and {Bacon}, Roland and {Brinchmann}, Jarle and {Richard}, Johan},
        title = "{MusE GAs FLOw and Wind (MEGAFLOW) VIII. Discovery of a MgII emission halo probed by a quasar sightline}",
      journal = {\mnras},
         year = 2021,
        month = nov,
       volume = {507},
       number = {3},
        pages = {4294-4315},
          doi = {10.1093/mnras/stab2165},
archivePrefix = {arXiv},
       eprint = {2105.14090},
 primaryClass = {astro-ph.GA},
       adsurl = {https://ui.adsabs.harvard.edu/abs/2021MNRAS.507.4294Z}
}

@article{Schroetter:2019es,
author = {Schroetter, Ilane and Bouch{\'e}, Nicolas F and Zabl, Johannes and Contini, Thierry and Wendt, Martin and Schaye, Joop and Mitchell, Peter and Muzahid, Sowgat and Marino, Raffaella A and Bacon, Roland and Lilly, Simon J and Richard, Johan and Wisotzki, Lutz},
title = {{MusE GAs FLOw and Wind (MEGAFLOW) - III. Galactic wind properties using background quasars}},
journal = {Monthly Notices of the Royal Astronomical Society},
year = {2019},
volume = {490},
number = {3},
pages = {4368--4381},
month = oct
}

@ARTICLE{Heckman2015,
       author = {{Heckman}, Timothy M. and {Alexandroff}, Rachel M. and {Borthakur}, Sanchayeeta and {Overzier}, Roderik and {Leitherer}, Claus},
        title = "{The Systematic Properties of the Warm Phase of Starburst-Driven Galactic Winds}",
      journal = {\apj},
         year = 2015,
        month = aug,
       volume = {809},
       number = {2},
          eid = {147},
        pages = {147},
          doi = {10.1088/0004-637X/809/2/147},
archivePrefix = {arXiv},
       eprint = {1507.05622},
 primaryClass = {astro-ph.GA},
       adsurl = {https://ui.adsabs.harvard.edu/abs/2015ApJ...809..147H}
}

@ARTICLE{Bezanson2024,
       author = {{Bezanson}, Rachel and {Labbe}, Ivo and {Whitaker}, Katherine E. and {Leja}, Joel and {Price}, Sedona H. and {Franx}, Marijn and {Brammer}, Gabriel and {Marchesini}, Danilo and {Zitrin}, Adi and {Wang}, Bingjie and {Weaver}, John R. and {Furtak}, Lukas J. and {Atek}, Hakim and {Coe}, Dan and {Cutler}, Sam E. and {Dayal}, Pratika and {van Dokkum}, Pieter and {Feldmann}, Robert and {F{\"o}rster Schreiber}, Natascha M. and {Fujimoto}, Seiji and {Geha}, Marla and {Glazebrook}, Karl and {de Graaff}, Anna and {Greene}, Jenny E. and {Juneau}, St{\'e}phanie and {Kassin}, Susan and {Kriek}, Mariska and {Khullar}, Gourav and {Maseda}, Michael and {Mowla}, Lamiya A. and {Muzzin}, Adam and {Nanayakkara}, Themiya and {Nelson}, Erica J. and {Oesch}, Pascal A. and {Pacifici}, Camilla and {Pan}, Richard and {Papovich}, Casey and {Setton}, David J. and {Shapley}, Alice E. and {Smit}, Renske and {Stefanon}, Mauro and {Taylor}, Edward N. and {Williams}, Christina C.},
        title = "{The JWST UNCOVER Treasury Survey: Ultradeep NIRSpec and NIRCam Observations before the Epoch of Reionization}",
      journal = {\apj},
         year = 2024,
        month = oct,
       volume = {974},
       number = {1},
          eid = {92},
        pages = {92},
          doi = {10.3847/1538-4357/ad66cf},
archivePrefix = {arXiv},
       eprint = {2212.04026},
 primaryClass = {astro-ph.GA},
       adsurl = {https://ui.adsabs.harvard.edu/abs/2024ApJ...974...92B}
}

@ARTICLE{Popesso2023,
       author = {{Popesso}, P. and {Concas}, A. and {Cresci}, G. and {Belli}, S. and {Rodighiero}, G. and {Inami}, H. and {Dickinson}, M. and {Ilbert}, O. and {Pannella}, M. and {Elbaz}, D.},
        title = "{The main sequence of star-forming galaxies across cosmic times}",
      journal = {\mnras},
         year = 2023,
        month = feb,
       volume = {519},
       number = {1},
        pages = {1526-1544},
          doi = {10.1093/mnras/stac3214},
archivePrefix = {arXiv},
       eprint = {2203.10487},
 primaryClass = {astro-ph.GA},
       adsurl = {https://ui.adsabs.harvard.edu/abs/2023MNRAS.519.1526P}
}

@ARTICLE{Rihtar2025,
       author = {{Rihtar{\v{s}}i{\v{c}}}, G. and {Brada{\v{c}}}, M. and {Desprez}, G. and {Harshan}, A. and {Noirot}, G. and {Estrada-Carpenter}, V. and {Martis}, N.~S. and {Abraham}, R.~G. and {Asada}, Y. and {Brammer}, G. and {Iyer}, K.~G. and {Matharu}, J. and {Mowla}, L. and {Muzzin}, A. and {Sarrouh}, G.~T.~E. and {Sawicki}, M. and {Strait}, V. and {Willott}, C.~J. and {Gledhill}, R. and {Markov}, V. and {Tripodi}, R.},
        title = "{CANUCS: Constraining the MACS J0416.1-2403 strong lensing model with JWST NIRISS, NIRSpec, and NIRCam}",
      journal = {\aap},
         year = 2025,
        month = apr,
       volume = {696},
          eid = {A15},
        pages = {A15},
          doi = {10.1051/0004-6361/202451117},
       adsurl = {https://ui.adsabs.harvard.edu/abs/2025A&A...696A..15R}
}

@ARTICLE{Shipley2018,
       author = {{Shipley}, Heath V. and {Lange-Vagle}, Daniel and {Marchesini}, Danilo and {Brammer}, Gabriel B. and {Ferrarese}, Laura and {Stefanon}, Mauro and {Kado-Fong}, Erin and {Whitaker}, Katherine E. and {Oesch}, Pascal A. and {Feinstein}, Adina D. and {Labb{\'e}}, Ivo and {Lundgren}, Britt and {Martis}, Nicholas and {Muzzin}, Adam and {Nedkova}, Kalina and {Skelton}, Rosalind and {van der Wel}, Arjen},
        title = "{HFF-DeepSpace Photometric Catalogs of the 12 Hubble Frontier Fields, Clusters, and Parallels: Photometry, Photometric Redshifts, and Stellar Masses}",
      journal = {\apjs},
         year = 2018,
        month = mar,
       volume = {235},
       number = {1},
          eid = {14},
        pages = {14},
          doi = {10.3847/1538-4365/aaacce},
archivePrefix = {arXiv},
       eprint = {1801.09734},
 primaryClass = {astro-ph.GA},
       adsurl = {https://ui.adsabs.harvard.edu/abs/2018ApJS..235...14S}
}

@ARTICLE{Pagul2021,
       author = {{Pagul}, A. and {S{\'a}nchez}, F.~J. and {Davidzon}, I. and {Mobasher}, Bahram},
        title = "{Hubble Frontier Field Clusters and Their Parallel Fields: Photometric and Photometric Redshift Catalogs}",
      journal = {\apjs},
         year = 2021,
        month = oct,
       volume = {256},
       number = {2},
          eid = {27},
        pages = {27},
          doi = {10.3847/1538-4365/abea9d},
archivePrefix = {arXiv},
       eprint = {2103.01952},
 primaryClass = {astro-ph.GA},
       adsurl = {https://ui.adsabs.harvard.edu/abs/2021ApJS..256...27P}
}

@ARTICLE{Barbani2025,
       author = {{Barbani}, Filippo and {Pascale}, Raffaele and {Marinacci}, Federico and {Torrey}, Paul and {Sales}, Laura V. and {Li}, Hui and {Vogelsberger}, Mark},
        title = "{Understanding the baryon cycle: Fueling star formation via inflows in Milky Way-like galaxies}",
      journal = {\aap},
         year = 2025,
        month = may,
       volume = {697},
          eid = {A121},
        pages = {A121},
          doi = {10.1051/0004-6361/202452608},
archivePrefix = {arXiv},
       eprint = {2504.01075},
 primaryClass = {astro-ph.GA},
       adsurl = {https://ui.adsabs.harvard.edu/abs/2025A&A...697A.121B}
}

@ARTICLE{Astropy2018,
       author = {{Astropy Collaboration} and {Price-Whelan}, A.~M. and {Sip{\H{o}}cz}, B.~M. and {G{\"u}nther}, H.~M. and {Lim}, P.~L. and {Crawford}, S.~M. and {Conseil}, S. and {Shupe}, D.~L. and {Craig}, M.~W. and {Dencheva}, N. and {Ginsburg}, A. and {VanderPlas}, J.~T. and {Bradley}, L.~D. and {P{\'e}rez-Su{\'a}rez}, D. and {de Val-Borro}, M. and {Aldcroft}, T.~L. and {Cruz}, K.~L. and {Robitaille}, T.~P. and {Tollerud}, E.~J. and {Ardelean}, C. and {Babej}, T. and {Bach}, Y.~P. and {Bachetti}, M. and {Bakanov}, A.~V. and {Bamford}, S.~P. and {Barentsen}, G. and {Barmby}, P. and {Baumbach}, A. and {Berry}, K.~L. and {Biscani}, F. and {Boquien}, M. and {Bostroem}, K.~A. and {Bouma}, L.~G. and {Brammer}, G.~B. and {Bray}, E.~M. and {Breytenbach}, H. and {Buddelmeijer}, H. and {Burke}, D.~J. and {Calderone}, G. and {Cano Rodr{\'\i}guez}, J.~L. and {Cara}, M. and {Cardoso}, J.~V.~M. and {Cheedella}, S. and {Copin}, Y. and {Corrales}, L. and {Crichton}, D. and {D'Avella}, D. and {Deil}, C. and {Depagne}, {\'E}. and {Dietrich}, J.~P. and {Donath}, A. and {Droettboom}, M. and {Earl}, N. and {Erben}, T. and {Fabbro}, S. and {Ferreira}, L.~A. and {Finethy}, T. and {Fox}, R.~T. and {Garrison}, L.~H. and {Gibbons}, S.~L.~J. and {Goldstein}, D.~A. and {Gommers}, R. and {Greco}, J.~P. and {Greenfield}, P. and {Groener}, A.~M. and {Grollier}, F. and {Hagen}, A. and {Hirst}, P. and {Homeier}, D. and {Horton}, A.~J. and {Hosseinzadeh}, G. and {Hu}, L. and {Hunkeler}, J.~S. and {Ivezi{\'c}}, {\v{Z}}. and {Jain}, A. and {Jenness}, T. and {Kanarek}, G. and {Kendrew}, S. and {Kern}, N.~S. and {Kerzendorf}, W.~E. and {Khvalko}, A. and {King}, J. and {Kirkby}, D. and {Kulkarni}, A.~M. and {Kumar}, A. and {Lee}, A. and {Lenz}, D. and {Littlefair}, S.~P. and {Ma}, Z. and {Macleod}, D.~M. and {Mastropietro}, M. and {McCully}, C. and {Montagnac}, S. and {Morris}, B.~M. and {Mueller}, M. and {Mumford}, S.~J. and {Muna}, D. and {Murphy}, N.~A. and {Nelson}, S. and {Nguyen}, G.~H. and {Ninan}, J.~P. and {N{\"o}the}, M. and {Ogaz}, S. and {Oh}, S. and {Parejko}, J.~K. and {Parley}, N. and {Pascual}, S. and {Patil}, R. and {Patil}, A.~A. and {Plunkett}, A.~L. and {Prochaska}, J.~X. and {Rastogi}, T. and {Reddy Janga}, V. and {Sabater}, J. and {Sakurikar}, P. and {Seifert}, M. and {Sherbert}, L.~E. and {Sherwood-Taylor}, H. and {Shih}, A.~Y. and {Sick}, J. and {Silbiger}, M.~T. and {Singanamalla}, S. and {Singer}, L.~P. and {Sladen}, P.~H. and {Sooley}, K.~A. and {Sornarajah}, S. and {Streicher}, O. and {Teuben}, P. and {Thomas}, S.~W. and {Tremblay}, G.~R. and {Turner}, J.~E.~H. and {Terr{\'o}n}, V. and {van Kerkwijk}, M.~H. and {de la Vega}, A. and {Watkins}, L.~L. and {Weaver}, B.~A. and {Whitmore}, J.~B. and {Woillez}, J. and {Zabalza}, V. and {Astropy Contributors}},
        title = "{The Astropy Project: Building an Open-science Project and Status of the v2.0 Core Package}",
      journal = {\aj},
         year = 2018,
        month = sep,
       volume = {156},
       number = {3},
          eid = {123},
        pages = {123},
          doi = {10.3847/1538-3881/aabc4f},
archivePrefix = {arXiv},
       eprint = {1801.02634},
 primaryClass = {astro-ph.IM},
       adsurl = {https://ui.adsabs.harvard.edu/abs/2018AJ....156..123A}
}

@ARTICLE{Astropy,
       author = {{Astropy Collaboration} and {Robitaille}, Thomas P. and {Tollerud}, Erik J. and {Greenfield}, Perry and {Droettboom}, Michael and {Bray}, Erik and {Aldcroft}, Tom and {Davis}, Matt and {Ginsburg}, Adam and {Price-Whelan}, Adrian M. and {Kerzendorf}, Wolfgang E. and {Conley}, Alexander and {Crighton}, Neil and {Barbary}, Kyle and {Muna}, Demitri and {Ferguson}, Henry and {Grollier}, Fr{\'e}d{\'e}ric and {Parikh}, Madhura M. and {Nair}, Prasanth H. and {Unther}, Hans M. and {Deil}, Christoph and {Woillez}, Julien and {Conseil}, Simon and {Kramer}, Roban and {Turner}, James E.~H. and {Singer}, Leo and {Fox}, Ryan and {Weaver}, Benjamin A. and {Zabalza}, Victor and {Edwards}, Zachary I. and {Azalee Bostroem}, K. and {Burke}, D.~J. and {Casey}, Andrew R. and {Crawford}, Steven M. and {Dencheva}, Nadia and {Ely}, Justin and {Jenness}, Tim and {Labrie}, Kathleen and {Lim}, Pey Lian and {Pierfederici}, Francesco and {Pontzen}, Andrew and {Ptak}, Andy and {Refsdal}, Brian and {Servillat}, Mathieu and {Streicher}, Ole},
        title = "{Astropy: A community Python package for astronomy}",
      journal = {\aap},
         year = 2013,
        month = oct,
       volume = {558},
          eid = {A33},
        pages = {A33},
          doi = {10.1051/0004-6361/201322068},
archivePrefix = {arXiv},
       eprint = {1307.6212},
 primaryClass = {astro-ph.IM},
       adsurl = {https://ui.adsabs.harvard.edu/abs/2013A&A...558A..33A}
}

@MISC{Prochaska2016,
       author = {{Prochaska}, J. Xavier and {Tejos}, Nicolas and {Crighton}, Neil and {jnburchett} and {Tuo-Ji} and {tiffanyhsyu} and {ktirimba} and {jhennawi} and {O'Meara}, John and {Werk}, Jessica},
        title = "{Linetools/Linetools: Second Major Release}",
 howpublished = {Zenodo},
         year = 2016,
        month = nov,
          eid = {10.5281/zenodo.168270},
          doi = {10.5281/zenodo.168270},
      version = {v0.2},
    publisher = {Zenodo},
       adsurl = {https://ui.adsabs.harvard.edu/abs/2016zndo....168270P}
}

@INPROCEEDINGS{PyMUSE,
       author = {{Pessa}, Ismael and {Tejos}, Nicolas and {Moya1}, Cristobal},
        title = "{PyMUSE: a Python Package for VLT/MUSE Data}",
    booktitle = {Astronomical Data Analysis Software and Systems XXVII},
         year = 2020,
       editor = {{Ballester}, Pascal and {Ibsen}, Jorge and {Solar}, Mauricio and {Shortridge}, Keith},
       series = {Astronomical Society of the Pacific Conference Series},
       volume = {522},
        month = apr,
        pages = {61},
       adsurl = {https://ui.adsabs.harvard.edu/abs/2020ASPC..522...61P}
}

@Article{Harris2020,
 title         = {Array programming with {NumPy}},
 author        = {Charles R. Harris and K. Jarrod Millman and St{\'{e}}fan J.
                 van der Walt and Ralf Gommers and Pauli Virtanen and David
                 Cournapeau and Eric Wieser and Julian Taylor and Sebastian
                 Berg and Nathaniel J. Smith and Robert Kern and Matti Picus
                 and Stephan Hoyer and Marten H. van Kerkwijk and Matthew
                 Brett and Allan Haldane and Jaime Fern{\'{a}}ndez del
                 R{\'{i}}o and Mark Wiebe and Pearu Peterson and Pierre
                 G{\'{e}}rard-Marchant and Kevin Sheppard and Tyler Reddy and
                 Warren Weckesser and Hameer Abbasi and Christoph Gohlke and
                 Travis E. Oliphant},
 year          = {2020},
 month         = sep,
 journal       = {Nature},
 volume        = {585},
 number        = {7825},
 pages         = {357--362},
 doi           = {10.1038/s41586-020-2649-2},
 publisher     = {Springer Science and Business Media {LLC}},
 url           = {https://doi.org/10.1038/s41586-020-2649-2}
}

@ARTICLE{Virtanen2020,
  author  = {Virtanen, Pauli and Gommers, Ralf and Oliphant, Travis E. and
            Haberland, Matt and Reddy, Tyler and Cournapeau, David and
            Burovski, Evgeni and Peterson, Pearu and Weckesser, Warren and
            Bright, Jonathan and {van der Walt}, St{\'e}fan J. and
            Brett, Matthew and Wilson, Joshua and Millman, K. Jarrod and
            Mayorov, Nikolay and Nelson, Andrew R. J. and Jones, Eric and
            Kern, Robert and Larson, Eric and Carey, C J and
            Polat, {\.I}lhan and Feng, Yu and Moore, Eric W. and
            {VanderPlas}, Jake and Laxalde, Denis and Perktold, Josef and
            Cimrman, Robert and Henriksen, Ian and Quintero, E. A. and
            Harris, Charles R. and Archibald, Anne M. and
            Ribeiro, Ant{\^o}nio H. and Pedregosa, Fabian and
            {van Mulbregt}, Paul and {SciPy 1.0 Contributors}},
  title   = {{{SciPy} 1.0: Fundamental Algorithms for Scientific
            Computing in Python}},
  journal = {Nature Methods},
  year    = {2020},
  volume  = {17},
  pages   = {261--272},
  adsurl  = {https://rdcu.be/b08Wh},
  doi     = {10.1038/s41592-019-0686-2},
}

@Article{Hunter2007,
  Author    = {Hunter, J. D.},
  Title     = {Matplotlib: A 2D graphics environment},
  Journal   = {Computing in Science \& Engineering},
  Volume    = {9},
  Number    = {3},
  Pages     = {90--95},
  publisher = {IEEE COMPUTER SOC},
  doi       = {10.1109/MCSE.2007.55},
  year      = 2007
}

@article{Zabl:2019ija,
author = {Zabl, Johannes and Bouch{\'e}, Nicolas F and Schroetter, Ilane and Wendt, Martin and Finley, Hayley and Schaye, Joop and Conseil, Simon and Contini, Thierry and Marino, Raffaella A and Mitchell, Peter and Muzahid, Sowgat and Pezzulli, Gabriele and Wisotzki, Lutz},
title = {{MusE GAs FLOw and Wind (MEGAFLOW) II. A study of gas accretion around z $\approx$ 1 star-forming galaxies with background quasars}},
journal = {Monthly Notices of the Royal Astronomical Society},
year = {2019},
volume = {485},
number = {2},
pages = {1961--1980},
month = may
}

@ARTICLE{Bruzual2003,
       author = {{Bruzual}, G. and {Charlot}, S.},
        title = "{Stellar population synthesis at the resolution of 2003}",
      journal = {\mnras},
         year = 2003,
        month = oct,
       volume = {344},
       number = {4},
        pages = {1000-1028},
          doi = {10.1046/j.1365-8711.2003.06897.x},
archivePrefix = {arXiv},
       eprint = {astro-ph/0309134},
 primaryClass = {astro-ph},
       adsurl = {https://ui.adsabs.harvard.edu/abs/2003MNRAS.344.1000B}
}

@ARTICLE{Chabrier2003,
       author = {{Chabrier}, Gilles},
        title = "{Galactic Stellar and Substellar Initial Mass Function}",
      journal = {\pasp},
         year = 2003,
        month = jul,
       volume = {115},
       number = {809},
        pages = {763-795},
          doi = {10.1086/376392},
archivePrefix = {arXiv},
       eprint = {astro-ph/0304382},
 primaryClass = {astro-ph},
       adsurl = {https://ui.adsabs.harvard.edu/abs/2003PASP..115..763C}
}

@ARTICLE{Dutta2020,
       author = {{Dutta}, Rajeshwari and {Fumagalli}, Michele and {Fossati}, Matteo and {Lofthouse}, Emma K. and {Prochaska}, J. Xavier and {Arrigoni Battaia}, Fabrizio and {Bielby}, Richard M. and {Cantalupo}, Sebastiano and {Cooke}, Ryan J. and {Murphy}, Michael T. and {O'Meara}, John M.},
        title = "{MUSE Analysis of Gas around Galaxies (MAGG) - II: metal-enriched halo gas around z {\ensuremath{\sim}} 1 galaxies}",
      journal = {\mnras},
         year = 2020,
        month = dec,
       volume = {499},
       number = {4},
        pages = {5022-5046},
          doi = {10.1093/mnras/staa3147},
archivePrefix = {arXiv},
       eprint = {2009.14219},
 primaryClass = {astro-ph.GA},
       adsurl = {https://ui.adsabs.harvard.edu/abs/2020MNRAS.499.5022D}
}

@ARTICLE{Dutta2023,
       author = {{Dutta}, Rajeshwari and {Fossati}, Matteo and {Fumagalli}, Michele and {Revalski}, Mitchell and {Lofthouse}, Emma K. and {Nelson}, Dylan and {Papini}, Giulia and {Rafelski}, Marc and {Cantalupo}, Sebastiano and {Arrigoni Battaia}, Fabrizio and {Dayal}, Pratika and {Longobardi}, Alessia and {P{\'e}roux}, Celine and {Prichard}, Laura J. and {Prochaska}, J. Xavier},
        title = "{Metal line emission from galaxy haloes at z {\ensuremath{\approx}} 1}",
      journal = {\mnras},
         year = 2023,
        month = jun,
       volume = {522},
       number = {1},
        pages = {535-558},
          doi = {10.1093/mnras/stad1002},
archivePrefix = {arXiv},
       eprint = {2302.09087},
 primaryClass = {astro-ph.GA},
       adsurl = {https://ui.adsabs.harvard.edu/abs/2023MNRAS.522..535D}
}

@ARTICLE{Dutta2026,
       author = {{Dutta}, Sayak and {Muzahid}, Sowgat and {Schaye}, Joop and {Johnson}, Sean and {Herenz}, Edmund Christian and {Pessa}, Ismael and {Augustin}, Ramona and {Bouch{\'e}}, Nicolas F. and {Braspenning}, Joey and {Cantalupo}, Sebastiano and {Das}, Sourav and {Wendt}, Martin},
        title = "{MUSEQuBES: Probing Anisotropies in Gas and Metal Distributions in the Circumgalactic Medium}",
      journal = {\apj},
         year = 2026,
        month = sep,
       volume = {1008},
       number = {1},
          eid = {93},
        pages = {93},
          doi = {10.3847/1538-4357/ae9544},
archivePrefix = {arXiv},
       eprint = {2602.17593},
 primaryClass = {astro-ph.GA},
       adsurl = {https://ui.adsabs.harvard.edu/abs/2026ApJ..1008...93D}
}

@ARTICLE{Veilleux2020,
       author = {{Veilleux}, Sylvain and {Maiolino}, Roberto and {Bolatto}, Alberto D. and {Aalto}, Susanne},
        title = "{Cool outflows in galaxies and their implications}",
      journal = {\aapr},
         year = 2020,
        month = apr,
       volume = {28},
       number = {1},
          eid = {2},
        pages = {2},
          doi = {10.1007/s00159-019-0121-9},
archivePrefix = {arXiv},
       eprint = {2002.07765},
 primaryClass = {astro-ph.GA},
       adsurl = {https://ui.adsabs.harvard.edu/abs/2020A&ARv..28....2V}
}

@ARTICLE{Pessa2023,
       author = {{Pessa}, I. and {Schinnerer}, E. and {Sanchez-Blazquez}, P. and {Belfiore}, F. and {Groves}, B. and {Emsellem}, E. and {Neumann}, J. and {Leroy}, A.~K. and {Bigiel}, F. and {Chevance}, M. and {Dale}, D.~A. and {Glover}, S.~C.~O. and {Grasha}, K. and {Klessen}, R.~S. and {Kreckel}, K. and {Kruijssen}, J.~M.~D. and {Pinna}, F. and {Querejeta}, M. and {Rosolowsky}, E. and {Williams}, T.~G.},
        title = "{Resolved stellar population properties of PHANGS-MUSE galaxies}",
      journal = {\aap},
         year = 2023,
        month = may,
       volume = {673},
          eid = {A147},
        pages = {A147},
          doi = {10.1051/0004-6361/202245673},
archivePrefix = {arXiv},
       eprint = {2303.13676},
 primaryClass = {astro-ph.GA},
       adsurl = {https://ui.adsabs.harvard.edu/abs/2023A&A...673A.147P}
}

@INPROCEEDINGS{Weilbacher2014,
       author = {{Weilbacher}, P.~M. and {Streicher}, O. and {Urrutia}, T. and {P{\'e}contal-Rousset}, A. and {Jarno}, A. and {Bacon}, R.},
        title = "{The MUSE Data Reduction Pipeline: Status after Preliminary Acceptance Europe}",
    booktitle = {Astronomical Data Analysis Software and Systems XXIII},
         year = 2014,
       editor = {{Manset}, N. and {Forshay}, P.},
       series = {Astronomical Society of the Pacific Conference Series},
       volume = {485},
        month = may,
        pages = {451},
          doi = {10.48550/arXiv.1507.00034},
archivePrefix = {arXiv},
       eprint = {1507.00034},
 primaryClass = {astro-ph.IM},
       adsurl = {https://ui.adsabs.harvard.edu/abs/2014ASPC..485..451W}
}

@ARTICLE{Urrutia2019,
       author = {{Urrutia}, T. and {Wisotzki}, L. and {Kerutt}, J. and {Schmidt}, K.~B. and {Herenz}, E.~C. and {Klar}, J. and {Saust}, R. and {Werhahn}, M. and {Diener}, C. and {Caruana}, J. and {Krajnovi{\'c}}, D. and {Bacon}, R. and {Boogaard}, L. and {Brinchmann}, J. and {Enke}, H. and {Maseda}, M. and {Nanayakkara}, T. and {Richard}, J. and {Steinmetz}, M. and {Weilbacher}, P.~M.},
        title = "{The MUSE-Wide Survey: survey description and first data release}",
      journal = {\aap},
         year = 2019,
        month = apr,
       volume = {624},
          eid = {A141},
        pages = {A141},
          doi = {10.1051/0004-6361/201834656},
archivePrefix = {arXiv},
       eprint = {1811.06549},
 primaryClass = {astro-ph.GA},
       adsurl = {https://ui.adsabs.harvard.edu/abs/2019A&A...624A.141U}
}

@ARTICLE{Weilbacher2020,
       author = {{Weilbacher}, Peter M. and {Palsa}, Ralf and {Streicher}, Ole and {Bacon}, Roland and {Urrutia}, Tanya and {Wisotzki}, Lutz and {Conseil}, Simon and {Husemann}, Bernd and {Jarno}, Aur{\'e}lien and {Kelz}, Andreas and {P{\'e}contal-Rousset}, Arlette and {Richard}, Johan and {Roth}, Martin M. and {Selman}, Fernando and {Vernet}, Jo{\"e}l},
        title = "{The data processing pipeline for the MUSE instrument}",
      journal = {\aap},
         year = 2020,
        month = sep,
       volume = {641},
          eid = {A28},
        pages = {A28},
          doi = {10.1051/0004-6361/202037855},
archivePrefix = {arXiv},
       eprint = {2006.08638},
 primaryClass = {astro-ph.IM},
       adsurl = {https://ui.adsabs.harvard.edu/abs/2020A&A...641A..28W}
}

@ARTICLE{Werk2013,
       author = {{Werk}, Jessica K. and {Prochaska}, J. Xavier and {Thom}, Christopher and {Tumlinson}, Jason and {Tripp}, Todd M. and {O'Meara}, John M. and {Peeples}, Molly S.},
        title = "{The COS-Halos Survey: An Empirical Description of Metal-line Absorption in the Low-redshift Circumgalactic Medium}",
      journal = {\apjs},
         year = 2013,
        month = feb,
       volume = {204},
       number = {2},
          eid = {17},
        pages = {17},
          doi = {10.1088/0067-0049/204/2/17},
archivePrefix = {arXiv},
       eprint = {1212.0558},
 primaryClass = {astro-ph.CO},
       adsurl = {https://ui.adsabs.harvard.edu/abs/2013ApJS..204...17W}
}

@article{Feltre:2018in,
author = {Feltre, Anna and Bacon, Roland and Tresse, Laurence and Finley, Hayley and Carton, David and Blaizot, J{\'e}r{\'e}my and Bouch{\'e}, Nicolas F and Garel, Thibault and Inami, Hanae and Boogaard, Leindert A and Brinchmann, Jarle and Charlot, St{\'e}phane and Chevallard, Jacopo and Contini, Thierry and Michel-Dansac, Leo and Mahler, Guillaume and Marino, Raffaella A and Maseda, Michael V and Richard, Johan and Schmidt, Kasper B and Verhamme, Anne},
title = {{The MUSE Hubble Ultra Deep Field Survey. XII. Mg II emission and absorption in star-forming galaxies}},
journal = {Astronomy {\&} Astrophysics},
year = {2018},
volume = {617},
pages = {A62},
month = sep
}

@ARTICLE{Schroetter2024,
       author = {{Schroetter}, Ilane and {Bouch{\'e}}, Nicolas F. and {Zabl}, Johannes and {Wendt}, Martin and {Cherrey}, Maxime and {Langan}, Ivanna and {Schaye}, Joop and {Contini}, Thierry},
        title = "{MusE GAs FLOw and Wind (MEGAFLOW). XI. Scaling relations between outflows and host galaxy properties}",
      journal = {\aap},
         year = 2024,
        month = jul,
       volume = {687},
          eid = {A39},
        pages = {A39},
          doi = {10.1051/0004-6361/202348725},
archivePrefix = {arXiv},
       eprint = {2404.03300},
 primaryClass = {astro-ph.GA},
       adsurl = {https://ui.adsabs.harvard.edu/abs/2024A&A...687A..39S}
}

@ARTICLE{Bouche2025,
       author = {{Bouch{\'e}}, Nicolas F. and {Wendt}, Martin and {Zabl}, Johannes and {Cherrey}, Maxime and {Schroetter}, Ilane and {Langan}, Ivanna and {Muzahid}, Sowgat and {Schaye}, Joop and {Epinat}, Beno{\^\i}t and {Wisotzki}, Lutz and {Contini}, Thierry and {Richard}, Johan and {Bacon}, Roland and {Weilbacher}, Peter M.},
        title = "{MusE GAs FLOw and Wind (MEGAFLOW): XII. Rationale and design of a Mg II survey of the cool circum-galactic medium with MUSE and UVES: The MEGAFLOW Survey}",
      journal = {\aap},
         year = 2025,
        month = feb,
       volume = {694},
          eid = {A67},
        pages = {A67},
          doi = {10.1051/0004-6361/202451093},
archivePrefix = {arXiv},
       eprint = {2411.07014},
 primaryClass = {astro-ph.GA},
       adsurl = {https://ui.adsabs.harvard.edu/abs/2025A&A...694A..67B}
}

@ARTICLE{Harvey2025,
       author = {{Harvey}, Zoe and {Krishna}, Sahyadri and {Wild}, Vivienne and {Tojeiro}, Rita and {Hewett}, Paul},
        title = "{Cool Gas in the Circumgalactic Medium of Massive Post Starburst Galaxies}",
      journal = {The Open Journal of Astrophysics},
         year = 2025,
        month = nov,
       volume = {8},
        pages = {47836},
          doi = {10.33232/001c.147836},
archivePrefix = {arXiv},
       eprint = {2506.22287},
 primaryClass = {astro-ph.GA},
       adsurl = {https://ui.adsabs.harvard.edu/abs/2025OJAp....847836H}
}

@ARTICLE{Weng2023,
       author = {{Weng}, Simon and {P{\'e}roux}, C{\'e}line and {Ramesh}, Rahul and {Nelson}, Dylan and {Sadler}, Elaine M. and {Zwaan}, Martin and {Bollo}, Victoria and {Casavecchia}, Benedetta},
        title = "{The physical origins of gas in the circumgalactic medium using observationally motivated TNG50 mocks}",
      journal = {\mnras},
         year = 2024,
        month = jan,
       volume = {527},
       number = {2},
        pages = {3494-3516},
          doi = {10.1093/mnras/stad3426},
archivePrefix = {arXiv},
       eprint = {2310.18310},
 primaryClass = {astro-ph.GA},
       adsurl = {https://ui.adsabs.harvard.edu/abs/2024MNRAS.527.3494W}
}

@article{Herenz:2017er,
author = {Herenz, Edmund Christian and Wisotzki, Lutz},
title = {{LSDCat: Detection and cataloguing of emission-line sources in integral-field spectroscopy datacubes}},
journal = {Astronomy {\&} Astrophysics},
year = {2017},
volume = {602},
pages = {A111},
month = jun
}

@article{Vazdekis:2016ho,
author = {Vazdekis, A and Koleva, M and Ricciardelli, E and R{\"o}ck, B and Falc{\'o}n-Barroso, J},
title = {{UV-extended E-MILES stellar population models: young components in massive early-type galaxies}},
journal = {Monthly Notice of the Royal Astronomical Society},
year = {2016},
volume = {463},
number = {4},
pages = {3409--3436},
month = sep
}

@ARTICLE{Castor1979,
       author = {{Castor}, J.~I. and {Lamers}, H.~J.~G.~L.~M.},
        title = "{An atlas of theoretical P Cygni profiles.}",
      journal = {\apjs},
         year = 1979,
        month = apr,
       volume = {39},
        pages = {481-511},
          doi = {10.1086/190583},
       adsurl = {https://ui.adsabs.harvard.edu/abs/1979ApJS...39..481C}
}

@ARTICLE{Choi2016,
       author = {{Choi}, Jieun and {Dotter}, Aaron and {Conroy}, Charlie and {Cantiello}, Matteo and {Paxton}, Bill and {Johnson}, Benjamin D.},
        title = "{Mesa Isochrones and Stellar Tracks (MIST). I. Solar-scaled Models}",
      journal = {\apj},
         year = 2016,
        month = jun,
       volume = {823},
       number = {2},
          eid = {102},
        pages = {102},
          doi = {10.3847/0004-637X/823/2/102},
archivePrefix = {arXiv},
       eprint = {1604.08592},
 primaryClass = {astro-ph.SR},
       adsurl = {https://ui.adsabs.harvard.edu/abs/2016ApJ...823..102C}
}

@ARTICLE{Vazdekis2015,
       author = {{Vazdekis}, A. and {Coelho}, P. and {Cassisi}, S. and {Ricciardelli}, E. and {Falc{\'o}n-Barroso}, J. and {S{\'a}nchez-Bl{\'a}zquez}, P. and {La Barbera}, F. and {Beasley}, M.~A. and {Pietrinferni}, A.},
        title = "{Evolutionary stellar population synthesis with MILES - II. Scaled-solar and {\ensuremath{\alpha}}-enhanced models}",
      journal = {\mnras},
         year = 2015,
        month = may,
       volume = {449},
       number = {2},
        pages = {1177-1214},
          doi = {10.1093/mnras/stv151},
archivePrefix = {arXiv},
       eprint = {1504.08032},
 primaryClass = {astro-ph.GA},
       adsurl = {https://ui.adsabs.harvard.edu/abs/2015MNRAS.449.1177V}
}

@ARTICLE{Salpeter1955,
       author = {{Salpeter}, Edwin E.},
        title = "{The Luminosity Function and Stellar Evolution.}",
      journal = {\apj},
         year = 1955,
        month = jan,
       volume = {121},
        pages = {161},
          doi = {10.1086/145971},
       adsurl = {https://ui.adsabs.harvard.edu/abs/1955ApJ...121..161S}
}

@ARTICLE{Lastennet2002,
       author = {{Lastennet}, E. and {Lejeune}, T. and {Oblak}, E. and {Westera}, P. and {Buser}, R.},
        title = "{BaSeL: A Library of Synthetic Spectra and Colours for GAIA}",
      journal = {\apss},
         year = 2002,
        month = apr,
       volume = {280},
        pages = {83-87},
          doi = {10.1023/A:1015552407031},
archivePrefix = {arXiv},
       eprint = {astro-ph/0112461},
 primaryClass = {astro-ph},
       adsurl = {https://ui.adsabs.harvard.edu/abs/2002Ap&SS.280...83L}
}

@ARTICLE{Conroy2009,
       author = {{Conroy}, Charlie and {Gunn}, James E. and {White}, Martin},
        title = "{The Propagation of Uncertainties in Stellar Population Synthesis Modeling. I. The Relevance of Uncertain Aspects of Stellar Evolution and the Initial Mass Function to the Derived Physical Properties of Galaxies}",
      journal = {\apj},
         year = 2009,
        month = jul,
       volume = {699},
       number = {1},
        pages = {486-506},
          doi = {10.1088/0004-637X/699/1/486},
archivePrefix = {arXiv},
       eprint = {0809.4261},
 primaryClass = {astro-ph},
       adsurl = {https://ui.adsabs.harvard.edu/abs/2009ApJ...699..486C}
}

@article{ForemanMackey:2013io,
author = {Foreman-Mackey, Daniel and Hogg, David W. and Lang, Dustin and Goodman, Jonathan},
title = {{emcee: The MCMC Hammer}},
journal = {Publications of the Astronomical Society of the Pacific},
year = {2013},
volume = {125},
number = {925},
pages = {306--312},
month = mar
}

@article{Martin:2012dx,
author = {Martin, Crystal L and Shapley, Alice E and Coil, Alison L and Kornei, Katherine A and Bundy, Kevin and Weiner, Benjamin J. and Noeske, Kai G and Schiminovich, David},
title = {{Demographics and Physical Properties of Gas Outflows/Inflows at 0.4 < z < 1.4}},
journal = {The Astrophysical Journal},
year = {2012},
volume = {760},
number = {2},
pages = {127},
month = dec
}

@article{Prochaska:2011eqa,
author = {Prochaska, J Xavier and Kasen, Daniel and Rubin, Kate},
title = {{Simple Models of Metal-line Absorption and Emission from Cool Gas Outflows}},
journal = {The Astrophysical Journal},
year = {2011},
volume = {734},
number = {1},
pages = {24},
month = jun
}

@article{Rubin:2011bm,
author = {Rubin, Kate H R and Prochaska, J Xavier and M{\'e}nard, Brice and Murray, Norman and Kasen, Daniel and Koo, David C and Phillips, Andrew C},
title = {{Low-ionization Line Emission from a Starburst Galaxy: A New Probe of a Galactic-scale Outflow}},
journal = {The Astrophysical Journal},
year = {2011},
volume = {728},
number = {1},
pages = {55},
month = feb
}

@inproceedings{Bacon:2010jn,
author = {Bacon, R and Accardo, M and Adjali, L and Anwand, H and Bauer, S and Biswas, I and Blaizot, J and Boudon, D and Brau-Nogue, S and Brinchmann, J and Caillier, P and Capoani, L and Carollo, C M and Contini, T and Couderc, P and Daguis{\'e}, E and Deiries, S and Delabre, B and Dreizler, S and Dubois, J and Dupieux, M and Dupuy, C and Emsellem, E and Fechner, T and Fleischmann, A and Fran{\c c}ois, M and Gallou, G and Gharsa, T and Glindemann, A and Gojak, D and Guiderdoni, B and Hansali, G and Hahn, T and Jarno, A and Kelz, A and Koehler, C and Kosmalski, J and Laurent, F and Le Floch, M and Lilly, S J and Lizon, J L and Loupias, M and Manescau, A and Monstein, C and Nicklas, H and Olaya, J C and Pares, L and Pasquini, L and P{\'e}contal-Rousset, A and Pell{\`o}, R and Petit, C and Popow, E and Reiss, R and Remillieux, A and Renault, E and Roth, M and Rupprecht, G and Serre, D and Schaye, J and Soucail, G and Steinmetz, M and Streicher, O and Stuik, R and Valentin, H and Vernet, J and Weilbacher, P and Wisotzki, L. and Yerle, N},
title = {{The MUSE second-generation VLT instrument}},
booktitle = {Astronomical Telescopes and Instrumentation},
year = {2010},
editor = {McLean, Ian S and Ramsay, Suzanne K and Takami, Hideki},
pages = {773508--773508--9},
publisher = {SPIE},
month = jul
}

@article{Kriek:2009cs,
author = {Kriek, Mariska and van Dokkum, Pieter G and Labb{\'e}, Ivo and Franx, Marijn and Illingworth, Garth D and Marchesini, Danilo and Quadri, Ryan F},
title = {{An Ultra-Deep Near-Infrared Spectrum of a Compact Quiescent Galaxy at z = 2.2}},
journal = {The Astrophysical Journal},
year = {2009},
volume = {700},
number = {1},
pages = {221--231},
month = jul
}

@ARTICLE{Cappellari2017,
       author = {{Cappellari}, Michele},
        title = "{Improving the full spectrum fitting method: accurate convolution with Gauss-Hermite functions}",
      journal = {\mnras},
         year = 2017,
        month = apr,
       volume = {466},
       number = {1},
        pages = {798-811},
          doi = {10.1093/mnras/stw3020},
archivePrefix = {arXiv},
       eprint = {1607.08538},
 primaryClass = {astro-ph.GA},
       adsurl = {https://ui.adsabs.harvard.edu/abs/2017MNRAS.466..798C}
}

@article{Cappellari:2004gm,
author = {Cappellari, Michele and Emsellem, Eric},
title = {{Parametric Recovery of Line-of-Sight Velocity Distributions from Absorption-Line Spectra of Galaxies via Penalized Likelihood}},
journal = {Publications of the Astronomical Society of the Pacific},
year = {2004},
volume = {116},
number = {816},
pages = {138--147},
month = feb
}

@article{Calzetti:2000iy,
author = {Calzetti, Daniela and Armus, Lee and Bohlin, Ralph C and Kinney, Anne L and Koornneef, Jan and Storchi-Bergmann, Thaisa},
title = {{The Dust Content and Opacity of Actively Star-forming Galaxies*}},
journal = {The Astrophysical Journal},
year = {2000},
volume = {533},
number = {2},
pages = {682--695},
month = apr
}

\begin{appendix}

\section{Galactic wind kinematics versus specific star formation rates}
\begin{figure}
\centering
    \begin{minipage}{0.45\textwidth}
        \centering
        \includegraphics[width=0.85\columnwidth]{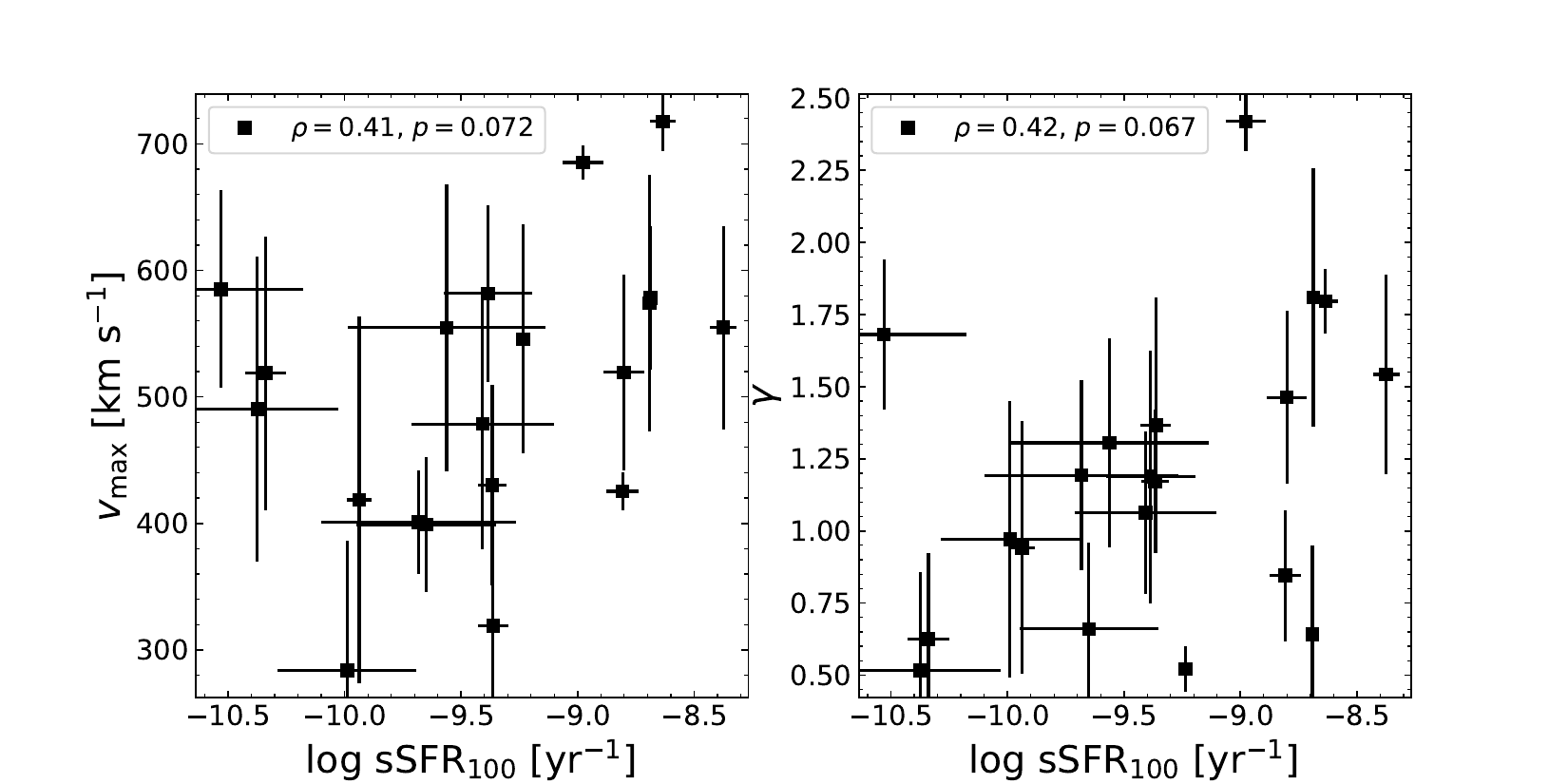}
        \includegraphics[width=0.85\columnwidth]{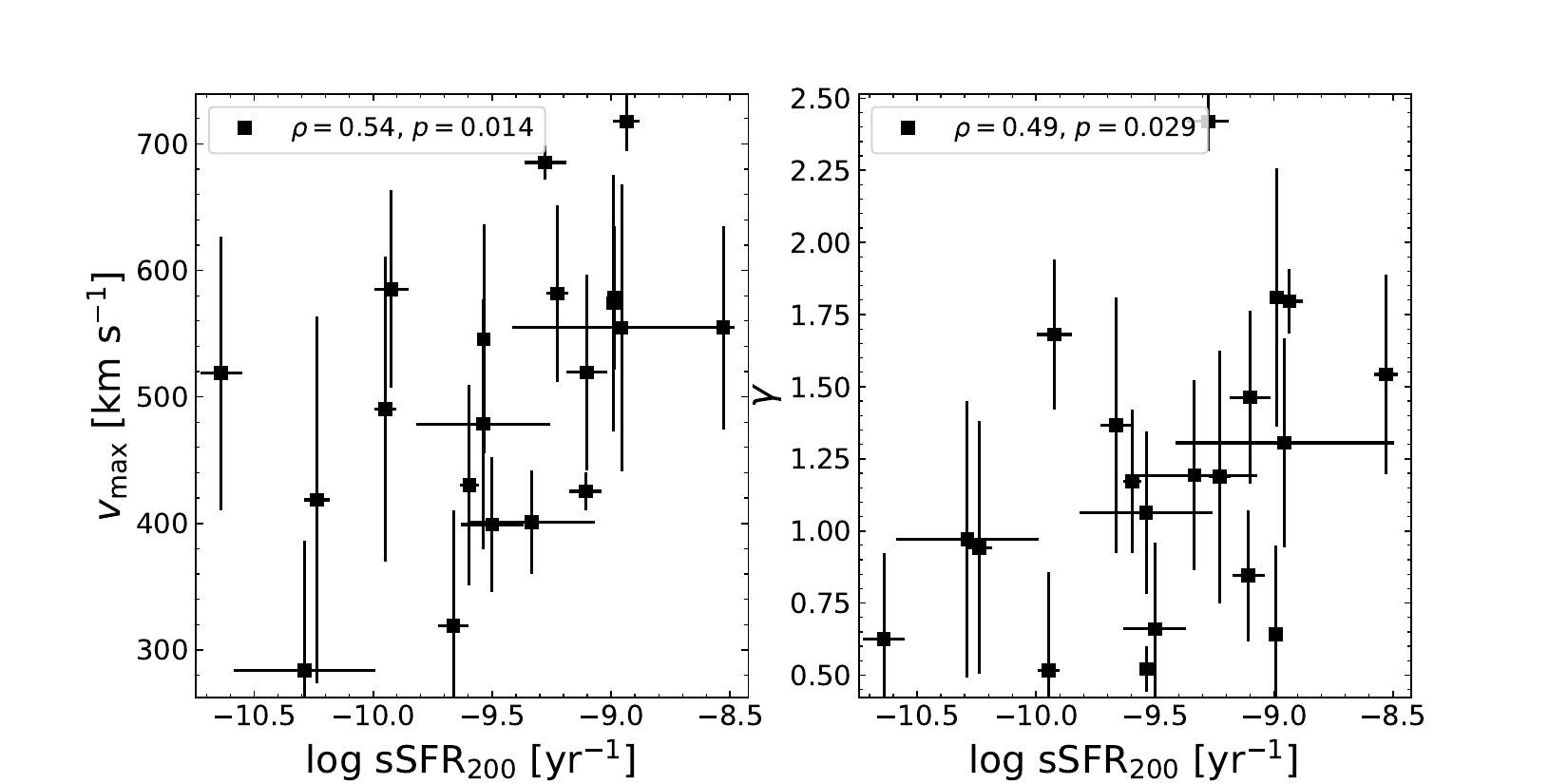}
        \includegraphics[width=0.85\columnwidth]{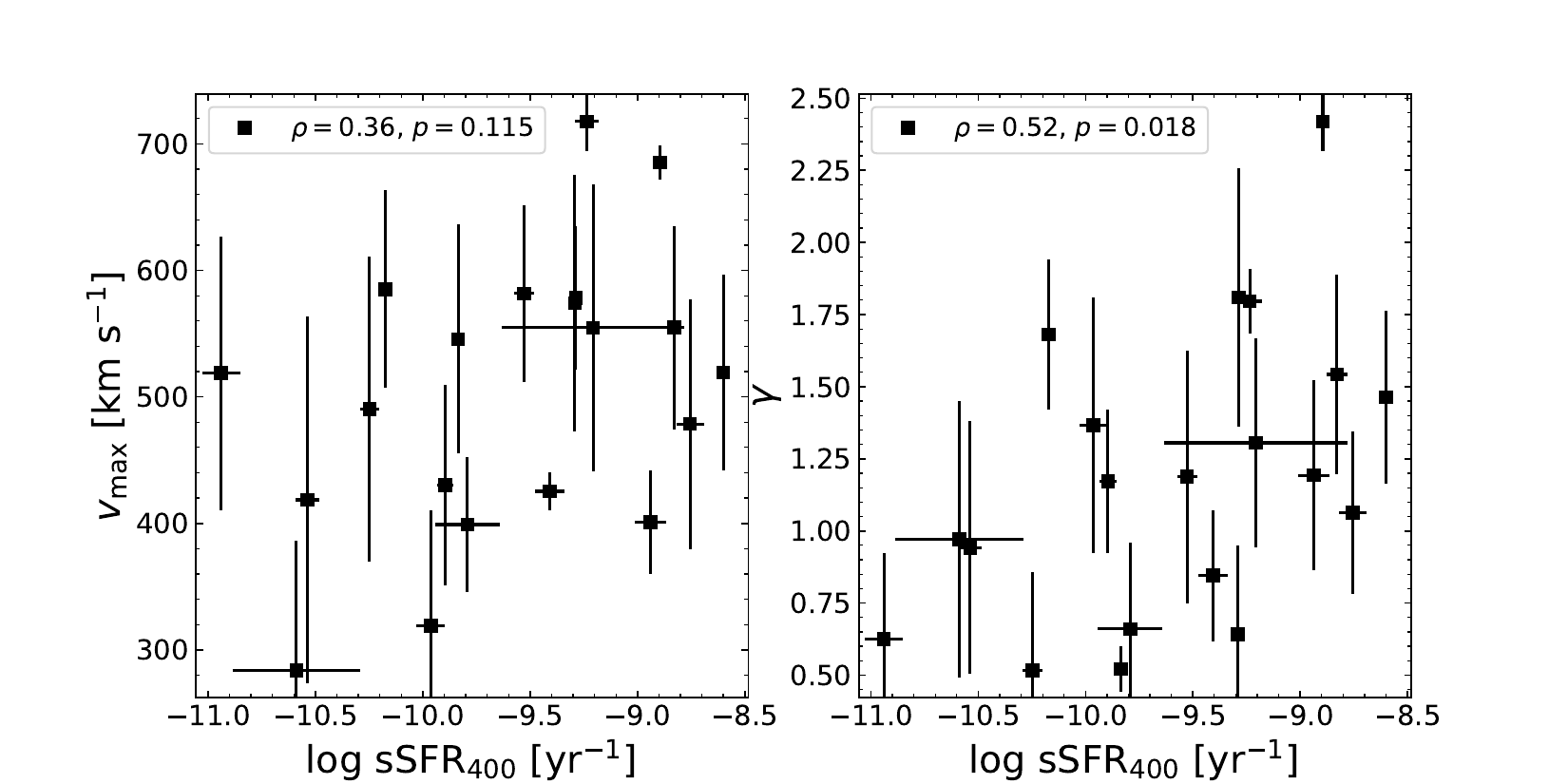}
    \end{minipage}

\caption{Same as the left column of Fig.~\ref{fig:SFR_SFH_vs_wind}, but for the sSFR of our sample galaxies. The panels show the maximum outflow velocity ($v_{\rm max}$) and acceleration rate ($\gamma$) for our sample galaxies, as a function of the sSFR computed on timescales of 100 (top), 200 (middle), and 400 (bottom) Myr from the non-parametric SFHs as described in Sec.~\ref{sec:SFR_measurements}. The Spearman rank correlation coefficient and $p$-value of each correlation are given in the corner of each panel.}
\label{fig:sSFR_SFH_vs_wind}
\end{figure}

\FloatBarrier
In this appendix, we show the correlations between the outflow maximum velocity ($v_{\rm max}$) and acceleration rate ($\gamma$) and the sSFR of our sample galaxies. The correlations are shown in Fig.~\ref{fig:sSFR_SFH_vs_wind}, and are similar to those computed for the SFR, shown in the left column of Fig.~\ref{fig:SFR_SFH_vs_wind}, in terms of their correlation coefficient and $p$-values, although somewhat stronger for $\gamma$ and somewhat weaker for $v_{\rm max}$.
\label{app:wind_vs_sSFR}

\section{Star formation histories versus the outflow launching times of our sample galaxies}
\label{app:SFH_and_launching_times}
In this appendix, we describe how we compute the SFR in timescales designed to capture the star formation episodes that triggered the galactic outflows, and briefly discuss the only galaxy in our sample for which we do not identify a clear star formation episode in this timescale, that is, we do not find a peak in the SFH in the time window when we find likely that the outflow was launched, according to our inferred wind kinematics. This time window is defined as a time interval around the time it took for the outflow to reach its HLR. The borders of the time window are defined by the time it took the outflow to reach half of the HLR, and $1.5$ HLR. 

\begin{figure}
\centering
        \includegraphics[width=0.7\columnwidth]{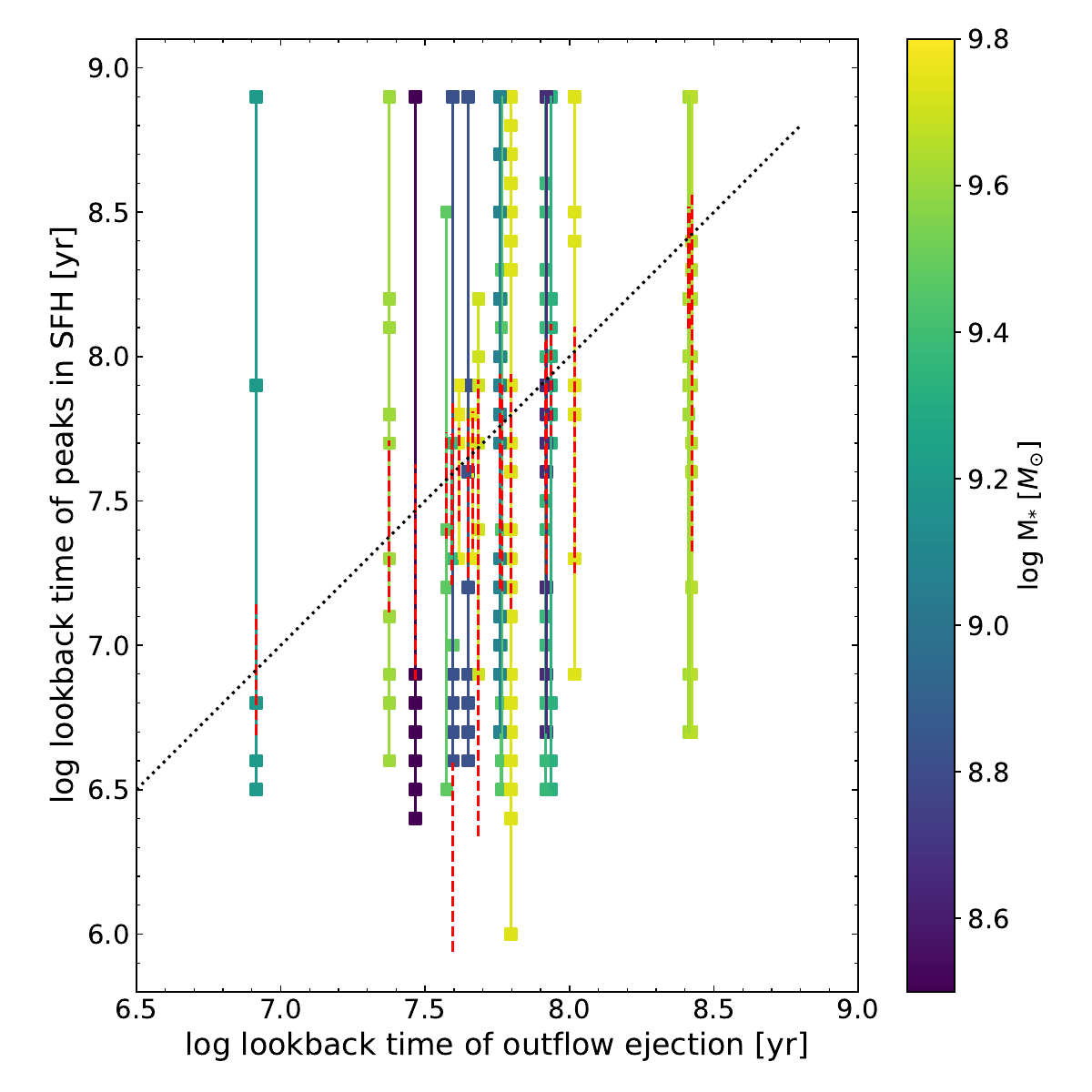}

\caption{Recent SFH of our sample galaxies (last $\sim1$ Gyr) as a function of the inferred outflow launch lookback time. Each vertical line represents a galaxy, and each square represents a star formation episode in its SFH. The color of the line scales with the galaxy's stellar mass, as indicated with the colorbar. The black diagonal line marks the identity, that is, a square coinciding with the black line would indicate a star formation episode that occurred a time ago that corresponds to the outflow launch lookback time. The vertical red dashed lines indicate the time interval in the SFH, where star-forming episodes could be causally connected to the extended \ion{Mg}{II} emission, given the measured traveling times and extent of the \ion{Mg}{II} halos. Thus, star formation episodes that fall inside the red lines are considered when calculating the SFR values described in Sec.~\ref{sec:SFR_halo_size}.}
\label{fig:launch_time_vs_SFH}
\end{figure}

\begin{figure}
\centering
        \includegraphics[width=0.7\columnwidth]{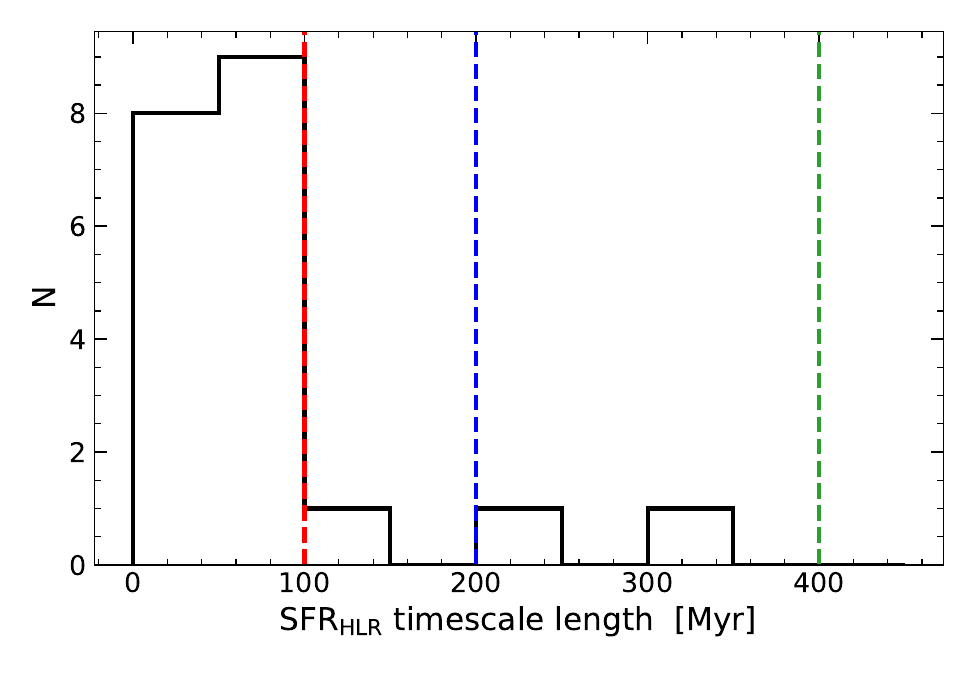}

\caption{Histogram of the time-window lengths used to compute the SFR$_{\rm HLR}$ values for the galaxies in our sample (see Sec.~\ref{sec:SFR_halo_size}). The three vertical lines indicate, for reference, the 100, 200, and 400 Myr timescales adopted for the SFH-based SFR measurements described in Sec.~\ref{sec:SFR_measurements}.}
\label{fig:HLR_timescale_duration_dist}
\end{figure}

Figure~\ref{fig:launch_time_vs_SFH} shows the peaks in the recent SFH (last one gigayear) of each sample galaxy in the vertical axis, versus their outflow ejection lookback time (i.e., traveling time of the outflows to the HLR), and the time window defined as described above (vertical red dashed lines), based on the HLR of each \ion{Mg}{II} halo. It shows that for most galaxies (all except one), there is at least one peak in the SFH that overlaps with the timescale defined by the size and velocity of the \ion{Mg}{II} halo. That is, there is at least one peak in the recovered SFH that occurs at a lookback time consistent with the inferred launching time of the galactic wind, for all except one galaxy in our sample. For an easier comparison with the fixed timescales used throughout the paper, Fig.~\ref{fig:HLR_timescale_duration_dist} shows the distribution of the time-window lengths derived from the \ion{Mg}{II} halo HLRs of our sample galaxies. The median and mean of the distribution are $\sim65$ and $80$ Myr, respectively. Although these values are relatively close to the 100 Myr timescale, there is an important difference between the two definitions: the HLR-based time window does not necessarily start at the present time (i.e., at a lookback time of zero), as is the case for SFR$_{100}$. Instead, both the beginning and end of the HLR-based window are determined from the inferred outflow travel time (see Sec.~\ref{sec:SFR_halo_size} for details).

We have further examined the galaxy for which we do not find any SF episode at the timescales that match the size and kinematics of the \ion{Mg}{II} halo (M0416-DF-024), and we do not find a priori anything abnormal with the full spectral fitting or with the outflow modeling. The estimated launching time window for this galaxy ranges between $20$ and $70$ Myr ago, and its recent SFH shows strong peaks only earlier than 10 Myr ago. This implies that, for this galaxy, the outflow must necessarily have accelerated faster than what is predicted by our fiducial modeling, where we employ a radially accelerating power-law to model the velocity field. 

While using a monotonically (radially) increasing velocity field is a basic assumption required for the implementation of the Sobolev approximation, such as the simple power law used \citep[see extensive discussion about our model limitations in][]{Pessa2024}, an initially faster acceleration, followed by a slower acceleration, and eventually reaching a deceleration regime would be a more physically-driven model, consistent with momentum-driven outflows \citep{Heckman2015}. Thus, while implementing a more realistic velocity field in our model is outside the scope of this paper, such a disagreement between the predicted launching timescales and the computed SFHs can be used to put additional constraints on the kinematics of the galactic wind, and test the use of more realistic prescriptions for the velocity field. In the meantime, we keep this galaxy outside of the analysis carried out in this section. In future work, we plan on implementing a more flexible model (through the use of more flexible velocity prescriptions, or via employing radiative transfer simulations), with particular emphasis on M0416-DF-024 \citep[and other galaxies where the simple outflow model employed seems to lack the complexity required to properly model the data, see section 6.4 of][]{Pessa2026}.

\section{Comparison of SFR values versus using eMILES templates}
\label{sec:comparison_templates}
In this appendix, we assess how the SFRs obtained by integrating the derived SFHs of our sample galaxies over different timescales change when adopting a different set of SSP models. In particular, we compare our fiducial measurements against those obtained using the eMILES \citep{Vazdekis2015,Vazdekis:2016ho} templates.

For the eMILES templates, we assumed a Salpeter IMF \citep{Salpeter1955}, Padova isochrones \citep{Bressan2012}, ages between 63 Myr and the age of the Universe at the redshift of each galaxy, logarithmically spaced in steps of 0.1 dex, and six metallicity bins of [$M/H$] = [-1.71, -1.31, -0.71, -0.40, 0.00, 0.22].

\begin{figure}
\centering
        \includegraphics[width=0.4\columnwidth]{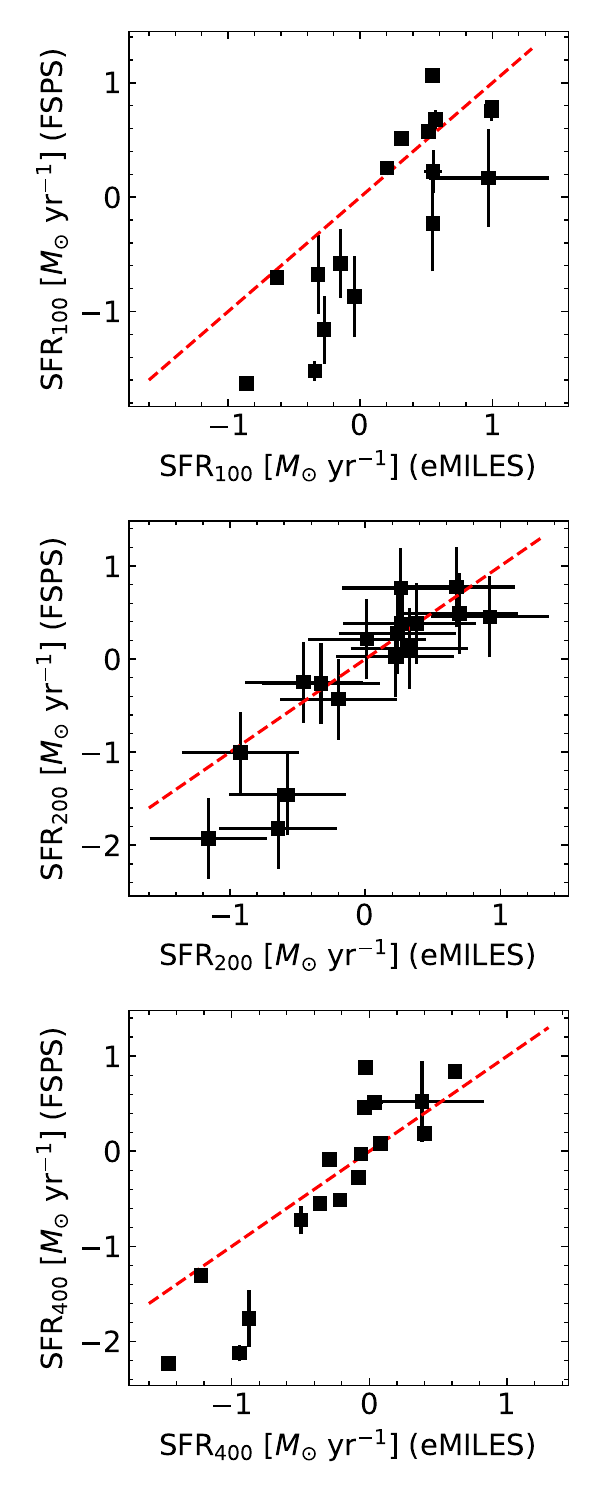}

\caption{Comparison of the inferred SFR values, assuming different sets of SSP models. The $y$-axis shows our fiducial values, computed using models from the FSPS library. The $x$-axis shows the corresponding values for the same galaxies, using the eMILES templates. The red dashed line shows the identity, for reference.}
\label{fig:comparison_models}
\end{figure}

Figure~\ref{fig:comparison_models} compares the SFRs obtained using the two sets of templates for the same galaxies. For most objects, the measurements are very similar and lie close to the identity line for all three averaging timescales. The largest differences are found for SFR$_{100}$, which is expected because it probes the shortest timescale and is therefore more sensitive to uncertainties in the recovered SFHs. Moreover, the galaxies that lie farthest from the identity line generally exhibit larger error bars, indicating that the larger differences occur mostly for those systems whose recovered SFHs are subject to greater uncertainties, as reflected in the estimated error bars. Nevertheless, the two measurements remain strongly correlated. The median difference between the two estimates is $0.4$, $0.02$ and $0.01$ dex for SFR$_{100}$, SFR$_{200}$, and SFR$_{400}$, respectively. Overall, we find that although the detailed shape of the recovered SFHs is known to depend on the assumptions underlying the fitting procedure, including the adopted SSP models \citep[e.g.,][]{Ge2019}, the time-averaged SFRs employed throughout this work are comparatively robust. While small systematic differences are present, they do not significantly affect the inferred SFRs on the timescales considered here.

\FloatBarrier
\section{Spectral and photometric fits and recovered SFHs for the full sample}

\label{sec:all_fits_appendix}

In this section, we present the spectral and photometric fits for all galaxies in our sample, together with their recovered SFHs. Figure~\ref{fig:all_SFH_fits} shows, for each galaxy, the observed spectrum and, where available, the photometric data together with their best-fitting models, as well as the age--metallicity grid of the SSP models used to reproduce the observations and the weights assigned to each model in the best-fitting linear combination of SSP spectra, following the same format as Fig.~\ref{fig:fit_example}. We note that the weights shown for each galaxy correspond to those obtained from the fit to the observed data, rather than the mean weights obtained from the MC iterations performed for each spectrum (see Sec.~\ref{sec:SFH_det} for details).

\begin{figure*}[p]
    \centering

    \begin{subfigure}[t]{0.49\textwidth}
        \centering
        M0416-DF-005
        \par\smallskip
        \includegraphics[width=\linewidth]{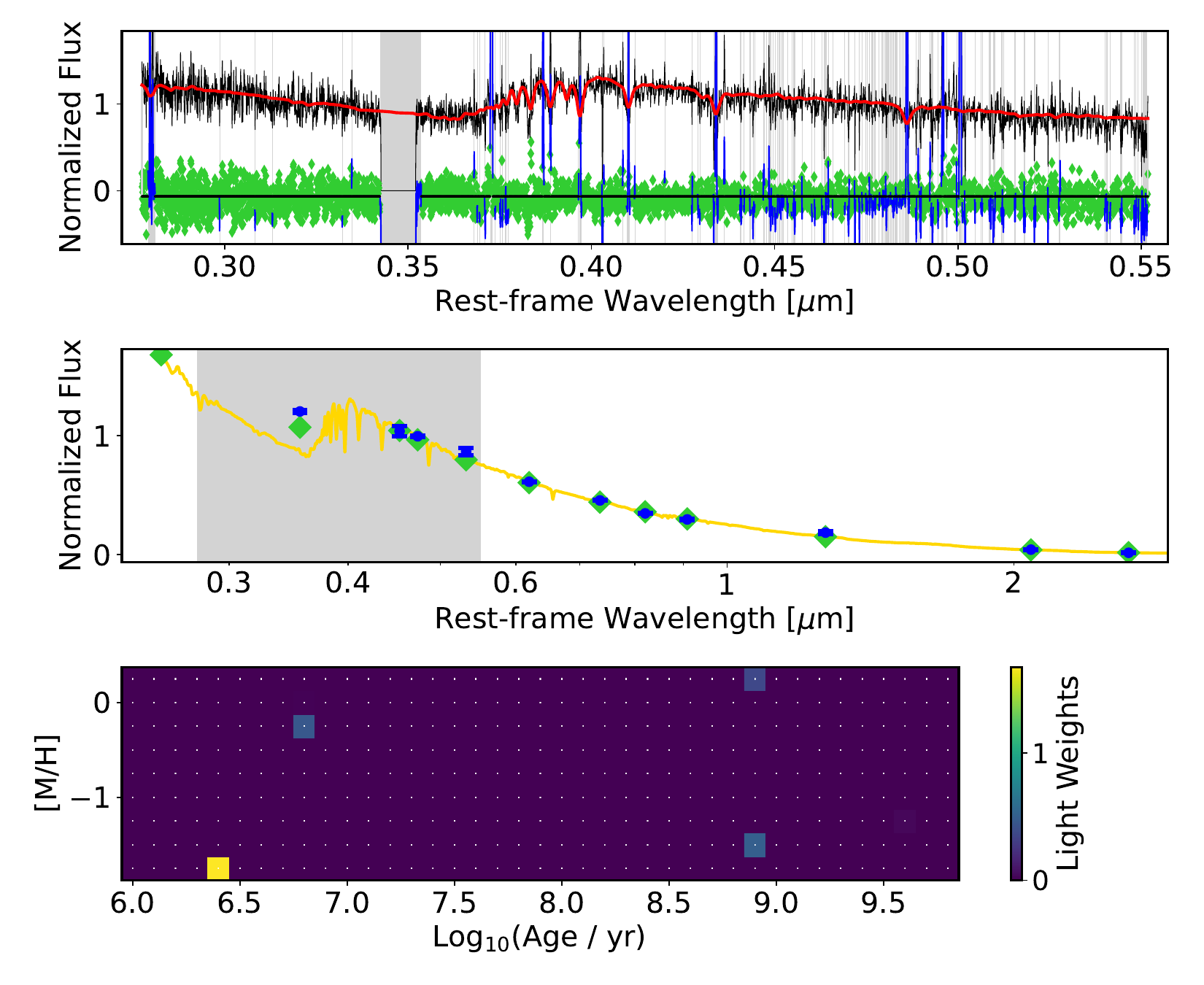}
    \end{subfigure}
    \hfill
    \begin{subfigure}[t]{0.49\textwidth}
        \centering
        AS1063-MF3-091
        \par\smallskip
        \includegraphics[width=\linewidth]{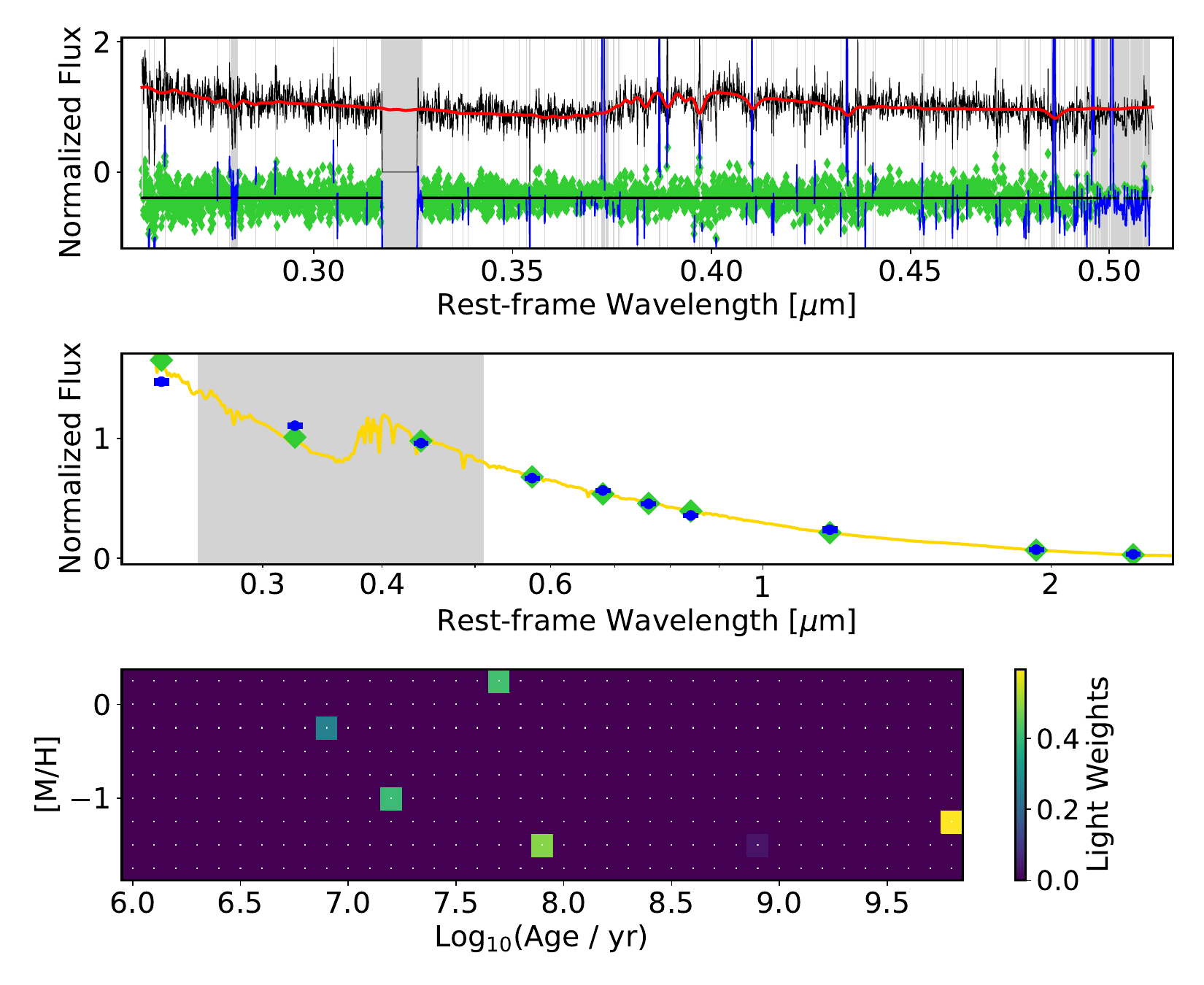}
    \end{subfigure}

    \vspace{0.3em}

    \begin{subfigure}[t]{0.49\textwidth}
        \centering
        M0416-DF-024
        \par\smallskip
        \includegraphics[width=\linewidth]{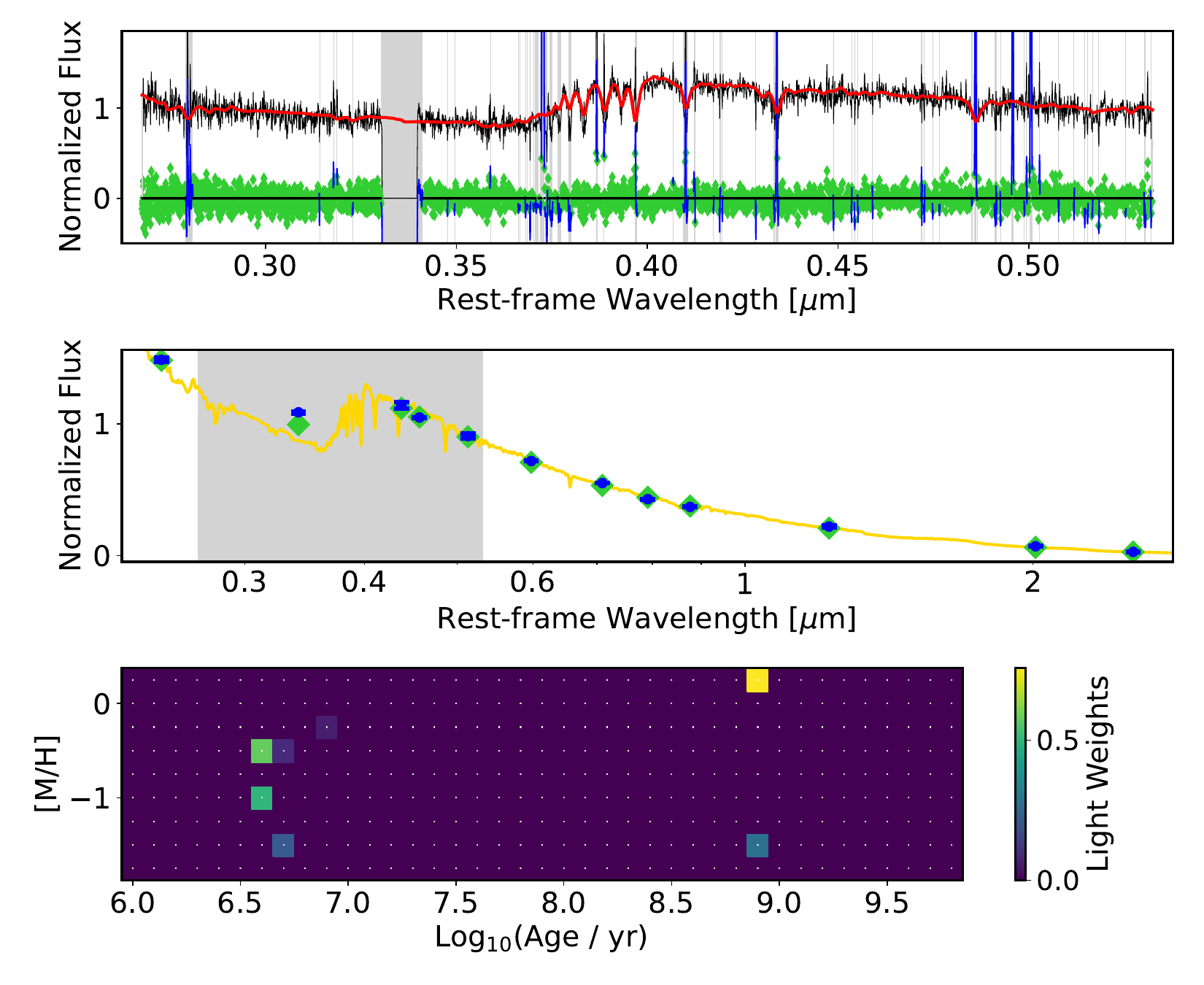}
    \end{subfigure}
    \hfill
    \begin{subfigure}[t]{0.49\textwidth}
        \centering
        A370-DF-099
        \par\smallskip
        \includegraphics[width=\linewidth]{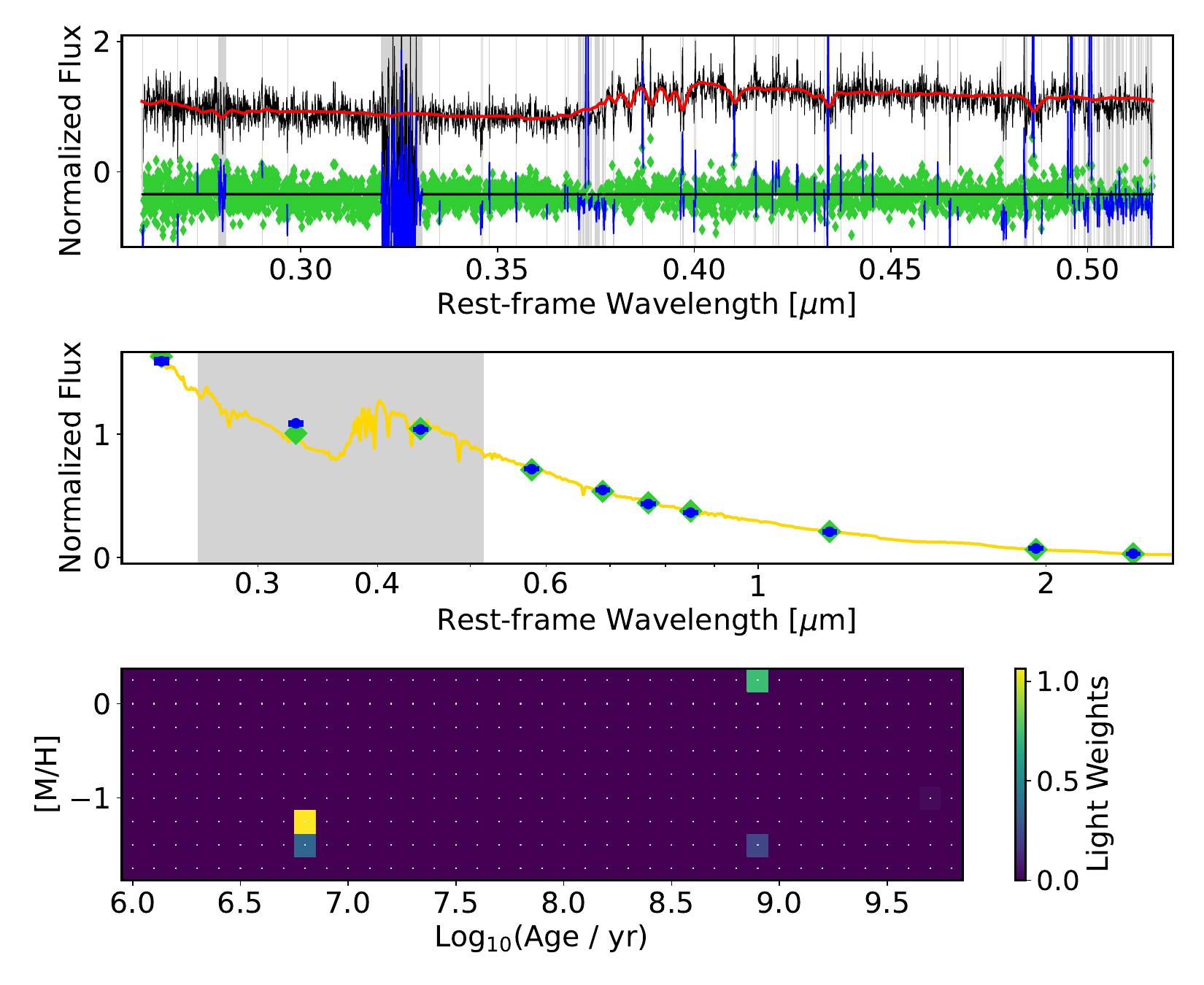}
    \end{subfigure}

    \vspace{0.3em}

    \begin{subfigure}[t]{0.49\textwidth}
        \centering
        M0416-DF-295
        \par\smallskip
        \includegraphics[width=\linewidth]{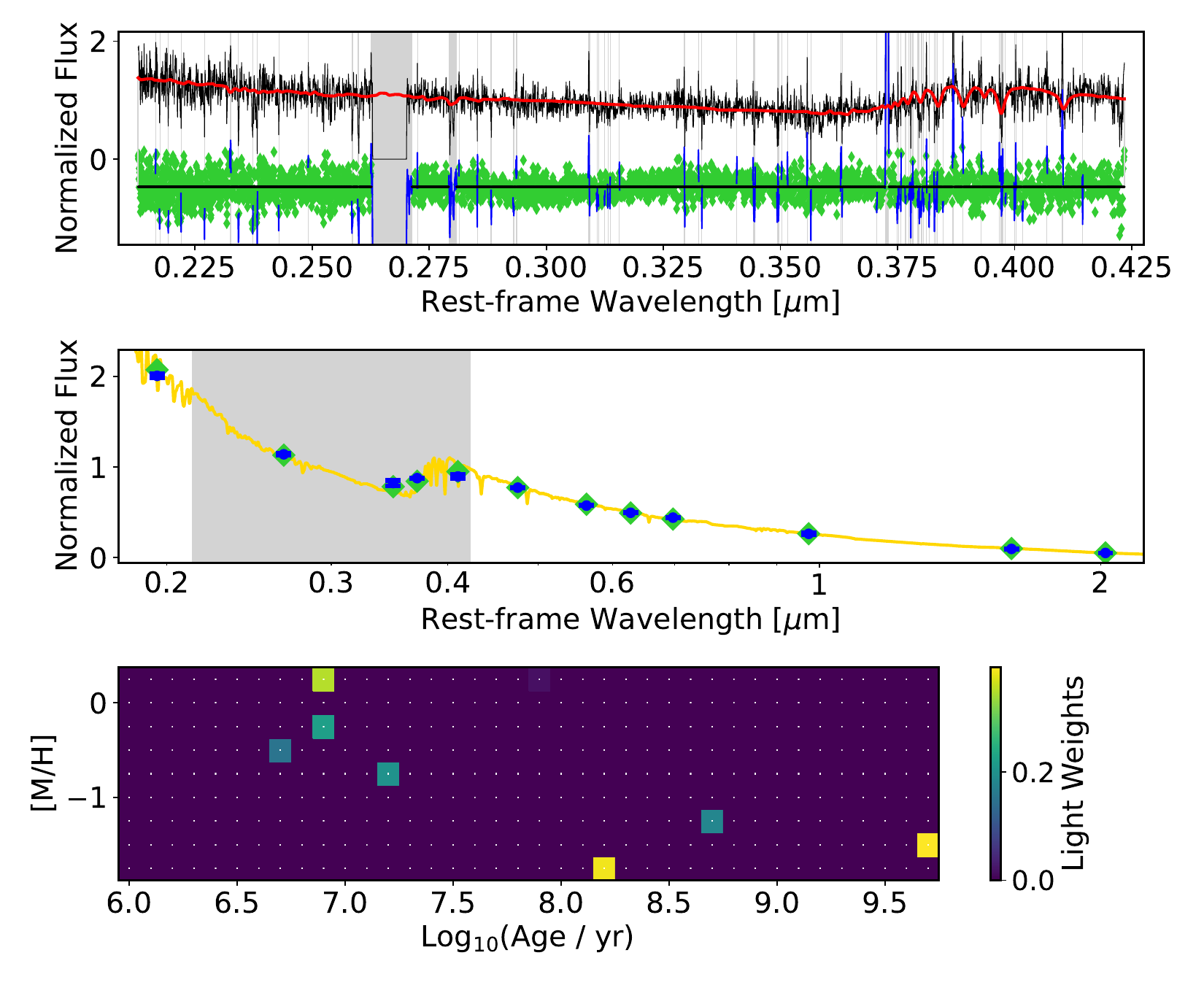}
    \end{subfigure}
    \hfill
    \begin{subfigure}[t]{0.49\textwidth}
        \centering
        A2744-DF-002
        \par\smallskip
        \includegraphics[width=\linewidth]{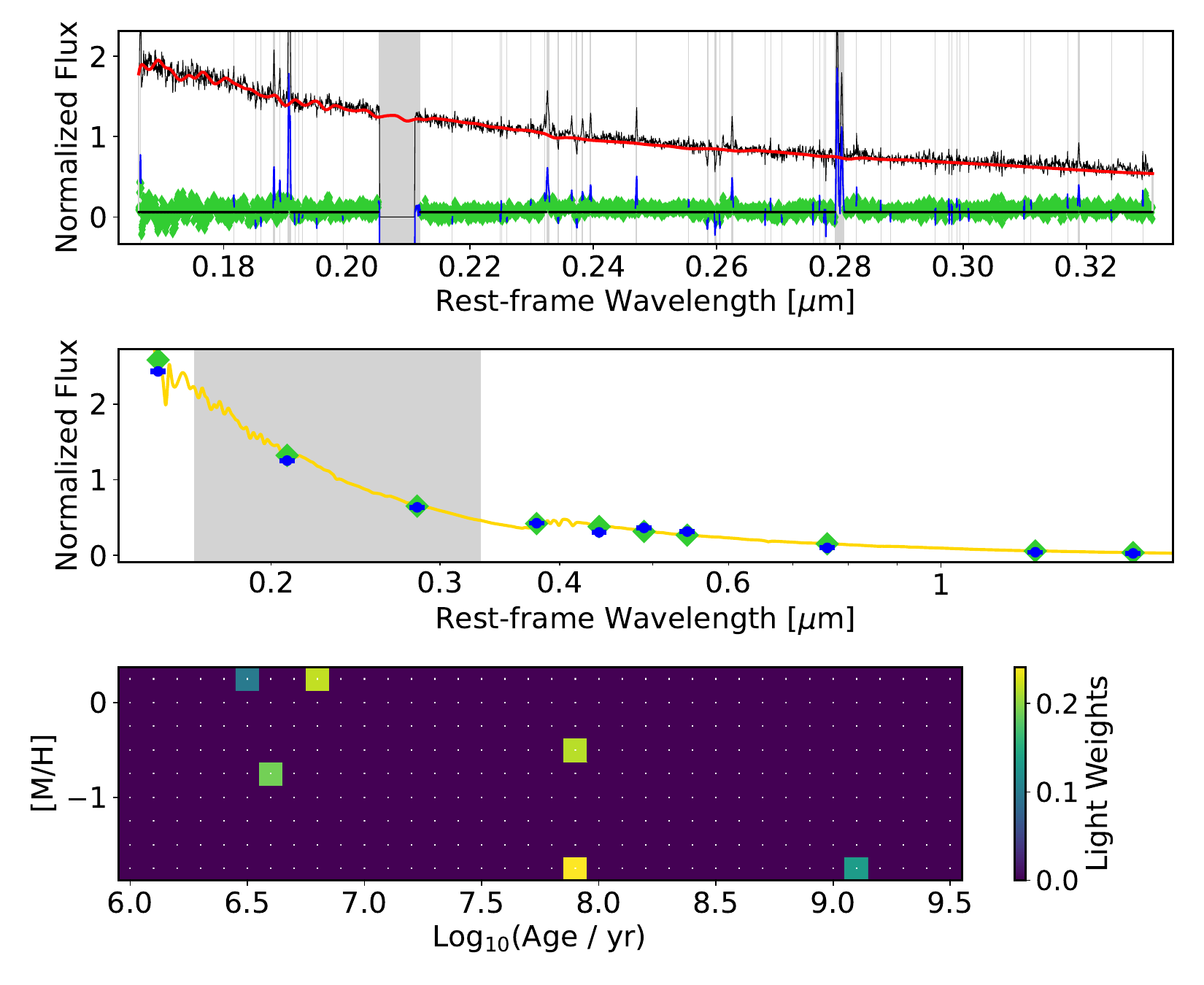}
    \end{subfigure}

    \caption{Spectral and photometric fits for the galaxies in our final sample, ordered by stellar mass in ascending order, as in Table~\ref{tab:sample}. The panels for each galaxy
    follow the same format as Fig.~\ref{fig:fit_example}.}
    \label{fig:all_SFH_fits}
\end{figure*}

\begin{figure*}[p]
    \ContinuedFloat
    \centering

    \begin{subfigure}[t]{0.49\textwidth}
        \centering
        A2744-DF-202
        \par\smallskip
        \includegraphics[width=\linewidth]{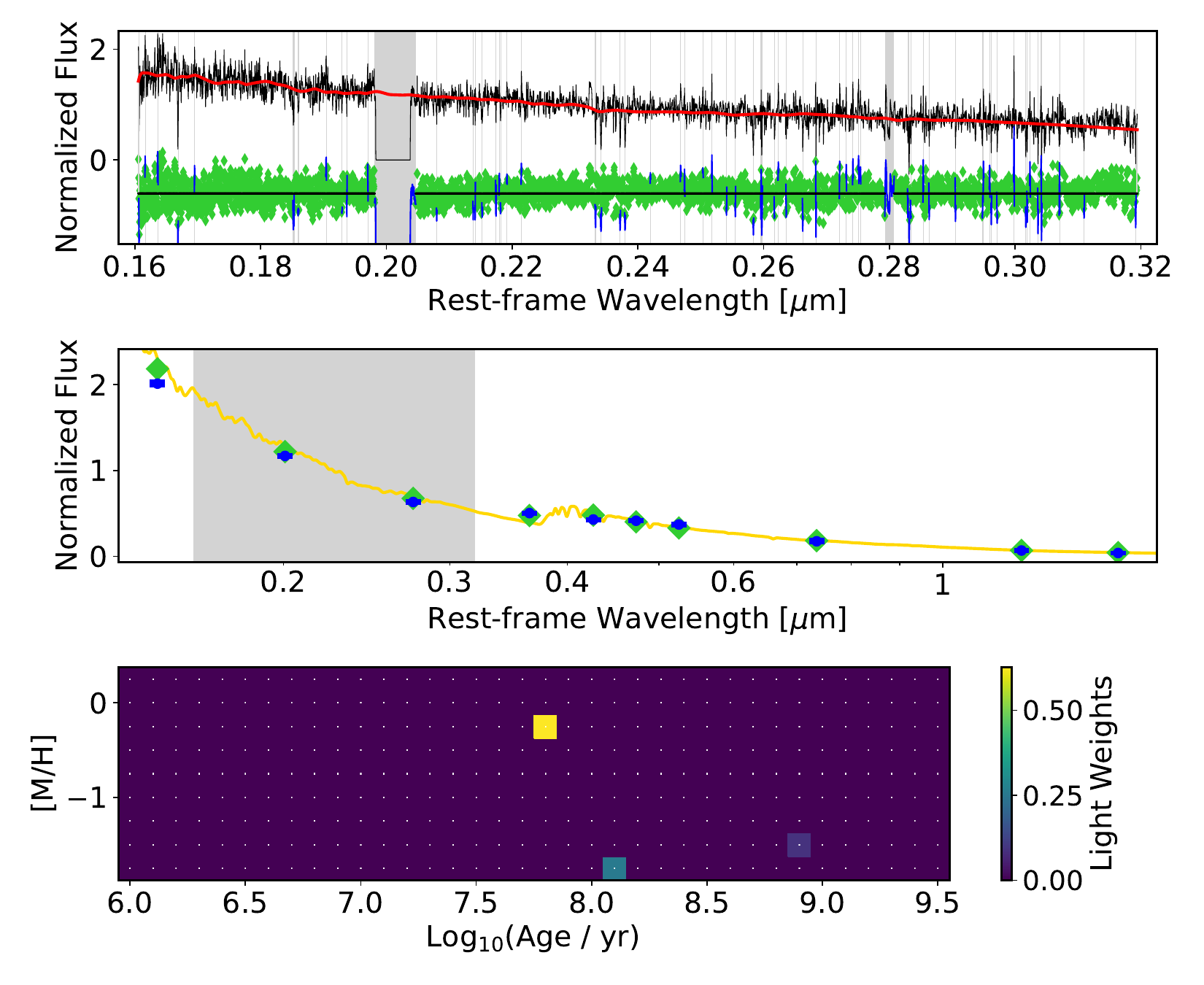}
    \end{subfigure}
    \hfill
    \begin{subfigure}[t]{0.49\textwidth}
        \centering
        AS1063-DF-240
        \par\smallskip
        \includegraphics[width=\linewidth]{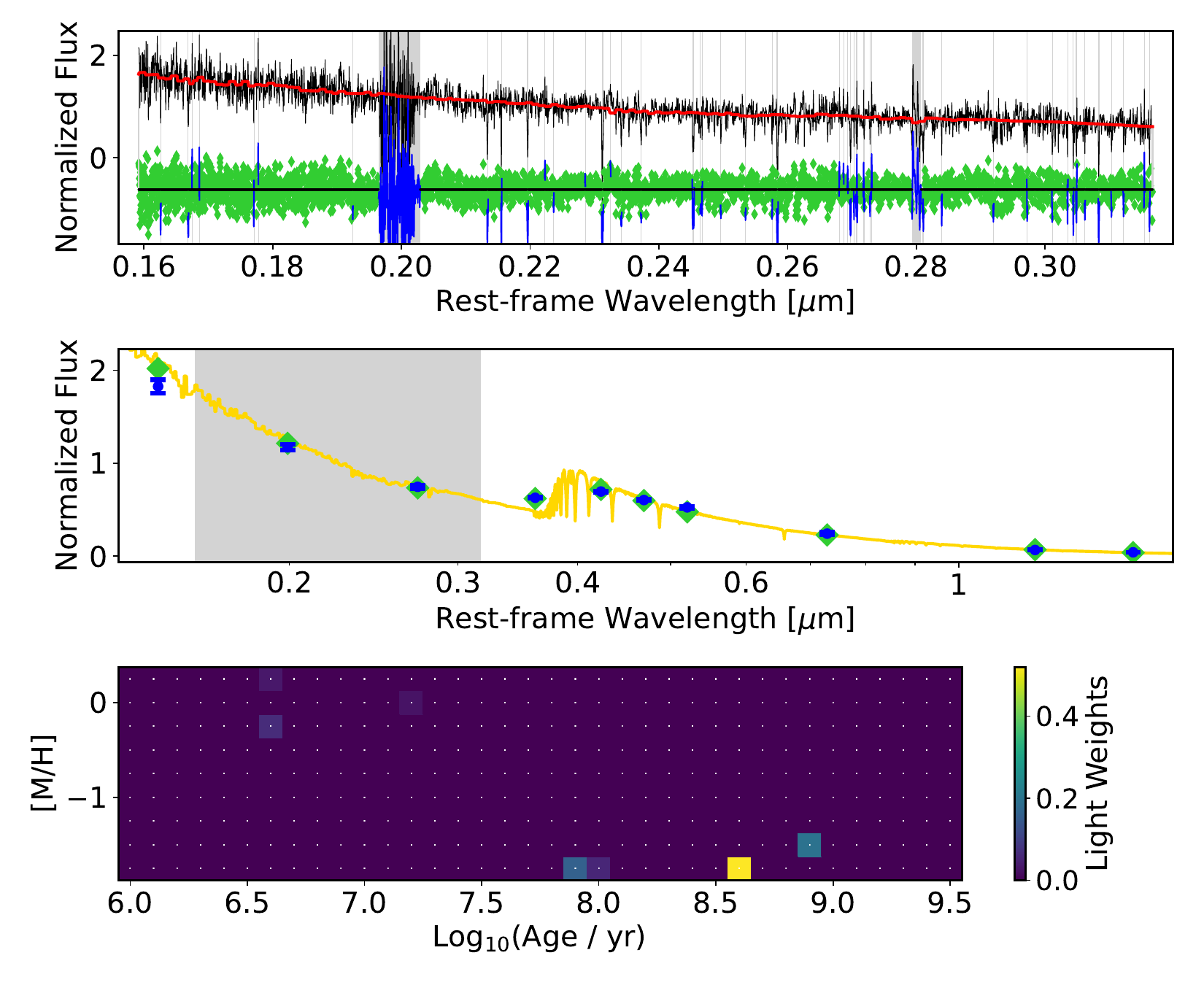}
    \end{subfigure}

    \vspace{0.3em}

    \begin{subfigure}[t]{0.49\textwidth}
        \centering
        M0416-SF7-012
        \par\smallskip
        \includegraphics[width=\linewidth]{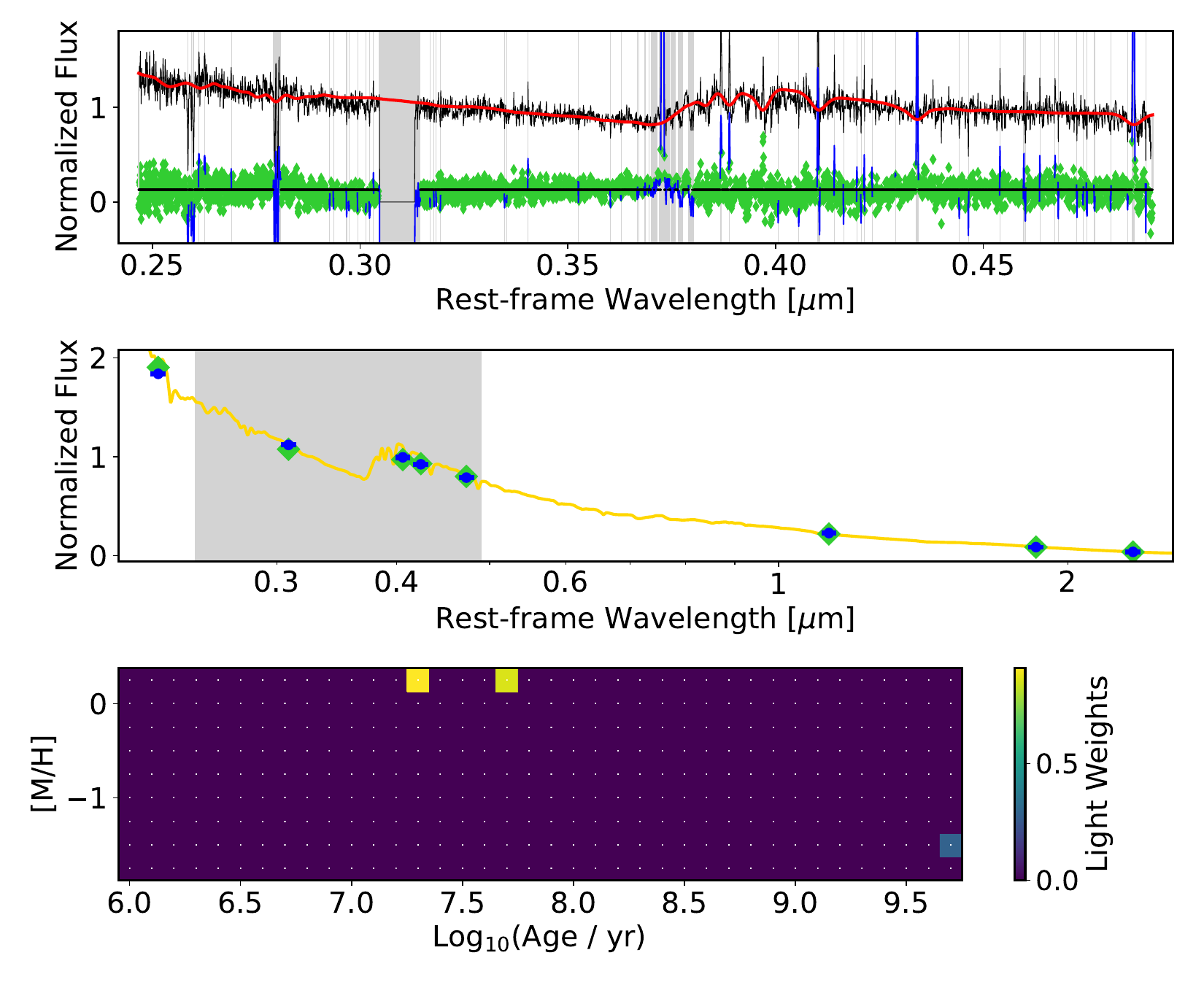}
    \end{subfigure}
    \hfill
    \begin{subfigure}[t]{0.49\textwidth}
        \centering
        A2744-DF-227
        \par\smallskip
        \includegraphics[width=\linewidth]{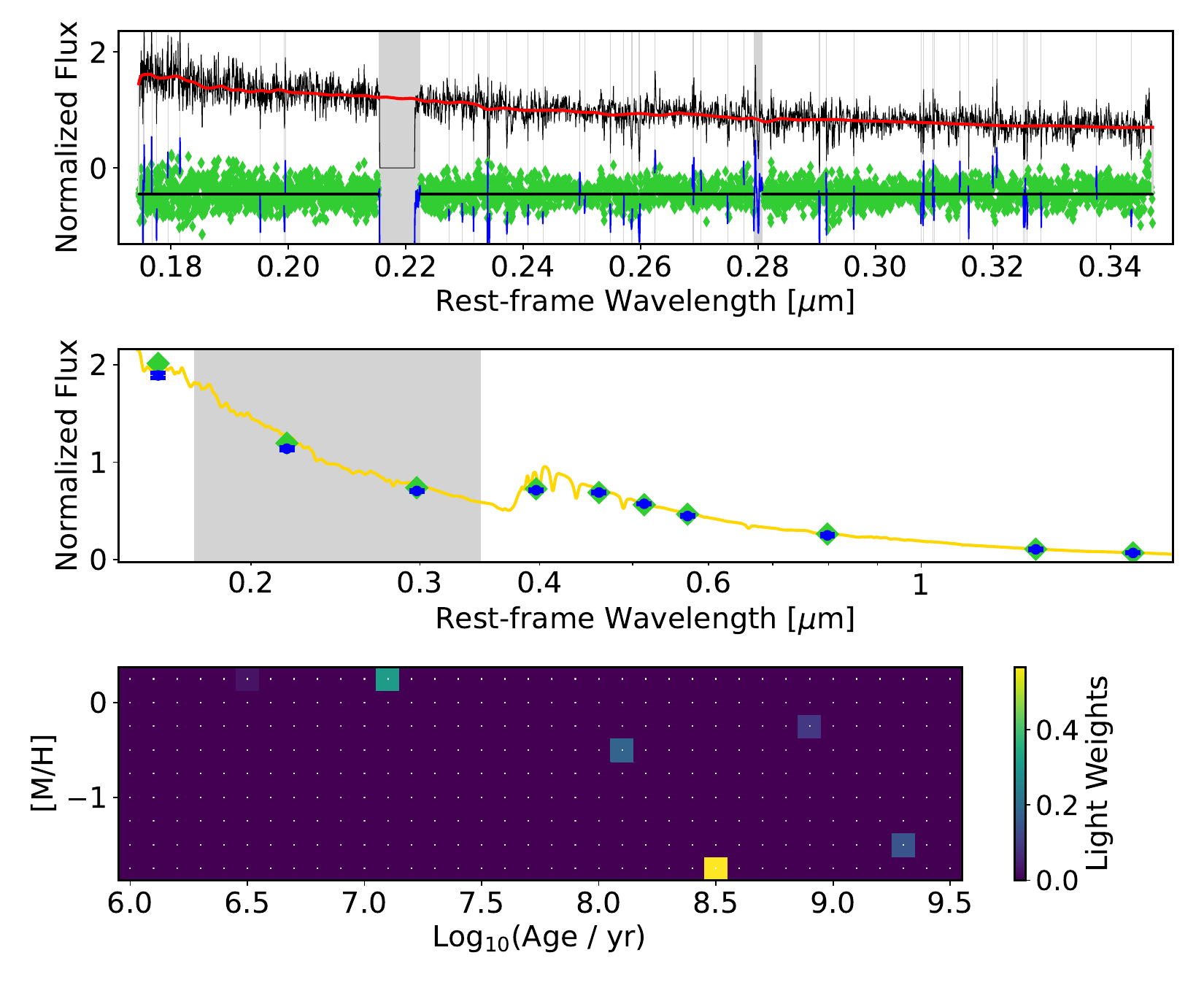}
    \end{subfigure}

    \vspace{0.3em}

    \begin{subfigure}[t]{0.49\textwidth}
        \centering
        AS1063-MF4-024
        \par\smallskip
        \includegraphics[width=\linewidth]{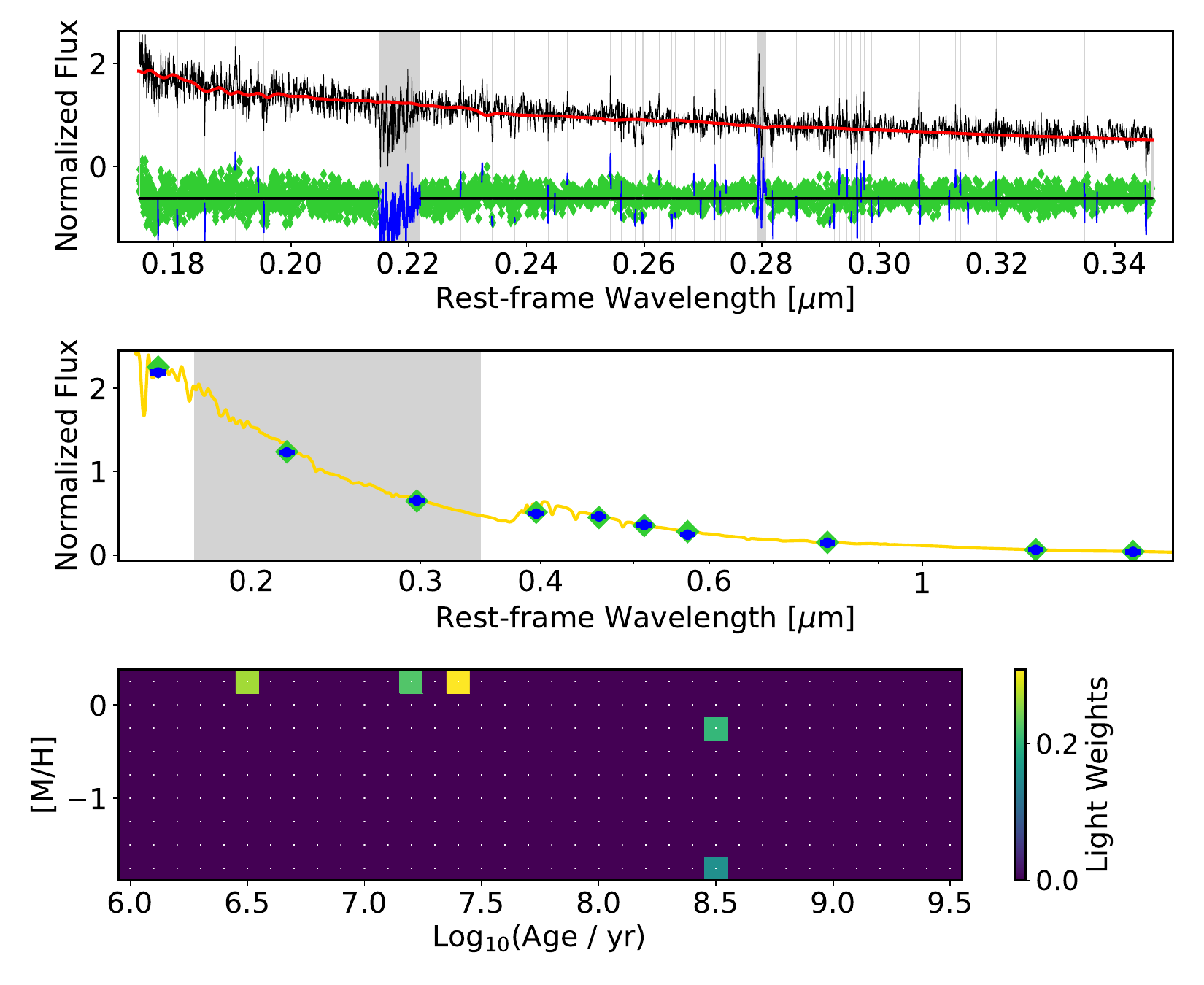}
    \end{subfigure}
    \hfill
    \begin{subfigure}[t]{0.49\textwidth}
        \centering
        AS1063-SF1-015
        \par\smallskip
        \includegraphics[width=\linewidth]{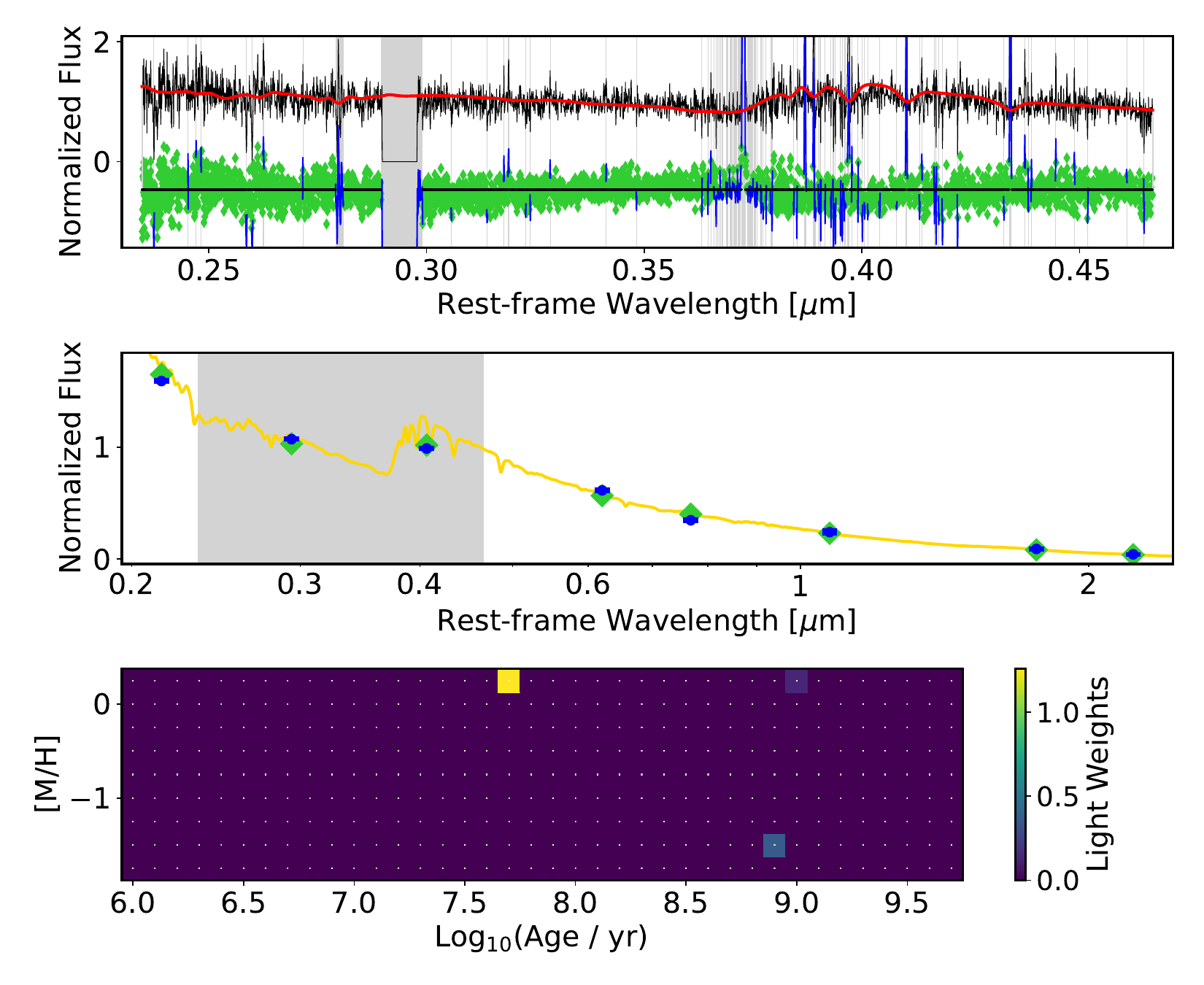}
    \end{subfigure}

    \caption{Continued.}
\end{figure*}

\begin{figure*}[p]
    \ContinuedFloat
    \centering

    \begin{subfigure}[t]{0.49\textwidth}
        \centering
        M0416-MF3-043
        \par\smallskip
        \includegraphics[width=\linewidth]{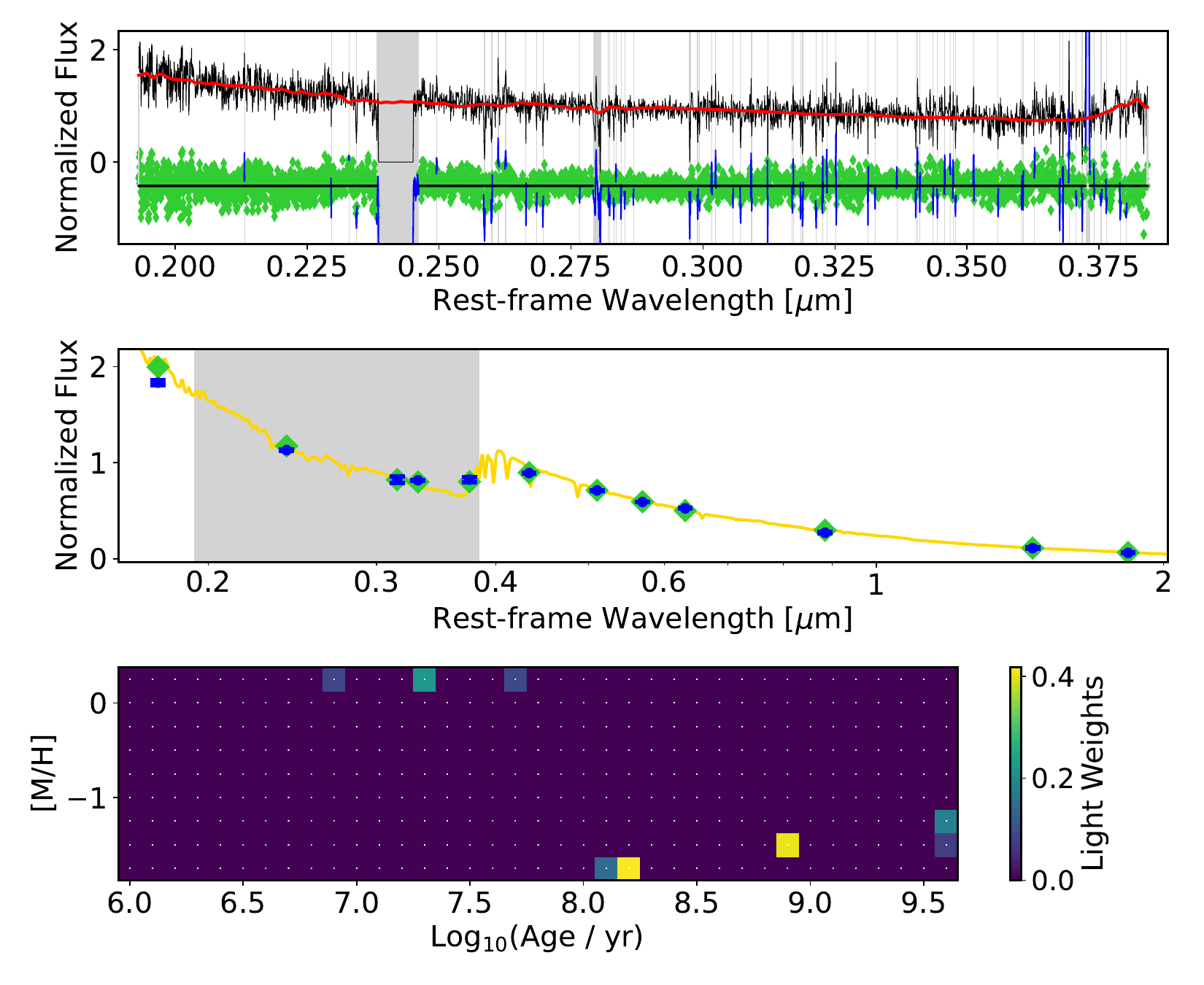}
    \end{subfigure}
    \hfill
    \begin{subfigure}[t]{0.49\textwidth}
        \centering
        AS1063-MF3-004
        \par\smallskip
        \includegraphics[width=\linewidth]{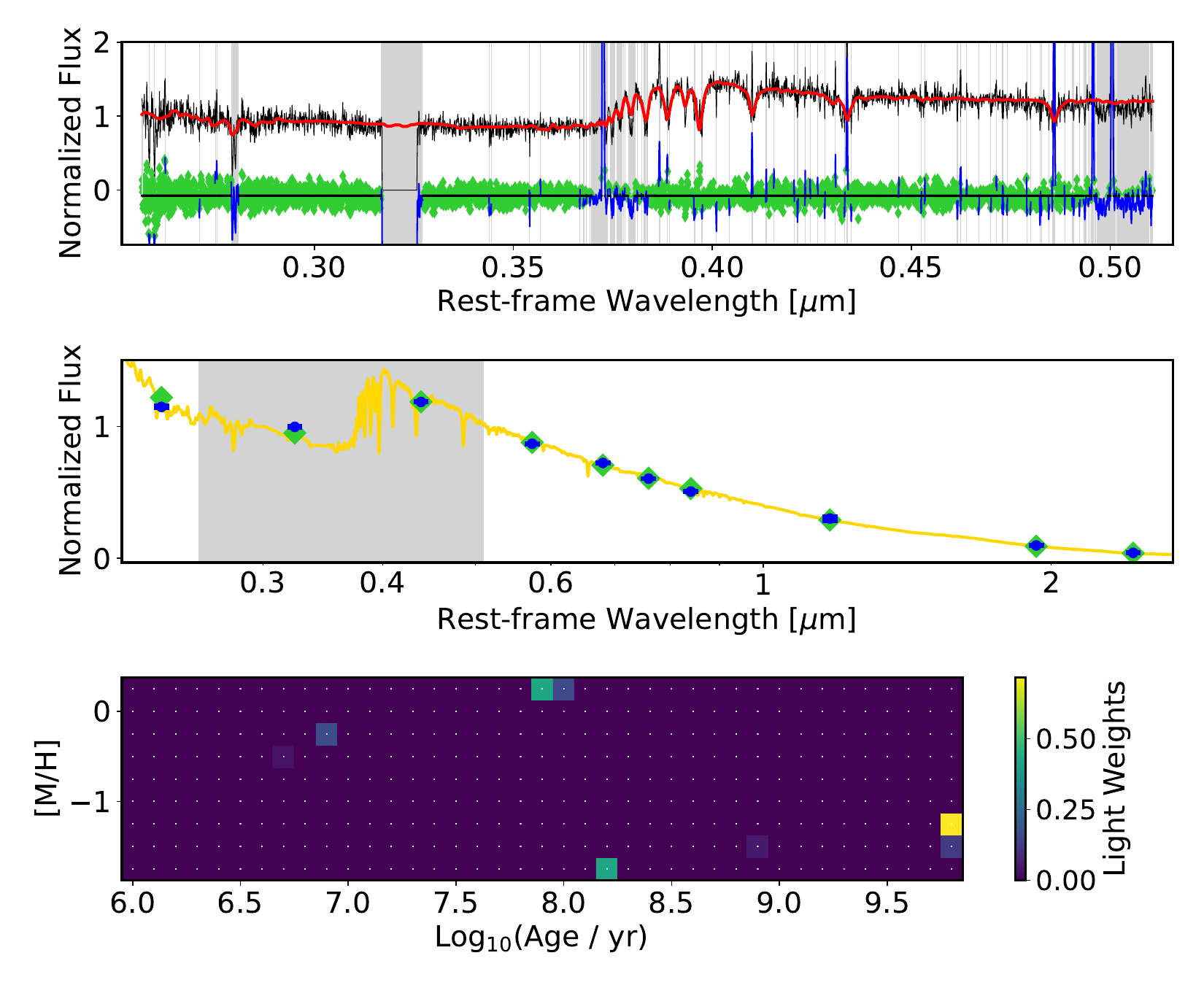}
    \end{subfigure}

    \vspace{0.3em}

    \begin{subfigure}[t]{0.49\textwidth}
        \centering
        M0416-DF-032
        \par\smallskip
        \includegraphics[width=\linewidth]{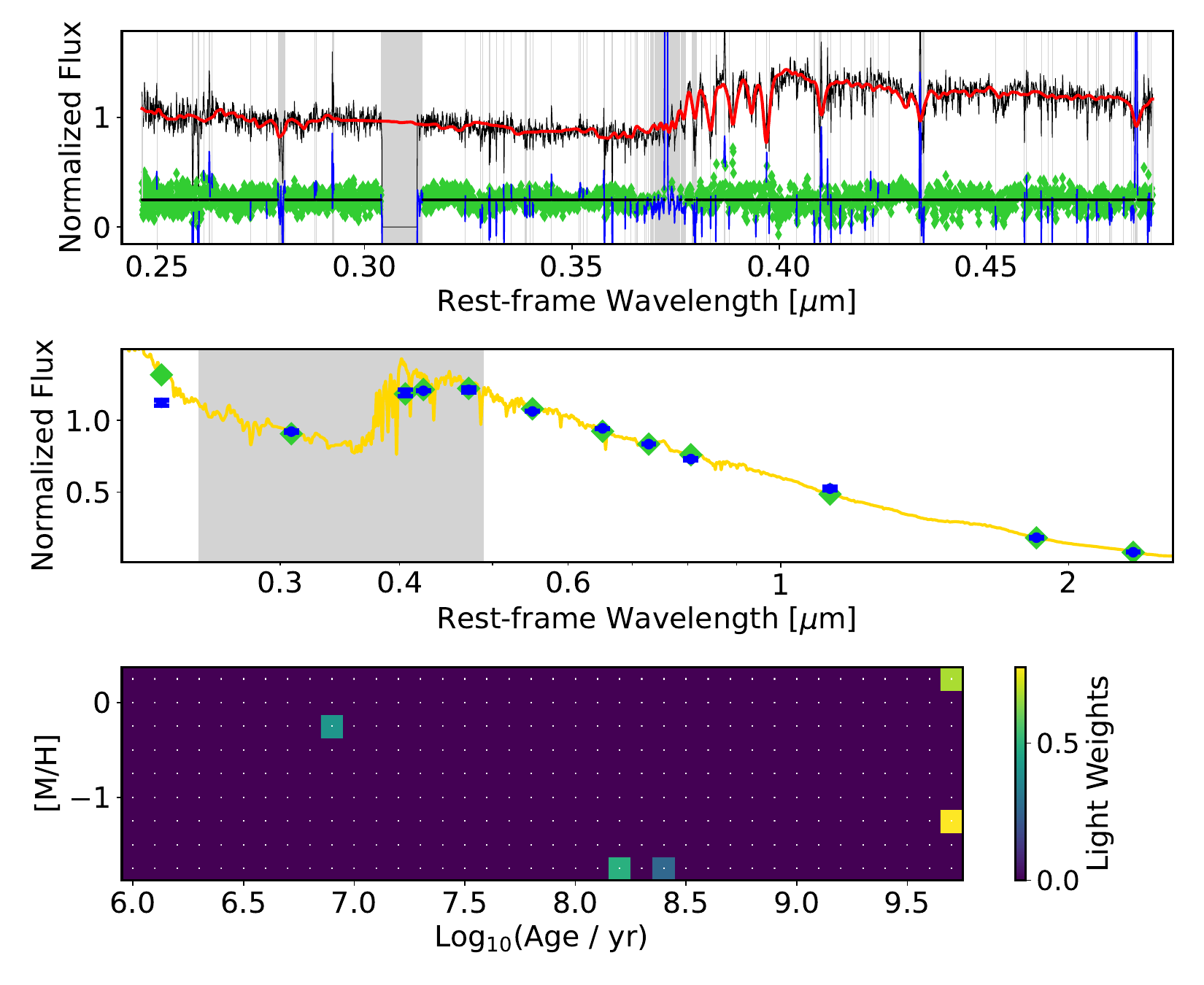}
    \end{subfigure}
    \hfill
    \begin{subfigure}[t]{0.49\textwidth}
        \centering
        AS1063-MF3-009
        \par\smallskip
        \includegraphics[width=\linewidth]{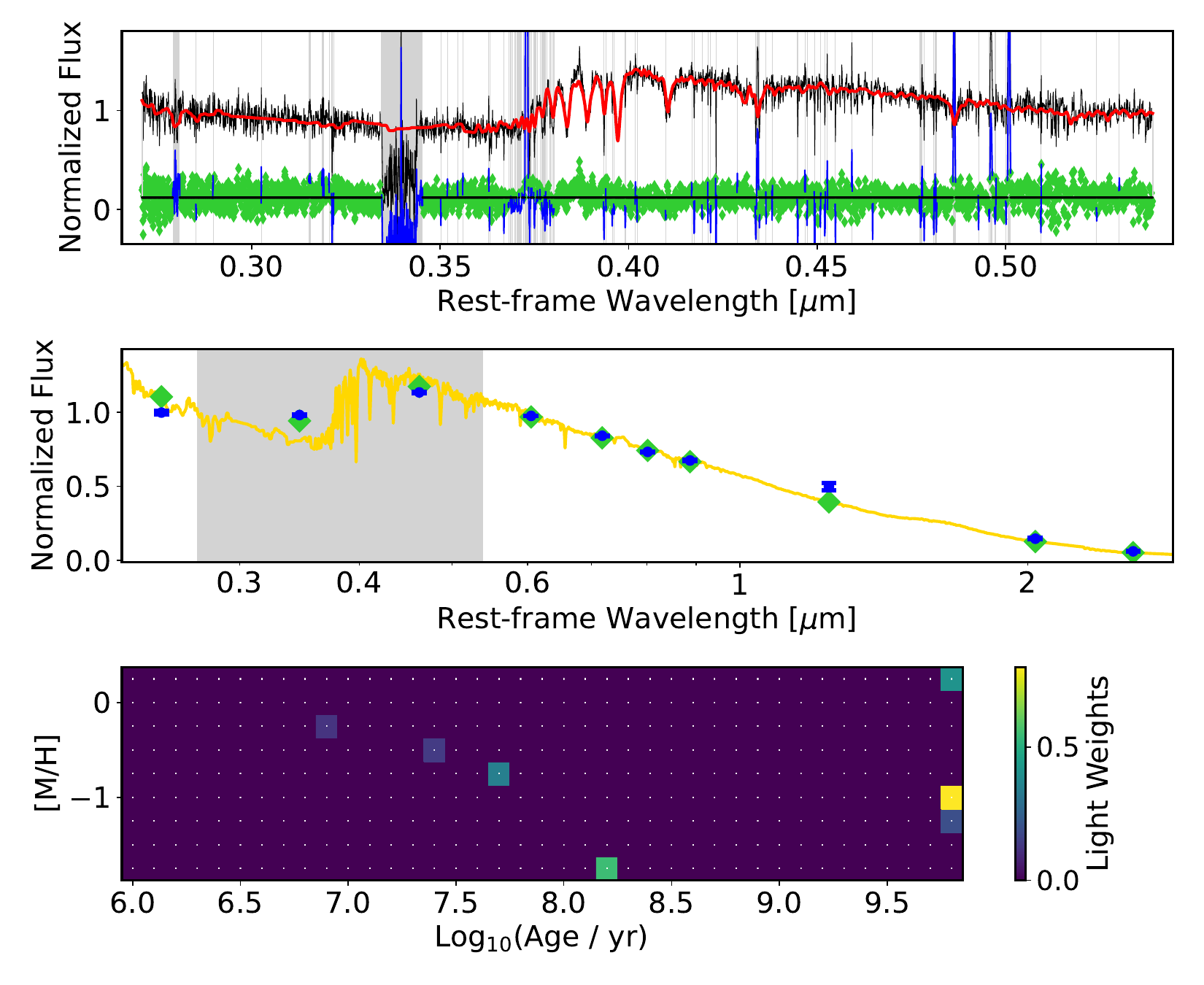}
    \end{subfigure}

    \vspace{0.3em}

    \begin{subfigure}[t]{0.49\textwidth}
        \centering
        AS1063-DF-029
        \par\smallskip
        \includegraphics[width=\linewidth]{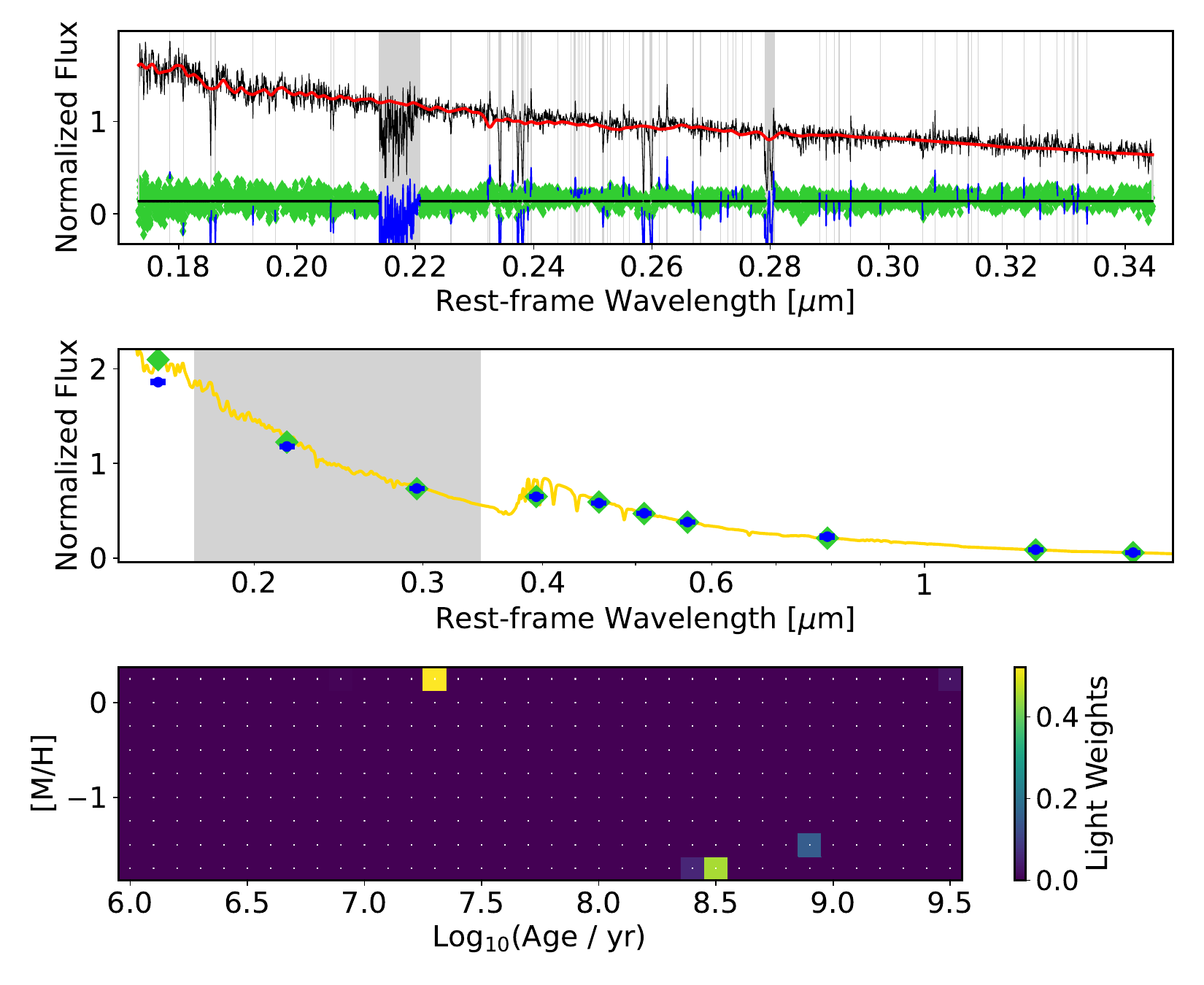}
    \end{subfigure}
    \hfill
    \begin{subfigure}[t]{0.49\textwidth}
        \centering
        M0416-DF-147
        \par\smallskip
        \includegraphics[width=\linewidth]{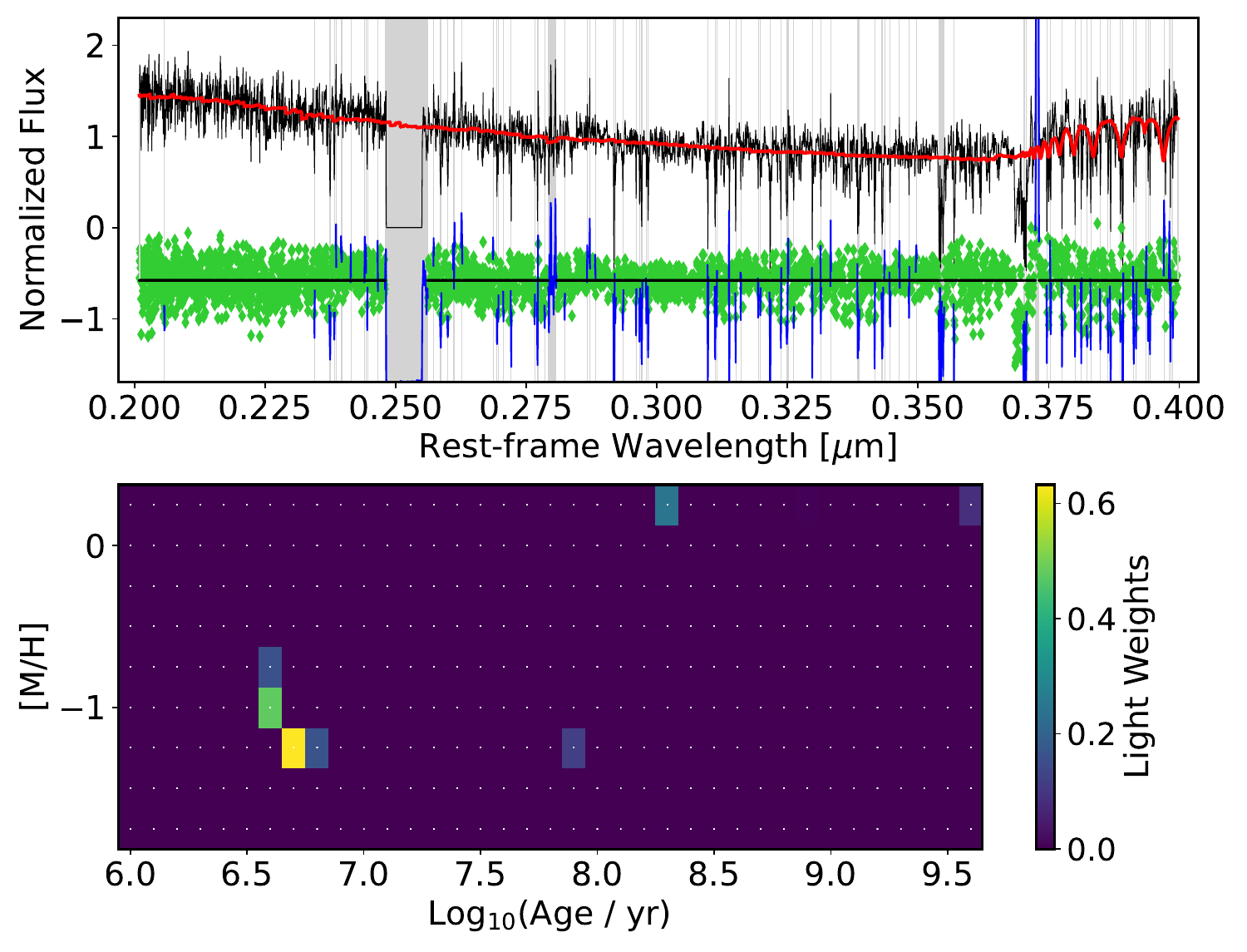}
    \end{subfigure}

    \caption{Continued.}
\end{figure*}

\begin{figure*}[p]
    \ContinuedFloat
    \centering

    \begin{subfigure}[t]{0.49\textwidth}
        \centering
        AS1063-MF2-011
        \par\smallskip
        \includegraphics[width=\linewidth]{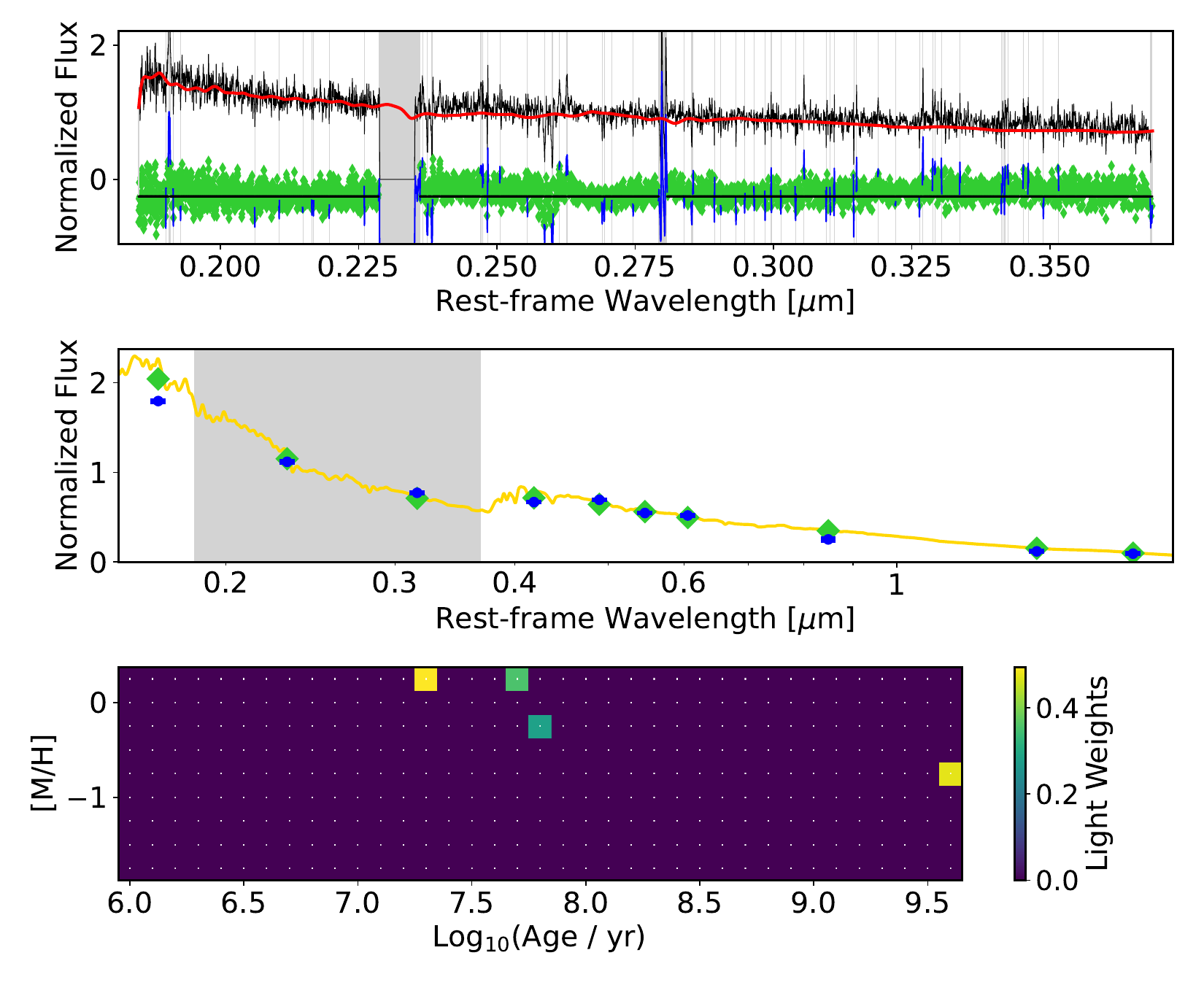}
    \end{subfigure}
    \hfill
    \begin{subfigure}[t]{0.49\textwidth}
        \centering
        AS1063-MF3-055
        \par\smallskip
        \includegraphics[width=\linewidth]{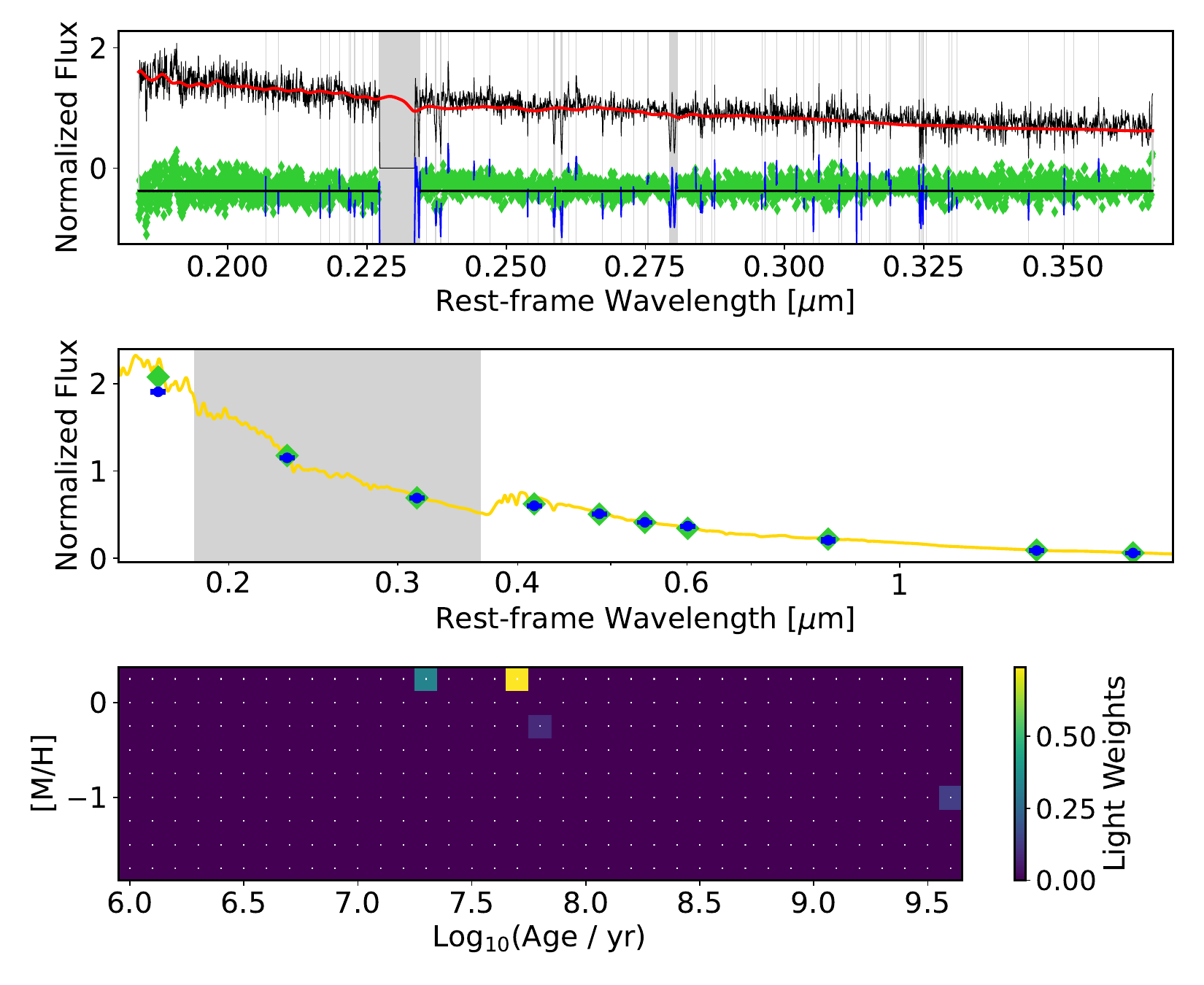}
    \end{subfigure}

    \caption{Continued.}
\end{figure*}

\end{appendix}

\end{document}